\documentclass[10pt,twocolumn,letterpaper]{article}

\usepackage[pagenumbers]{cvpr} 
\usepackage{algorithm}
\usepackage{amssymb}
\usepackage[T1]{fontenc}
\usepackage{algpseudocode}
\usepackage{tabularx}
\usepackage{listings}
\usepackage{subcaption} 

\definecolor{paperblue}{rgb}{0.21,0.49,0.74}
\usepackage[breaklinks,colorlinks,allcolors=paperblue]{hyperref} 

\usepackage{bibunits}
\defaultbibliographystyle{ieeenat_fullname}
\defaultbibliography{references} 

\title{THOR: A Versatile Foundation Model for Earth Observation Climate and Society Applications}

\author{
Theodor Forgaard\\
Norwegian Computing Center \\
Oslo, Norway\\
{\tt\small tforgaard@nr.no}
\and
Jarle Reksten \\
Norwegian Computing Center \\
Oslo, Norway\\
{\tt\small jarlebh@nr.no} 
\and
Anders Waldeland\\
Norwegian Computing Center \\
Oslo, Norway\\
{\tt\small andersuw@nr.no} 
\and
Valerio Marsocci\\
European Space Agency $\Phi$-lab\\
Frascati, Italy\\
{\tt\small valerio.marsocci@esa.int}
\and
Nicolas Long\'ep\'e\\
European Space Agency $\Phi$-lab\\
Frascati, Italy\\
{\tt\small nicolas.longepe@esa.int}
\and
Michael Kampffmeyer \\
UiT - The Arctic University of Tromsø \\
Tromsø, Norway\\
{\tt\small michael.c.kampffmeyer@uit.no} 
\and
Arnt-Børre Salberg \\
Norwegian Computing Center \\
Oslo, Norway\\
{\tt\small salberg@nr.no}
}

\begin{document}
\maketitle
\begin{abstract}
Current Earth observation foundation models are architecturally rigid, struggle with heterogeneous sensors and are constrained to fixed patch sizes. This limits their deployment in real-world scenarios requiring flexible compute-accuracy trade-offs. We propose THOR, a "compute-adaptive" foundation model that solves both input heterogeneity and deployment rigidity. THOR is the first architecture to unify data from Copernicus Sentinel-1, -2, and -3 (OLCI \& SLSTR) satellites, processing their native 10 m to 1000 m resolutions in a single model. We pre-train THOR with a novel randomized patch and input image size strategy. This allows a single set of pre-trained weights to be deployed at inference with any patch size, enabling a dynamic trade-off between computational cost and feature resolution without retraining. We pre-train THOR on THOR Pretrain, a new, large-scale multi-sensor dataset and demonstrate state-of-the-art performance on downstream benchmarks, particularly in data-limited regimes like the PANGAEA 10\% split, validating that THOR's flexible feature generation excels for diverse climate and society applications.

\end{abstract}

\begin{bibunit}

\section{Introduction}
Earth Observation (EO) enables large-scale monitoring of Earth's systems (e.g., \cite{hollmann2013esa, mccabe2017hydrology, hansen2013forest, reichstein2019deep}), but this presents a monumental computer vision challenge. 
Foundation models (FM) promise to solve EO \cite{Longepe_2025}, but simply applying models pre-trained on standard natural images is often sub-optimal \cite{rolf2024position} as one must ingest a vast, heterogeneous data stream from diverse sensors (e.g., optical, SAR) at scales from meters to kilometers ground sampling distance (GSD). 

Most current EO-specific FMs  (e.g., \cite{szwarcman2024prithvi, jakubik2025terramind, fuller2023croma, xiong_dofa_neural_2024, wang_copernicus_fm_2025}), often built on Vision Transformers (ViT), are architecturally rigid. They are trained using a fixed input image size and a fixed patch size (e.g., $16 \times 16$), which creates a critical bottleneck for data-efficient adaptation: coarse patching produces a low-resolution token sequence. Consequently, dense pixel-level tasks like segmentation require large, complex decoders (e.g., UperNet \cite{ruiping2024vitupernet}) to upsample the features. These decoders often demand significant amounts of labeled data for fine-tuning, undermining the core data-efficiency promise of FMs. 

To address these shortcomings, we propose THOR (Transformer based foundation model for Heterogeneous Observation and Resolution), a versatile multi-modal foundation model designed for flexibility. THOR is the first architecture to both unify the 10 m - 1000 m GSD range of Sentinel-1, -2, and -3 (including the SLSTR sensor) and integrate a compute-adaptive patching strategy, solving both input heterogeneity and deployment rigidity simultaneously. By incorporating a randomized patch size and input image size during pre-training, THOR becomes "compute-adaptive". A single set of weights can be deployed with various patch sizes and input image sizes. This allows a user to select a smaller patch size at inference time, producing a denser, higher-resolution token sequence that can be processed by simpler, less data-hungry decoders. 
This increased detail is crucial for tasks requiring high-resolution understanding, such as fine-grained classification, and allows the dense representations to be paired with simpler, lightweight decoders. Such lightweight decoders are especially useful for cases with limited training data, as they reduce the risk of overfitting compared to heavier decoder architectures. Conversely, selecting a lower token density significantly decreases the ViT memory and compute requirements, making it more applicable for global-scale tasks like climate trend analysis and ocean monitoring, or scenarios where sufficient training data is available to support larger, more complex decoders.
Moreover, the multi-sensor integration allows THOR to leverage synergistic information: the all-weather radar sensing capability from Sentinel-1, the rich spectral detail of optics from Sentinel-2, and the broad-scale climate context from the Sentinel-3 OLCI and SLSTR instruments, all within a single, cohesive model. 

To enable a model to learn this compute-adaptive, multi-resolution capability, we created the THOR Pretrain dataset, a new, large-scale dataset of 22TB that has been aligned spatio-temporally and across modalities. It is the first to unify data from Sentinel-1, -2, and Sentinel-3 (both OLCI and SLSTR) satellites, processing their data at native resolutions from 10 m to 1000 m. THOR Pretrain also contains diverse land cover products, digital elevation models (DEM), and ERA5-Land variables.

In summary, our key contributions are as follows:
\begin{itemize}
    \item A flexible, multi-sensor architecture that is the first FM to unify Sentinel-1 SAR, Sentinel-2 MSI, and Sentinel-3 OLCI \& SLSTR data from 10 m - 1000 m GSD, built on a compute-adaptive backbone (Sec.~\ref{sec:encoder}).
    \item A novel multi-modal pre-training framework that extends the flexible patching to a MAE setup, combining pixel-level reconstruction with pretext tasks for land cover and climate variables (Sec.~\ref{sec:loss_formulation}).
    \item THOR Pretrain: A new, large-scale and diverse multi-modal EO dataset, curated with a novel  sampling strategy to ensure geographic and thematic diversity (Sec.~\ref{sec:thor_pretrain}).
\end{itemize}
We demonstrate that this co-design achieves state-of-the-art performance in limited training data regimes (Sec.~\ref{sec:experiments}).

\section{Related work}
\textbf{Self-supervised pre-training strategies in EO.}
Recent EO FMs leverage self-supervised learning (SSL) \cite{wang2022self}, primarily via Masked Autoencoders (MAE) (e.g., Prithvi-EO-2.0 \cite{szwarcman2024prithvi}, MMEarth \cite{nedungadi2024mmearth}, SatMAE \cite{cong2022satmae}) and hybrid contrastive methods (e.g., CROMA \cite{fuller2023croma}). While powerful, these models are architecturally rigid. Prithvi-EO-2.0 is pre-trained exclusively on 30 m GSD data \cite{szwarcman2024prithvi}, MMEarth adopts a "resample-to-grid" strategy, harmonizing all data to a 10 m grid and discarding native resolution information \cite{nedungadi2024mmearth}, and CROMA adopts a contrastive objective for radar-optical sensor invariance with an MAE reconstruction objective \cite{fuller2023croma}. They are all built on fixed patch sizes (e.g., $16 \times 16$ or $8 \times 8$). This locks in a specific computational profile and, as argued in the introduction, necessitates large, data-hungry decoders for dense pixel-level tasks. 

\noindent\textbf{Architectural solutions for input heterogeneity.}
State-of-the-art models like TerraMind \cite{jakubik2025terramind}, DOFA \cite{xiong_dofa_neural_2024}, and Copernicus-FM \cite{wang_copernicus_fm_2025} employ sophisticated input data processing strategies, such as TerraMind's "dual-scale early fusion" for nine modalities or DOFA's wavelength dependent "dynamic weight generator" that functions as a flexible translation layer for heterogeneous sensor data. 
AnySat \cite{astruc2025anysat} achieves this versatility by utilizing scale-adaptive spatial encoders and introducing Joint Embedding Predictive Architecture \cite{assran2023jepa} for multi-modal EO data and leverages the spatial alignment of multiple modalities as a source of self-supervision.
Copernicus-FM unifies all major Copernicus Sentinel missions (Sentinel-1 SAR, Sentinel-2 MSI, Sentinel-3 OLCI, Sentinel-5P), spanning the full 10 m to 1000 m GSD range \cite{wang_copernicus_fm_2025}. Its "extended dynamic hypernetwork" generates weights based on sensor metadata, creating an "input-flexible" model. However, this flexibility is primarily focused on handling diverse inputs rather than deployment versatility. 
Scale-MAE \cite{reed2023scale} modifies the positional encoding to be "scale-aware" by scaling its positional encoding by the image's GSD. However, it only handles one modality at a time, and its fixed patch size leads to inconsistent sequence lengths and computational loads for the same ground area. USat \cite{irvin2023usat} uses separate patch projection layers for different bands. While this effectively ingests multi-modal data, USat's architecture remains rigid. Its "superpositional encoding" scheme is structurally constrained and, most importantly, incompatible with a flexible patching strategy required for compute-adaptive inference.
While these models represent the best efforts to handle diverse inputs, they are "deployment-rigid", with a fixed patch size and input image size during pre-training. For instance, Copernicus-FM is trained with a fixed image footprint, limiting Sentinel-5P images to only a few pixels.

\noindent\textbf{Architectural rigidity and adaptive models.}
FlexiViT \cite{beyer2023flexivit} demonstrated that by randomizing the patch size during pre-training, a single set of ViT weights can perform compute-adaptive inference, allowing users to select the preferred patch size during inference.
This flexible-patching concept is only beginning to be adopted in EO-specific FMs. 
Galileo \cite{tseng2025galileo} incorporates resizable patch embeddings from FlexiViT, and pairs this architectural flexibility with a dual-objective training strategy to ensure the learned features capture both the high-level semantic context (global) and fine-grained detail (local) critical for diverse EO tasks. Similarly, FlexiMo \cite{li2025fleximo} utilizes the FlexiViT strategy and includes a "wavelength-guided channel adaptation" module to handle multi-sensor inputs, allowing the pre-trained model to adapt to arbitrary spatial resolutions and maintain multi-scale feature fidelity. While these models are a key step towards deployment versatility, they are focused on Sentinel-1 and -2, without scaling to the full multi-resolution challenge (10 m – 1000 m) posed by sensors like Sentinel-3 OLCI \& SLSTR.

\noindent\textbf{The gap: synthesizing input and deployment.}
The related work reveals two powerful, yet until now, parallel lines of research. On one side, models like Copernicus-FM and USat solve input heterogeneity but are deployment-rigid. On the other, models like Galileo and FlexiMo solve deployment versatility but have not been scaled to the full multi-resolution (10 m - 1000 m) challenge. A critical gap therefore exists: no single architecture has solved both input heterogeneity and deployment versatility simultaneously.
Our work, THOR, is designed to be the first to fill this gap. This challenge is non-trivial, as it requires co-designing the positional encoding, per-band patch projection, and MAE loss function to be mutually compatible across a 100x GSD range (10 m to 1000 m).
We propose a new architecture that synthesizes state-of-the-art approaches for multi-sensor input with compute-adaptive patching, enabling a single model to operate efficiently across the full 10 m - 1000 m GSD range. We detail this architecture in Sec. \ref{sec:thor_model}.

\section{THOR Pretrain}\label{sec:thor_pretrain}
THOR is pre-trained on a new, diverse, and large-scale dataset named THOR Pretrain. This dataset is curated to learn representations that are robust to variations in global land cover, ocean phenomena, and cloud conditions.

THOR Pretrain unifies data from four major Copernicus Sentinel sensors: Sentinel-1 SAR, Sentinel-2 MSI, Sentinel-3 OLCI and SLSTR. These sensors provide diverse image modalities, including radar, multispectral and thermal, with resolutions ranging from 10 m to 1000 m. 

Instead of stacking millions of small image crops, we sample EO data using the Sentinel-2 tiles ($110 \times 110$ km) as the sampling grid. This grounds the samples in a well-known geographic unit. 
To ensure a rich, diverse dataset not biased towards common land covers, we employed a stratified sampling strategy based on k-means clustering of land cover and RGB features. This actively over-samples rare geographic and thematic classes. A total of 6273 globally distributed locations were sampled. 
For a sampled grid location and time, we acquired the corresponding Sentinel-2 data (Level 2A) along with overlapping Sentinel-1 SAR (GRD), Sentinel-3 OLCI (Level 1C) and SLSTR data. 
To ensure temporal consistency across modalities, we restrict the acquisition window for Sentinel-1 and Sentinel-3 imagery to the be within $\pm$ 1 days of the Sentinel-2 anchor timestamp for land areas and the same day for ocean areas.
Sentinel-3 data is selected from a nine times larger area to account for the coarser resolution. For each location, we also include the digital elevation model (DEM), and diverse land cover maps: WorldCover \cite{WorldCover2020_v100}, GlobCover \cite{GlobCover2009}, MODIS \cite{MCD12Q1_v061}, and ERA5-Land climate variables \cite{ERA5Land_dataset, MunozSabater2021_ERA5Land_ESSD}. 

To obtain temporal diversity in the dataset, each location is sampled a random number of times, leading to a total number of 18332 tile and date combinations, designed to support compute-adaptive pre-training and downstream generalization across diverse climate and social applications.  The total size of the dataset is approximately 22 TB. 
Full details on data processing, spatio-temporal alignment, and the exact sampling weights are provided in the Supplementary Material.

\section{THOR foundation model} 
\label{sec:thor_model}
As established in the related work, THOR is the first architecture designed to simultaneously solve input heterogeneity and deployment versatility. The core novelty of THOR is the first successful extension, synthesis and scaling of three state-of-the-art concepts: 1) Per-band patch projection strategy inspired by USat to handle heterogeneous sensor data. 2) Extension of the flexible patching and weight-resizing strategy from FlexiViT to an MAE framework with random input image sizes, enabling dynamic input image sizes and patch sizes during inference. 3) GSD-aware 2D ALiBi encoding, inspired by CROMA, to maintain spatial context across varying GSDs and patch sizes. This section details the integration of these components and the multi-pretext learning framework. 

\subsection{Encoder architecture and flexible patching}\label{sec:encoder}
\begin{figure}[t]
    \centering
    \includegraphics[width=1.0\linewidth]{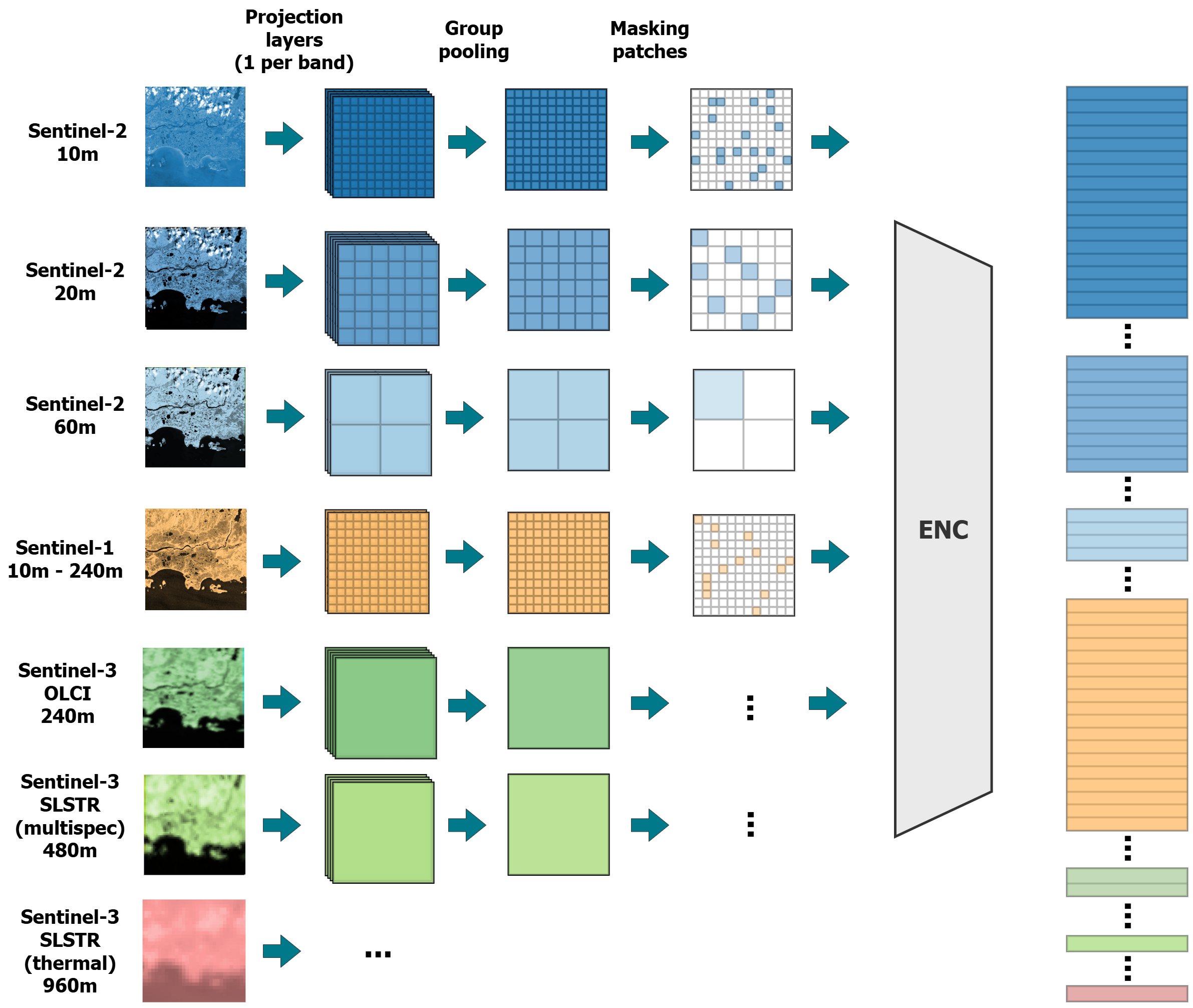}
    \caption{THOR encoder uses a single ViT. Data is processed using a band-wise patch projection layer and group average pooling.}\vspace{-0.3cm}
    \label{fig:thor_encoder}
\end{figure}

The core of THOR is a modified vision transformer (ViT) \cite{dosovitskiy2020image}, built to solve both input heterogeneity (multi-sensor, multi-resolution) and deployment rigidity (fixed patch size and flexible input image size) simultaneously (Fig.~\ref{fig:thor_encoder}).

\subsubsection{Multi-sensor integration}
To handle the highly diverse sensor inputs, the model is inspired by the USat architecture \cite{irvin2023usat}. As for USat, THOR employs a separate patch projection layer for each input band. This flexibility allows the model to process any subset of bands during fine-tuning, accommodating computational constraints or missing data. The encoder supports grouping arbitrary sets of bands with the same GSD, allocating a larger number of patches for higher resolution bands to capture finer details, and fewer patches for coarser resolution bands. To reduce the resulting long token sequence length, an average pooling step is applied to aggregate corresponding patches from the same band group (Fig.~\ref{fig:thor_encoder}). 

\subsubsection{Compute-adaptive inference}
To make THOR "compute-adaptive", allowing dynamic trade-offs between computational cost and accuracy without retraining, we incorporate the FlexiViT approach by randomizing the patch size during pre-training. The patch embedding weights are resized accordingly during training, enabling the resulting ViT to adapt to various patch sizes (e.g., from $4\times4$ to $32\times32)$ at inference time using a single set of pre-trained weights. This flexibility has a crucial downstream benefit: a user can opt for a smaller patch size at inference time, producing a denser, higher-resolution token sequence. This dense representation can be more effectively processed by simpler, more lightweight decoders, potentially reducing the amount of labeled data needed for fine-tuning pixel-level tasks and improving performance in data-limited scenarios.
We also randomize the input image size during pre-training, allowing THOR to extrapolate to larger input images than those used during fine-tuning.

\subsubsection{GSD-aware positional encoding}
USat's superpositional encodings assumes fixed patch dimensions \cite{irvin2023usat}, and if you randomly change the patch size this scheme become impractical as they require all patch sizes to be multiples of the smallest possible patch size.
We therefore extend the 2D ALiBi (Attention by Linear Bias) approach by CROMA \cite{fuller2023croma} to be GSD-aware. 

\begin{figure}
    \centering
    \includegraphics[width=0.9\linewidth]{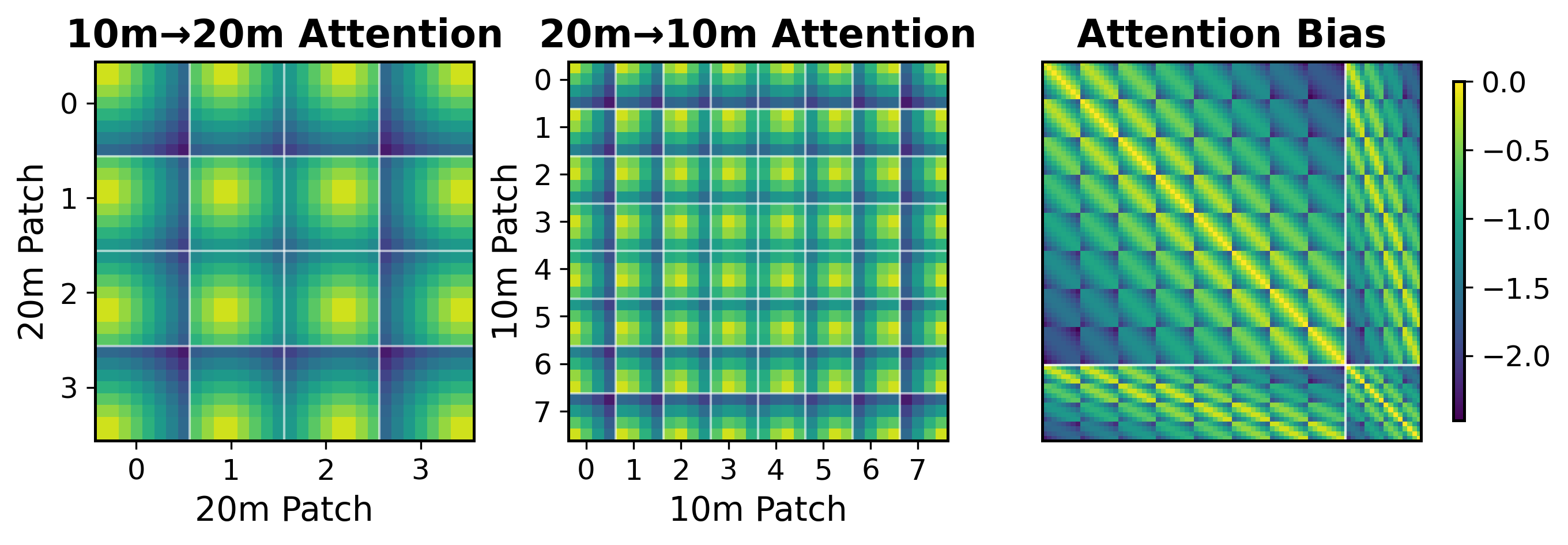}
    \caption{GSD-aware 2D-ALiBi for two groups: 10m GSD and 8x8 patches and 20m GSD and 4x4 patches. Left: Each sub-square is the ALiBi values (Eq.~\ref{eq:alibi}) between a 20 m GSD patch and all 10 m GSD patches. Mid: Each sub-square is the ALiBi values (Eq.~\ref{eq:alibi}) between a 10 m GSD patch and all 20 m GSD patches. Right: Full GSD-aware 2D-ALiBi matrix for the full sequence of 8x8 + 4x4 = 80 patches, where the off-diagonal and diagonal blocks are the intra product and inter product attention biases, respectively. }
    \label{fig:alibi_across_sensors}
\end{figure}

Let $a_{hij}$ denote  element $(h,i,j)$ of the attention matrix corresponding to the $i$th query \(\mathbf{q}_{hi}\in\mathbb{R}^d\) and the $j$th key \(\mathbf{k}_{hj}\in\mathbb{R}^d\), where $d$ is the head dimension. The attention bias is calculated based on the real-world ground distance between patch centers as 
\begin{equation} \label{eq:alibi}
    a_{hij} = \mathbf{q}^T_{hi} \mathbf{k}_{hj} / \sqrt{d} -\frac{{\rm dist}(\mathbf{x}_i, \mathbf{x}_j)}{\max(p)}\cdot m(h),
\end{equation}
where \( {\rm dist}(\mathbf{x}_i, \mathbf{x}_j) \) denotes the distance in meters between patch $\mathbf{x}_i$ and $\mathbf{x}_j$, \(\max(p) \) is the largest patch size (in meters), and $m(h)$ denote the strength of the positional biases to the $h$th self-attention head, called slopes $m$. We select slopes as in \cite{fuller2023croma}. 
The GSD-aware 2D-ALiBi is visually validated in Fig.~\ref{fig:alibi_across_sensors}, which compares the ALiBi values for two configurations of 10 m GSD with $8\times 8$ patches and 20 m GSD with $4\times 4$ patches, showcasing the relative positional encoding across tokens of different GSDs and patch dimensions.

For the lightweight decoder, which includes masked tokens for reconstruction, we modify the 2D sinusoidal positional encoding to be GSD-aware.
Let $g$ denote the GSD of the band we are reconstructing, and let $pos$ denote the center position of a patch. The encoding $v$ for that patch is:

\begin{equation}
    \begin{aligned}
v_{x}(p o s, 2 i) & =\sin \left( g \frac{pos+0.5}{10000^{2i/D}} \right) \\
v_{y}(p o s, 2 i+1) & =\cos \left(g \frac{pos+0.5}{10000^{2i/D}} \right)
\end{aligned}
\end{equation}

The GSD-aware 2D-ALiBi not only elegantly solves the problem of handling flexible patch sizes and relating products of various resolutions, but also allows for test-time extrapolation to input sizes much larger than used during training \cite{fuller2023croma}. 
As the decoder is discarded after pre-training, we opt for the simpler 2D sinusoidal positional encoding. 

\subsubsection{Randomized ground cover and patch size sampling}
To accommodate the large span in GSD between modalities, we devise a data sampling strategy where we sample a random ground cover from the range $~(1000, 50000)$m. We then extract samples for the available modalities, making sure to only include data within a valid input image size between $~(20, 500)$ pixels. 

We then select random patch sizes per modality GSD, makings sure not to exceed a predefined token budget. 
The actual resizing of patch sizes is implemented as a modified version of FlexiViT \cite{beyer2023flexivit}.

\subsection{Decoder architecture}
THOR is pre-trained using an extended MAE framework. This approach applies a self-supervised reconstruction objective combined with novel multi-modal prediction tasks.

Following the standard MAE framework \cite{he2022masked}, the decoder is substantially lighter than the encoder, focusing specifically on the reconstruction and land cover mapping task. This asymmetric design ensures that the high-quality feature representations reside solely within the heavier encoder, which is frozen for downstream tasks.

Unlike a standard MAE which, uses a linear projection layer, our decoder head projects tokens back to the patch space using a Conv2D-Transpose layer.

\subsection{Loss formulation}\label{sec:loss_formulation}
\begin{figure}
    \centering
    \includegraphics[width=1.0\linewidth]{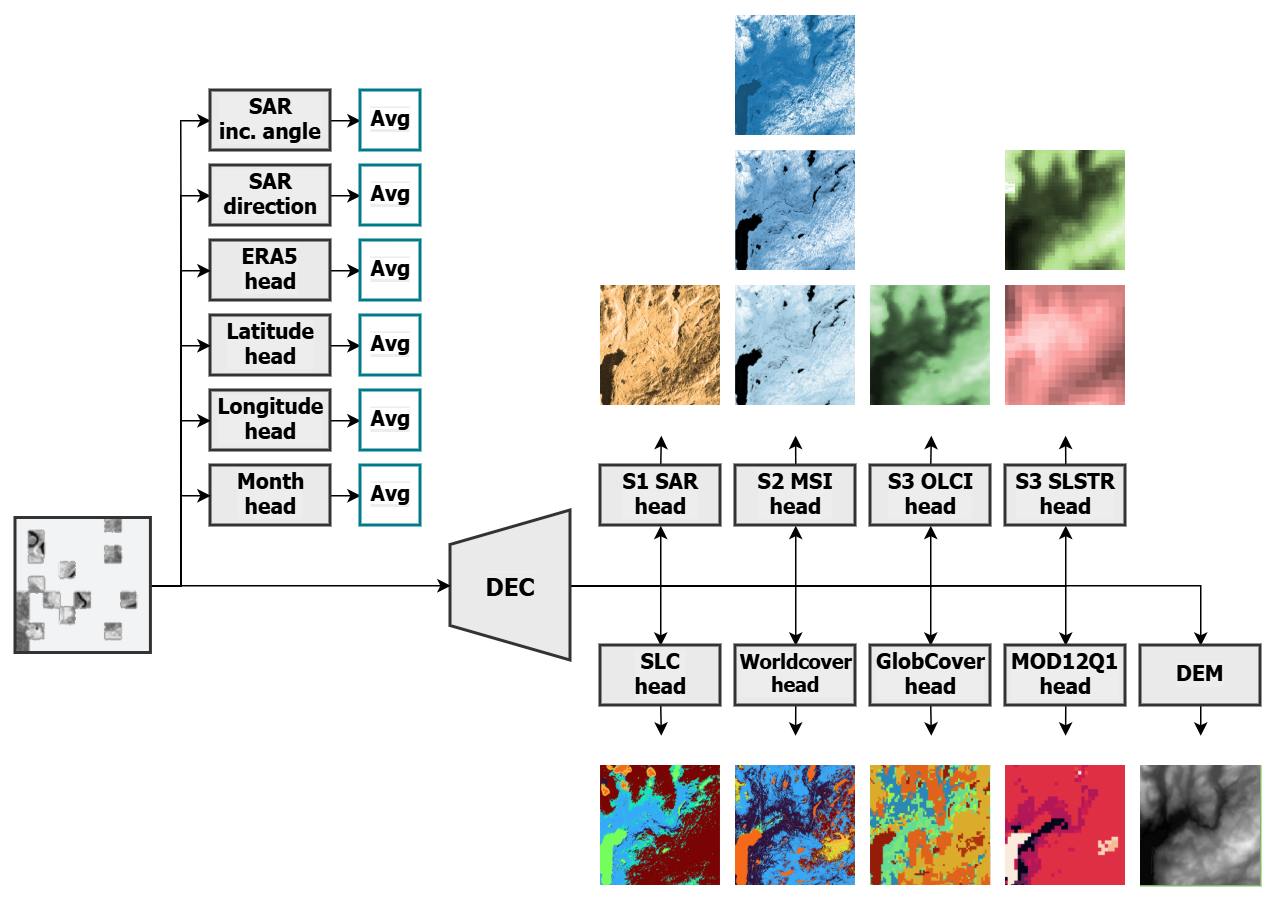}
    \caption{Pretext tasks used for learning THOR.}
    \label{fig:pretext_tasks}
\end{figure}

THOR is trained from multiple pretext tasks to ensure generality and robustness across different applications (Fig.~\ref{fig:pretext_tasks}):

\begin{itemize}
    \item Pixel-level reconstruction: Pixel-level input band reconstruction is performed using our proposed flexible VIT MAE loss (Sec.~\ref{sec:flexMAE}). This task forces the model to learn fine-grained details for pixel-level applications.
    \item Patch-level contrastive learning: We devise a patch level guided soft (multi label) contrastive loss to leverage rich semantic information in the available land cover products (Sec.~\ref{sec:patchLoss}).
    \item Pixel-level map prediction: The model predicts land cover maps, such as ESA WorldCover. This provides dense, geo-semantic supervision across different GSDs (e.g., WorldCover 10m maps predicted from Sentinel-1/-2 groups, and MOD12Q1 maps predicted from Sentinel-3 SLSTR). The model also predicts elevation and slope derived from a DEM at 10 m and 60m GSD (Sec. \ref{sec:map_prediction}). 
    \item Image-level prediction \cite{nedungadi2024mmearth, tseng2025galileo}: Predicts ERA5-Land variables (from daily statistics, e.g., soil water, temperature, precipitation), latitude, longitude, and month, providing coarse-grained, climate-relevant feature learning.
    \item  Image level SAR task: The learning includes Sentinel-1 SAR ascending/descending orbit direction classification task and incidence angle prediction task.
\end{itemize}

Similar to USat \cite{irvin2023usat} we use group specific projection layers to map decoder tokens from group $g$ to patches of shape $(P_g,P_g,C_g)$ where $C_g$ is the number of bands in this group. The input bands are then patchified to the same patch sizes and a channel-wise MSE loss over all the masked patches is used.

We add linear heads to the pooled encoder tokens for image-level tasks, using cyclic encoding for angular/temporal variables \cite{nedungadi2024mmearth}

\begin{equation}
\text{cyclic\_encoding}(\mathbf{x}, s) = 
\begin{bmatrix} 
\sin\left(\frac{2\pi}{s}\mathbf{x} \right) \\ 
\cos\left(\frac{2\pi}{s} \mathbf{x} \right) 
\end{bmatrix},
\end{equation} 
where the factors $\mathbf{x} \in \mathbb{R}^{B \times C}$ and $s \in \mathbb{R}^+$ serve as scaling parameters, with the output being of dimension $\in \mathbb{R}^{B \times 2C }$. 

\subsubsection{Flexible ViT MAE loss}\label{sec:flexMAE}
Using FlexiViT for processing the patches in the encoder introduces a challenge in the MAE framework: the decoder must reconstruct patches of varying sizes. Using arbitrary patch sizes for patchifying the target input bands is trivial and ensures that we have the same amount of decoder tokens as the number of target patches. However, the group specific decoder projection layers mapping decoder tokens is usually a linear layer resulting in a fixed output patch size. To address this, we replace the linear projection layer with a transposed Conv2D layer, enabling bilinear interpolation of the projection weights. Finally, to ensure that the loss stays the same, we introduce a scaling inspired by the FlexiViT derivations.

Formally, the standard MAE uses a pixel-wise MSE loss to reconstruct masked patches: 
 $\mathcal{L}_{mae} = (1/N)|| { \rm vec }(\mathbf{x}) - \langle \mathbf{v}, \mathbf{z},  \rangle ||^2$,  
 where \(\mathbf{x}\in \mathbb{R}^{p\times p}\) is the input patch,  $\mathbf{z} \in \mathbb{R}^{D_{d}}$ is the decoders predicted token embedding, and $\mathbf{v} \in \mathbb{R}^{D_{d} \times p \times p }$ is the weights mapping the decoder's embeddings to patch predictions.
 When the patch size $p$ changes to a new size $p^*$, the prediction weights $\mathbf{v} $ must also change to $\mathbf{v}^*$ to maintain the correct projection.  We use a bilinear interpolation to change the weights, such that
 $\mathbf{v}^* =\mathbf{v}\mathbf{B}^T$, where $\mathbf{B} \in \mathbb{R}^{p^2_*\times p^2}$. 
We prove that by scaling the entire loss term with the pseudo-inverse $\mathbf{B}^{+}$, the MSE loss for the resized patch is mathematically equivalent to the original loss. This guarantees that our normalized reconstruction target provides a consistent learning signal across all patch sizes: 
\begin{equation}
\begin{aligned}
 \mathcal{L}^*_{mae} & = \frac{1}{N} || \mathbf{B}^{+} (\mathbf{B}{\rm vec}(\mathbf{x}) -  \langle \mathbf{v}\mathbf{B}^T, \mathbf{z}) \rangle ||^2 \\
        & = \frac{1}{N} || {\rm vec}(\mathbf{x}) - \langle\mathbf{v}, \mathbf{z} \rangle ||^2 = \mathcal{L}_{mae}.
\end{aligned}
\end{equation}
This formulation allows THOR to seamlessly train with randomized patch sizes while maintaining a stable and mathematically consistent reconstruction objective.

\subsubsection{Patch-level contrastive loss}\label{sec:patchLoss}
Inspired by Galileo's approach of using a patch-wise contrastive loss to amplify local details and enforce discrimination between tokens in a single sample, improving the model's ability to handle fine-grained features \cite{tseng2025galileo}, we extend the multi-label guided approach from \cite{wang_multi-label_2024} to the patch level. 
We randomly partition the unmasked patch tokens into $K$ groups and compute an average embedding for each group. Concurrently, we extract an average land cover histogram from the corresponding patch locations for each of the $K$ groups, and soft similarity labels are then generated using cosine similarity between these $K$ normalized histograms. A contrastive loss (Eq.~\ref{eq:constrative_loss}) is applied to the $K$ average embeddings, forcing the model to produce similar representations for patch groups with similar land cover compositions. This loss is computed per band group, for all land cover tasks, adhering to the same viability constraints as our map prediction task.

The contrastive loss for group $g$ and task $t$ is  
\begin{equation}\label{eq:constrative_loss}
    \mathcal{L}^{g,t}_{con} = -\frac{1}{M}\sum_{i=1}^{M}\log\frac{\sum_{j\in P(i,j)}
    \exp\left(-h^{g,t}_{i,j}f(\mathbf{x}_j, \mathbf{x}_i)/\tau\right)}{\sum_{k\in Q(k,j)}\exp\left(-f(\mathbf{x}_k, \mathbf{x}_i)/\tau\right)},
\end{equation}
where $M=B_{g,t}\times K$ is the total batch size times the number of averaged tokens, 
$P(i,j)=\{j=1\dotsc M, j\neq i, y_j=y_i\}$,
$Q(i,k)=\{k=1\dotsc M, k\neq i,y_k\neq y_i\}$,
$f(\cdot )$ denotes the cosine similarity, $h^{g,t}_{i,j}$ is the $[0,1]$ normalized soft similarity label between the averaged set of patches $i$ and $j$, and $y_l$ is equal to the local GPU device. I.e., we only select positive pairs from the same local device, and select negatives from all other devices.

The total contrastive loss is thus
\begin{equation}
    \mathcal{L}_{con} = \frac{1}{G}\sum_{g=1}^G \frac{1}{|T_g|}\sum_{t \in T_g} \mathcal{L}_{con,g,t} 
\end{equation}
where $G$ is the total number of band groups, $T_g$ is the set of viable land cover tasks for the corresponding group

\subsubsection{Map prediction loss}\label{sec:map_prediction}
Following \cite{nedungadi2024mmearth, tseng2025galileo, jakubik2025terramind}, we incorporate pixel-level pretext tasks to provide dense, semantic supervision. These include land cover classification using multiple land cover products (e.g., WorldCover, GlobCover) at their native resolutions, as well as elevation and slope regression from a DEM.

We integrate these targets by treating them as additional output "bands". The decoder uses specialized projection heads to predict each map, and we adapt our FlexiViT resizing strategy to project the decoder's variable-GSD token embeddings to the fixed GSD of the target map. We enforce a compatibility constraint, permitting only projections that result in a target patch size within a $[4, 32]$ pixel range. For example, predicting a 10 m WorldCover map from a 16-pixel, 60 m GSD input patch is disallowed, as it would require an unstable $96 \times 96$ pixel projection.

Unlike the MAE reconstruction objective, which operates only on masked tokens, the map prediction loss is computed on all encoder patches (both masked and unmasked).
We apply a CE loss (0.1 label smoothing) for classification tasks and an MSE loss for regression, standardizing DEM targets (elevation and slope) using dataset statistics after a per-sample min-normalization of the elevation.
See Supplementary Material for details.

\subsubsection{Total loss}
Our final pre-training loss, $\mathcal{L}_{total}$, is a weighted sum of the MAE reconstruction loss ($\mathcal{L}_{mae}$), the map prediction losses ($\mathcal{L}_{map,t}$), the ERA5 land, month, coordinate and S1 incidence regression loss ($\mathcal{L}_{era5}, \mathcal{L}_{m}, \mathcal{L}_{coord}, \mathcal{L}_{inc}$), S1 orbit direction loss ($\mathcal{L}_{orb}$), and our novel contrastive loss ($\mathcal{L}_{con}$).
$\mathcal{L}_{total} = \lambda_1 \mathcal{L}_{mae} + \lambda_{2}\mathcal{L}_{con} + \sum_{t}\lambda_{3,t} \mathcal{L}_{map,t} + \lambda_4 \mathcal{L}_{era5} + \lambda_5 \mathcal{L}_{m} + \lambda_6 \mathcal{L}_{coord} + \lambda_7 \mathcal{L}_{inc} + \lambda_8
\mathcal{L}_{orb} + \lambda_9\mathcal{L}_{fft},$
where $t \in \{WC, GC, MCD, DEM, SCL \}$ is the land cover map prediction tasks. $\mathcal{L}_{fft}$ is a L1 MAE reconstruction loss in the Fourier domain, included for stability during training \cite{kraus_masked_2024}. The loss weights can be found in the Supplementary Material.

\section{Experiments}\label{sec:experiments}
Our experiments are designed to validate THOR's core hypotheses: \textbf{1)} Is THOR more data-efficient in low-label regimes, validating our "data-hungry decoder" hypothesis? \textbf{2)} Does our architecture successfully synthesize and learn from the full 10 m - 1000 m S1, S2, and S3 sensor suite?, and \textbf{3)} Is the compute-adaptive mechanism the key driver of this performance?

\subsection{Experimental setup}
We evaluate on two main benchmarks: PANGAEA \cite{marsocci2024pangaea} and Copernicus-Bench \cite{wang_copernicus_fm_2025}. PANGAEA consists of a diverse suite of 9 semantic segmentation tasks, including HLS, MADOS, PASTIS, and Sen1Floods11. We follow its standard protocol, evaluating on the 10\%, 50\%, and 100\% labeled data splits to test our data-efficiency hypothesis.
To specifically validate THOR's ability to process Sentinel-3 data, we evaluate on the four Sentinel-3 OLCI-specific tasks (Cloud-S3, LC100Cls-S3, LC100Seg-S3, Biomass-S3) from the Copernicus-FM benchmark. 

We compare THOR against a wide range of state-of-the-art (SOTA) FMs, including Copernicus-FM \cite{wang_copernicus_fm_2025}, DOFA \cite{xiong_dofa_neural_2024}, and TerraMind \cite{jakubik2025terramind}, using the scores reported in \cite{marsocci2024pangaea}.

\subsection{Implementation details}
We pre-train a family of THOR models (Tiny, Small, Base, Large) from scratch on the THOR Pretrain dataset for 400 epochs. All models are trained using the AdamW optimizer with a base learning rate of 3e-4 for the base and large model, and 4e-4 for small and tiny, a weight decay of 0.05, and a linear warmup of 40, 20, 10, 10 epochs for  ViT-Large, Base, Small and Tiny model, respectively warmup followed by a cosine decay schedule.

Training was conducted on 16 AMD MI250X GPUs, with a total batch size of 1024. 
During pre-training, we randomly sample both input resolution ($32 \times 32$ – $1024 \times 1024$) and patch size ($4 \times 4$ – $32 \times 32$) for every sample.
To manage computational constraints introduced by randomized patch sizes and input image sizes across multiple band groups, we implemented a simple token budget heuristic. During training, a  threshold for the maximum number of tokens is enforced. When sampling a product group, a patch size is drawn such that the resulting number of tokens does not exceed the remaining budget, ensuring efficient use of memory across heterogeneous inputs.

\subsection{Main result: State-of-the-art in data-limited scenarios}
Our central hypothesis, introduced in the Introduction, is that THOR's flexible patching overcomes the "data-hungry decoder" problem faced by rigid models. We test this directly on the 10\% Pangaea benchmark split using an $6\times 6$ patch size.

In this low training data regime, THOR-B (Base) achieve the best average rank across all datasets (Tab.~\ref{tab:10_perc_training_data}). THOR-B outperforms all other published models, including a +1.9 mIoU gain over the next-best model, TerraMind. This strong performance, especially on fine-grained tasks like sen1floods11 (86.29 mIoU), validates that THOR is significantly more data-efficient than its fixed-patch counterparts.

\begin{table*}[t]
\centering
\caption{Pangaea results with 10\% training data in mIoU. Bold/underline mark best/second-best per column.}
\resizebox{0.8\textwidth}{!}{%
\fontsize{9}{10.8}\selectfont
\begin{tabular}{lccccccccccc}
\toprule
Model & HLS Burns & MADOS & PASTIS & Sen1Floods11 & FBP & DynEarthNet & CropMap & SN7 & AI4Farms & Avg. Rank \\
\midrule
CROMA & 76.44 & 32.44 & 32.80 & \underline{87.22} & 37.39 & 36.08 & 36.77 & 42.15 & 38.48 & 6.11 \\
DOFA & 71.98 & 23.77 & 27.68 & 82.84 & 27.82 & \textbf{39.15} & 29.91 & 46.10 & 27.74 & 10.22 \\
GFM-Swin & 67.23 & 28.19 & 21.47 & 62.57 & 55.58 & 28.16 & 27.21 & 39.48 & 32.88 & 12.56 \\
Prithvi & 77.73 & 21.24 & 33.56 & 86.28 & 29.98 & 32.28 & 27.71 & 36.78 & 35.04 & 10.22 \\
RemoteCLIP & 69.40 & 20.57 & 17.19 & 62.22 & \underline{56.23} & 34.43 & 19.86 & 43.11 & 23.85 & 12.33 \\
SatlasNet & 74.79 & 29.87 & 16.76 & 83.92 & 37.86 & 34.64 & 29.08 & 49.78 & 13.91 & 10.22 \\
Scale-MAE & 75.47 & 21.47 & 22.86 & 64.74 & 48.75 & 35.27 & 13.44 & 49.68 & 26.66 & 10.78 \\
SpectralGPT & \textbf{83.35} & 20.29 & 34.53 & 83.12 & 39.51 & 35.33 & 31.06 & 36.31 & 37.35 & 8.56 \\
S12-MoCo & 73.11 & 19.47 & 32.51 & 79.58 & 35.57 & 32.24 & 36.54 & 49.46 & 37.97 & 10.67 \\
S12-DINO & 75.93 & 23.47 & 36.62 & 84.95 & 34.63 & 32.78 & \underline{38.44} & 41.15 & 37.91 & 8.33 \\
S12-MAE & 76.60 & 18.44 & 31.06 & 84.81 & 35.56 & 30.59 & 35.29 & 40.51 & 23.60 & 11.44 \\
S12-Data2Vec & 74.38 & 17.86 & 33.09 & 81.91 & 37.27 & 33.63 & 34.11 & 40.66 & 22.85 & 12.11 \\
Terramind-B & 77.39 & \textbf{44.06} & \textbf{39.96} & 84.43 & 54.00 & \underline{37.35} & 35.65 & 43.21 & 38.59 & \underline{4.00} \\
\midrule
UNet Baseline & \underline{79.46} & 24.30 & 29.53 & \textbf{88.55} & 52.58 & 35.59 & 13.88 & 46.08 & 34.84 & 7.11 \\
ViT Baseline & 75.92 & 10.18 & 38.44 & 81.85 & \textbf{56.53} & 35.39 & 27.76 & 36.01 & \textbf{39.20} & 8.78 \\
\midrule
THOR-B & 76.90 & 40.67 & \underline{38.93} & 86.29 & 42.80 & 35.21 & \textbf{42.23} & \underline{55.94} & \underline{38.90} & \textbf{3.78} \\
THOR-T & 75.98 & \underline{41.65} & 36.26 & 82.70 & 42.81 & 34.03 & 37.82 & \textbf{58.52} & 38.56 & 5.78 \\
\bottomrule
\end{tabular}
}
\label{tab:10_perc_training_data}
\end{table*}

\subsection{Full-data benchmarking}
We next evaluate THOR's performance in Pangaea using full training data availability and for the Sentinel-3 OLCI scenarios from Copernicus-Bench \cite{wang_copernicus_fm_2025}.

For 100\% training data, THOR-B remains state-of-the-art or highly competitive in the full-data regime. It achieves the top rank on PASTIS (40.76\% mIoU) and CropMap (56.78\% mIoU), demonstrating that its architecture scales effectively with more data. 
A full comparison against all baseline models is provided in the Supplementary Material.

\begin{table}[t]
\caption{Benchmark results on selected Copernicus-Bench benchmarks. $\dagger$: We use a patch size of 10 for Cloud-S3,  8 for LC100Cls-S3, 6 for LC100Seg-S3 and 4 for Biomass-S3.}
\resizebox{1.05\columnwidth}{!}{%
\begin{tabular}{lccccccc}
\hline
 & \textbf{Metric} & \textbf{Supervised} & \textbf{Supervised} & \textbf{Random} & \textbf{DOFA} & \textbf{Copernicus} & \textbf{THOR}\\
\hline
Backbone & -- & ViT-S/16 & ViT-B/16 & ViT-B/16 &  ViT-B/16 & ViT-B/8 & ViT-B/10$\dagger$ \\
Modality & -- & -- & -- & -- & All (spectral) & All & All (spectral) \\
\hline
Cloud-S3 & mIoU & 61.7 ± 0.7 & 63.0 ± 0.8 & 60.9 ± 0.0 & 58.2 ± 0.1 & 62.0 ± 0.7 & \textbf{66.1 ± 0.1} \\
LC100Cls-S3 & mAP & 91.3 ± 0.3 & 91.4 ± 0.5 & 88.9 ± 0.1 &  89.5 ± 0.0 & \textbf{93.3 ± 0.4} & 91.0 ± 0.0 \\
LC100Seg-S3 & mIoU & 20.1 ± 0.4 & 19.3 ± 0.5 & 18.2 ± 0.1 &  16.5 ± 0.1 & \textbf{24.1 ± 0.0} & 19.9 ± 0.0 \\
Biomass-S3 & RMSE ↓ & 68.1 ± 0.3 & 68.3 ± 0.4 & 68.7 ± 0.5 & 74.1 ± 0.1 & 66.3 ± 0.1 & \textbf{59.7 ± 0.0}\\
\hline
\end{tabular}
}
\label{tab:copernicus_bench}
\end{table}

The Copernius-Bench validates THOR's ability to synthesize Sentinel-3 OLCI data (Tab.~\ref{tab:copernicus_bench}). 
Our model outperforms all baselines on two of the four OLCI specific tasks, including a +4.1 mIoU gain on Cloud-S3 and a -6.6 RMSE (improvement) on Biomass-S3 over the Copernicus-FM baseline. This confirms that THOR effectively learns from the challenging 10 m - 1000 m GSD range.

\subsection{Ablation studies}

\noindent\textbf{Value of compute-adaptivity.}
\begin{figure}[t]
    \centering
    \includegraphics[width=0.8\linewidth]{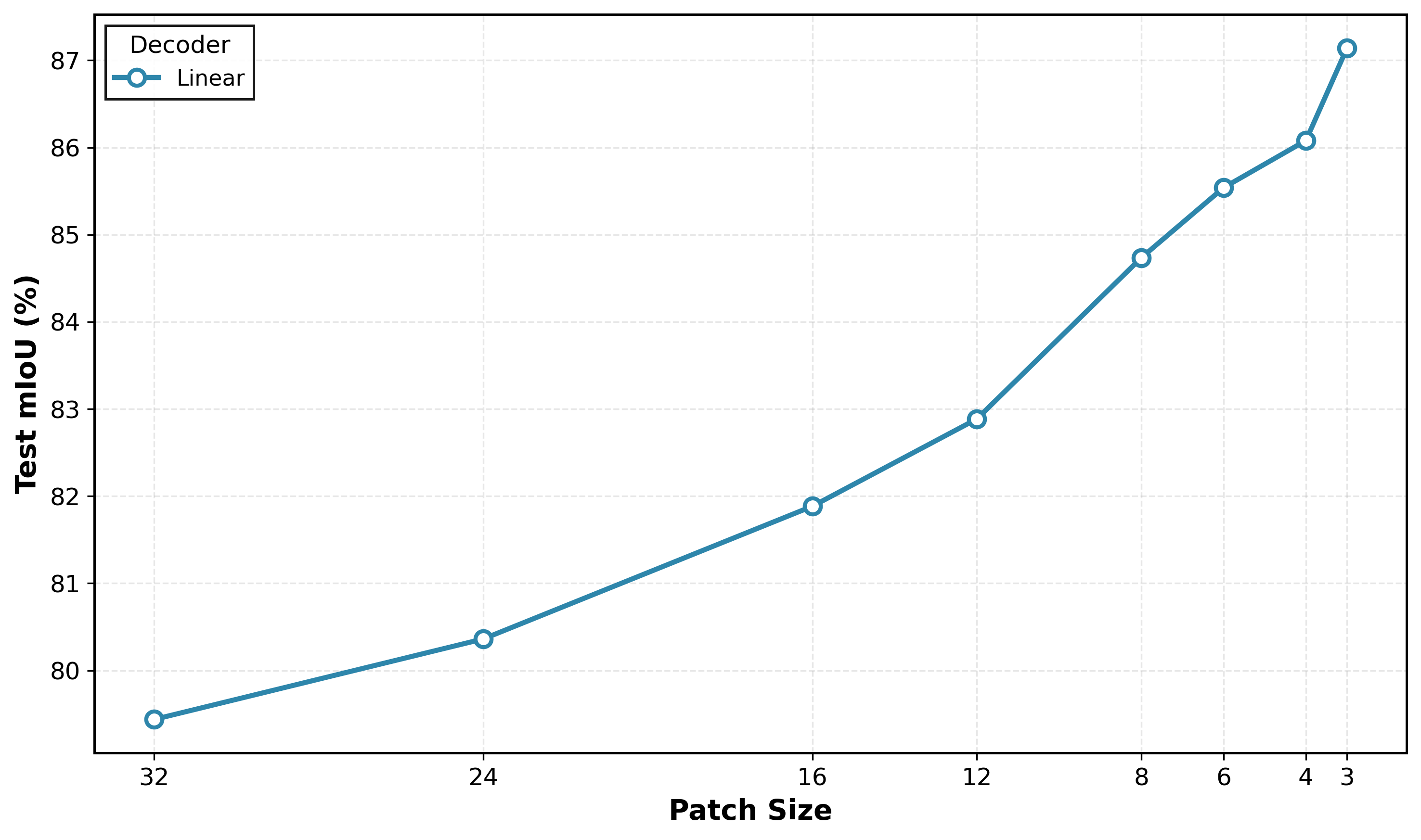}
    \caption{Test mIoU results for THOR-B model with varying patch sizes using a fixed number of tokens equal to 18 with linear probing segmentation on the Sen1Floods11 dataset using Sentinel 1 and Sentinel 2 data, 10\% of the training data, with mean aggregation of features.}
    \label{fig:sen1floods11_patch_size_ablation}
\end{figure}
We fine-tuned a single THOR-B model on Sen1Floods11 (10\% data) and evaluated it at multiple patch sizes (Fig.~\ref{fig:sen1floods11_patch_size_ablation}) using linear probing. The results shows simply by shrinking the patch size at inference time, from a coarse $16\times16$ (61.9 mIoU) to a fine $4\times4$ (81.1 mIoU), we gain nearly 20 mIoU points. This confirms our hypothesis that a single model can be dynamically deployed, and smaller patches (producing denser tokens) are critical for fine-grained tasks.

\noindent\textbf{Value of multi-sensor synthesis.}
A core design principle of THOR is its ability to ingest and synthesize synergistic information from heterogeneous sensors, such as Sentinel-1's radar and Sentinel-2's spectral details. To validate this capability, we conducted an ablation study on the Sen1floods11 benchmark using the 10\% data split. We fine-tuned the THOR-B model with an UperNet decoder on three different input modality configurations: S1 only, S2 only, and the combined S1 + S2 inputs. 
The results, presented in Tab.~\ref{tab:modality_configs}, demonstrate the value of this multi-modal fusion.

\begin{table}[h]
    \centering
    \fontsize{9}{10.8}\selectfont
    \caption{Sen1floods11 $10\%$ test mIoU results for different modality configurations using THOR-B model with Upernet decoder.}
    \label{tab:modality_configs}
    \begin{tabular}{lc}
        \toprule
        Modalities & mIoU \\
        \midrule
        Sentinel-1 & 78.09\\
        Sentinel-2 & \underline{87.25} \\
        Sentinel-1 + Sentinel-2 & \textbf{87.70} \\
        \bottomrule
    \end{tabular}
\end{table}

\subsection{Test-time extrapolation to larger images}
We validated our GSD-aware 2D-ALiBi's ability to extrapolate to larger input sizes than seen during training, a key property of relative positional encodings \cite{fuller2023croma}. We fine-tuned a UperNet decoder on the Sen1floods11 (10\% split) dataset using a frozen THOR-B encoder with S1 + S2 inputs. The model was trained only on $108 \times 108$ pixel crops with a patch size of $6$. 

\begin{figure}
    \centering
    \includegraphics[width=0.8\linewidth]{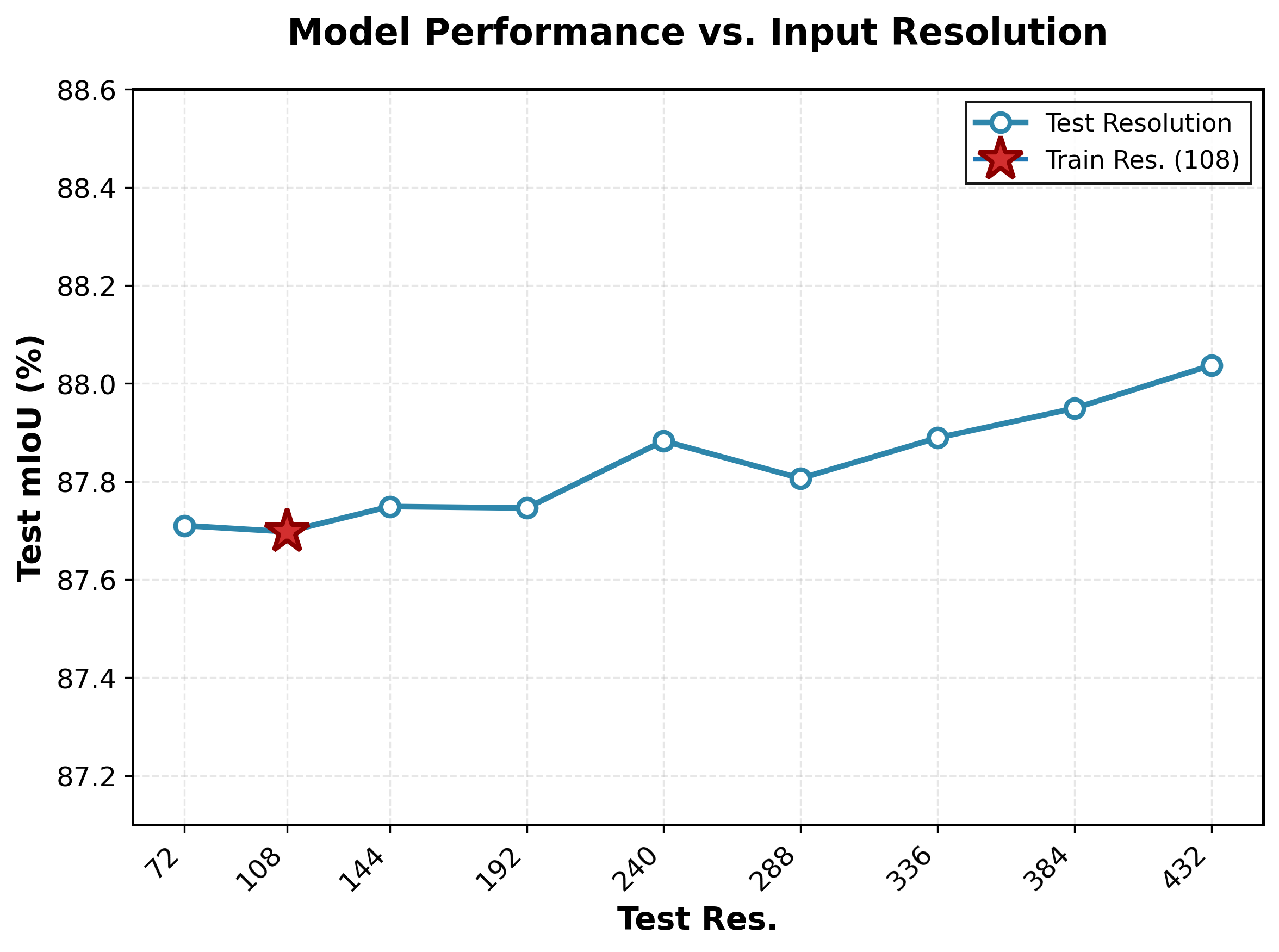}
    \caption{Test mIoU results of a THOR-B (w/UperNet), trained on a $108 \times 108$ image size, evaluated on increasingly larger images.}
    \label{fig:test_time_extrapolation}
\end{figure}

During evaluation, we tested this fixed model on the test set using various input sizes, applying Pangaea's sliding window inference up to the full $512 \times 512$ image size. As shown in Fig~\ref{fig:test_time_extrapolation}, performance does not degrade at larger scales; on the contrary, it shows a consistent improvement as the input size increases, confirming the robust extrapolation capability of our positional encoding.

\subsection{Use case: mapping of snow cover }
We validate THOR's compute-adaptive capability on a data-scarce climate task: snow cover fraction regression using 500 m / 1000 m GSD Sentinel-3 SLSTR data. 

We fine-tune THOR-B and compare a simple linear decoder against a UPerNet \cite{ruiping2024vitupernet}. The results in Tab.~\ref{tab:snow_results} provide two key insights:
\begin{itemize}
    \item Deploying the same pre-trained THOR model with an UPerNet decoder, but changing the inference patch size from $16\times16$ to $4\times4$, reduced the RMSE from 12.4 to 9.90. This 29\% reduction confirms that a denser token sequence is beneficial, even for coarse-resolution data.
    \item A simple linear decoder with $4\times4$ patches (9.99 RMSE) performs identically to the much larger UPerNet with $4\times4$ patches (9.90 RMSE). This confirms our hypothesis: the complex decoder was a crutch to compensate for a "token-starved" encoder. By providing a dense token sequence, THOR's flexible patching facilitates simpler decoders and validating its superior data-efficiency.
\end{itemize}

\begin{table}{}   
    \centering
    \fontsize{9}{10.8}\selectfont
    \caption{RMSE snow cover fraction. Image size $128\times 128$ and concatenated the tokens of the 500 m and 1000 m bands.}
    \begin{tabular}{lcc}
    \toprule
         Decoder & Patch size & RMSE \\ %
     \midrule
         UNet &  & 12.4 \\
         UPerNet & 16x16 &  14.0 \\ 
                 & 8x8  &  12.4 \\
                 & 4x4 &  \underline{9.90} \\
         Linear decoder &  4x4 & \textbf{9.88} \\  
     \bottomrule
    \end{tabular}
    \label{tab:snow_results}
\end{table}

\begin{figure}
    \centering
    \includegraphics[width=0.4\linewidth]{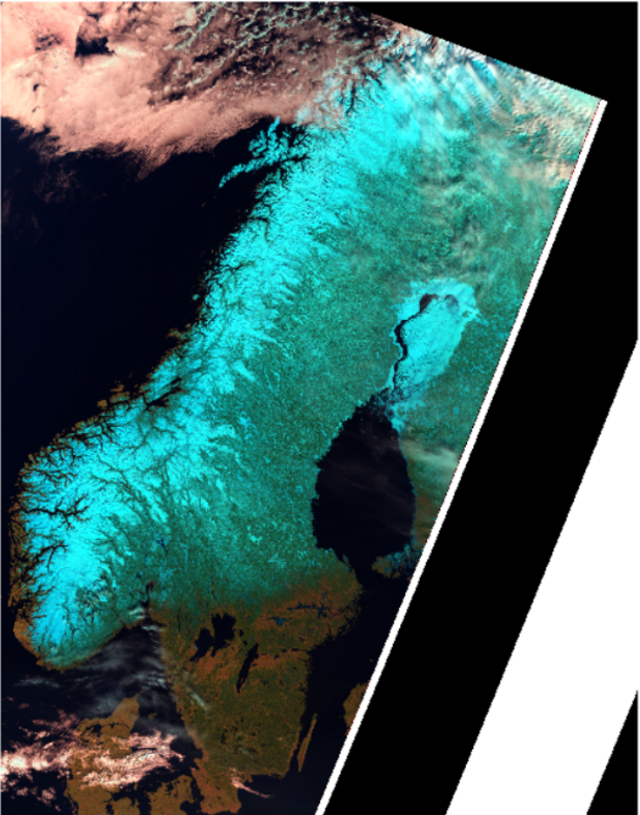}
    \includegraphics[width=0.4\linewidth]{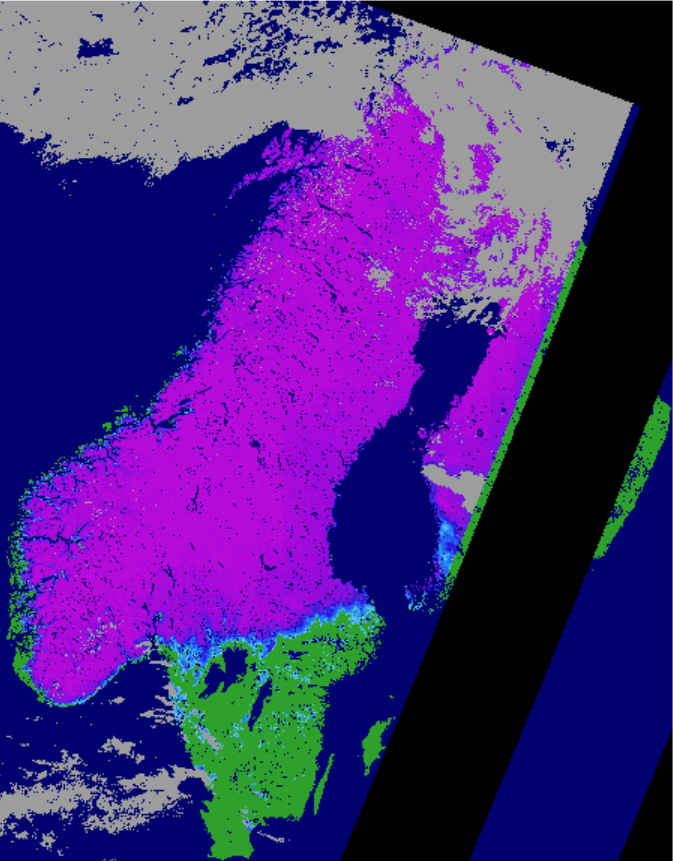}
    \caption{Left: False color SLSTR image. Right: Snow cover fraction, where green is less than 15\% and  purple is 100\%.}
    \label{fig:scf_results}
\end{figure}

\section{Conclusion}\label{sec:conclusions}
In this work, we addressed a weakness of current EO foundation models: their architectural rigidity. We argued that fixed patch sizes lead to data-hungry decoders, limiting their utility in data-scarce scenarios. We proposed THOR, the first FM to synthesize a compute-adaptive patching strategy with a multi-sensor architecture that unifies Sentinel-1, -2, and -3 (OLCI \& SLSTR) data.

Our experiments validate our central hypothesis. THOR achieves state-of-the-art performance in the Pangaea 10\% benchmark, demonstrating superior data efficiency. This confirms that the ability to use smaller patch sizes at inference provides a denser token sequence that is more effective for fine-tuning on limited data. We also proved our complex multi-sensor synthesis was successful, with THOR outperforming baselines on two of four Sentinel-3 OLCI specific tasks, validating its unique 10 m - 1000 m GSD capability.

THOR, while versatile, has limitations that open clear avenues for future work. While our dataset includes temporal samples, the architecture itself does not explicitly model time. 
Future work will focus on extending this flexible-patching concept to an explicit spatio-temporal backbone and integrating other key modalities like Sentinel-5P (air quality) or passive microwave data (climate) to further strengthen THOR applicability to climate and society challenges.

\section*{Acknowledgments}
This activity was funded and supported by European Space Agency (ESA) $\Phi$-lab (FM4CS project, contract no. 4000143489/24/I-DT), and the Research Council of Norway (KnowEarth project no. 337481).

{
    \putbib
}
\end{bibunit}

\clearpage
\setcounter{page}{1}
\appendix
\maketitlesupplementary
\setcounter{figure}{0} 
\renewcommand{\thefigure}{S.\arabic{figure}}
\setcounter{table}{0} 
\renewcommand{\thetable}{S.\arabic{table}}

\begin{bibunit}
\makeatletter
\renewcommand{\@biblabel}[1]{[S#1]}
\makeatother
\renewcommand{\citenumfont}[1]{S#1}

\section{THOR Pretrain}
The THOR FM is pre-trained on a new, diverse, and large-scale dataset named THOR Pretrain. This dataset is curated to learn representations that are robust to variations in global land cover, ocean phenomena, and cloud conditions.

THOR Pretrain unifies data from four major Copernicus Sentinel missions: Sentinel-1 SAR, Sentinel-2 MSI, Sentinel-3 OLCI, and Sentinel-3 SLSTR. These sensors provide diverse image modalities, including radar, multispectral and thermal sensors, with resolutions ranging from 10 m to 1000 m. In addition to the satellite data, the dataset includes a digial elevation model (DEM), diverse land cover maps, and ERA5-Land data. The dataset consists of 22TB of data from globally distributed locations (Fig.~\ref{fig:s2_locations}). 

\subsection{Data, pre-processing and alignment}
Instead of stacking millions of small image crops, we sample EO data using the Sentinel-2 tiles ($110 \times 110$ km) as the sampling grid. For a given grid location and time, we sample the Sentinel-2 tile  along with overlapping Sentinel-1 SAR, Sentinel-3 OLCI, and Sentinel-3 SLSTR data. Sentinel-3 data is selected from a 25 times larger area, centered at the Sentinel-2 tile,  to account for its coarser resolution.

\begin{figure}[t]
    \centering
    \includegraphics[width=1.0\linewidth]{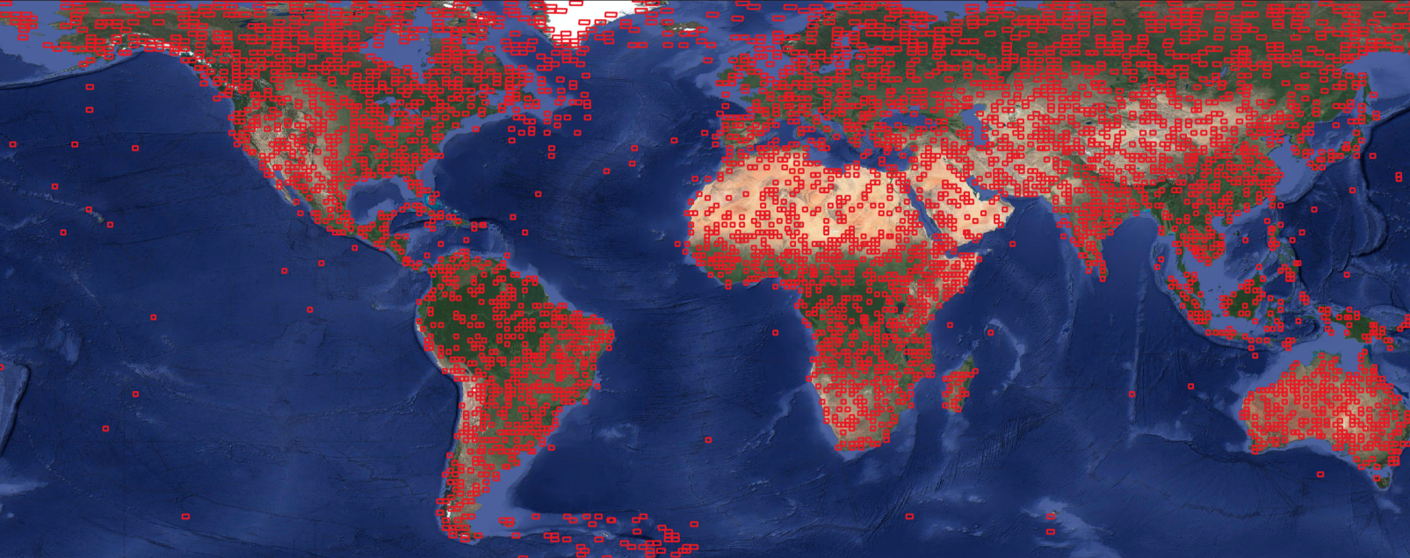}
    \caption{Overview of THOR Pretrain sampled locations.}
    \label{fig:s2_locations}
\end{figure}

To ensure a diverse dataset of global land covers, ocean phenomena, and cloud conditions, we employ a stratified sampling strategy utilizing land cover and RGB maps of the world (see Sec. \ref{sec:stratified} for details).  This methodology is crucial to balance the dataset by actively prioritizing locations with high thematic and geographic diversity (e.g., \cite{ordonez2024towards, al2024pretraining}). A total of 6273 globally distributed locations were sampled (Fig.~\ref{fig:s2_locations}). 

\subsubsection{Sensor data pre-processing}
The Sentinel data are downloaded from Copernicus Data Space Ecosystem 
and preprocessed into netCDF files along with relevant metadata.

\paragraph{Sentinel-1 SAR.} The SAR data is processed to sigma-naught, corrected for thermal noise, geocoded using the Range Doppler algorithm. Two different resolutions of Sentinel-1 are constructed: 10 m and 60 m GSD. The 10 m GSD Sentinel-image is aligned with the corresponding Sentinel-2 data, whereas the 60 m GSD is processed to a larger area, bounded by the Sentinel-3 footprint. 

\paragraph{Sentinel-2 MSI.} Level 2A Sentinel-2 data are acquired and the reflectance bands are collected into a single netCDF file along with metadata. The Scene Classification Map (SCL) product, which includes various land cover classes and a cloud mask, is also collected into the same netCDF file.

\paragraph{Sentinel-3 OLCI.} Level 1 OLCI data are acquired. The top of atmosphere radiance ($R_{TOA}$) bands are converted to reflectance ($L_{TOA}$) using 
\begin{align}
    R_{TOA}(\lambda) &= \frac{\pi L_{TOA}(\lambda)}{E_0(\lambda) cos(\phi)},
    \label{eq:rad2refl}
\end{align}
where $E_0$ is the solar spectral irradiance and $\phi$ is the sun zenith angle, both provided in the downloaded Sentinel-3 OLCI product file.

Further, the bands are resampled into the same UTM projection as the corresponding Sentinel-2 tile, but resampled to a GSD of 250 m and a geographic extent of 25 times larger area than the Sentinel-2 tile. This is done using the bilinear algorithm implemented in the \textit{pyresample} Python library.

\paragraph{Sentinel-3 SLSTR.} Level 1 SLSTR data are acquired. The Sentinel-3 SLSTR files are processed in the same manner as for OLCI: First, the top of atmosphere radiance bands are converted to reflectance using Eq.~\eqref{eq:rad2refl}. Then the reflectance and brightness temperature bands are resampled to UTM projection and geographic extent similar to the OLCI product, except that the GSDs are 500 m and 1000 m for the reflectance and brightness temperature bands, respectively.
 
For SLSTR, cloud detection is performed using the SCDA version 2.0 algorithm \cite{metsamaki2015globsnow}. 

\subsubsection{Auxiliary geospatial modalities (pretext targets)}
The dataset also includes auxiliary geospatial modalities for reconstruction and prediction pretext tasks:
\begin{itemize}
    \item Digital Elevation Model (pixel-level targets): DEMs are included, and the model reconstructs both slope and elevation at 10~m and 60~m GSD as part of the MAE reconstruction objective from Sentinel-1 and Sentinel 2 bands.
    \item Land cover maps (pixel-level targets): Several land cover products are incorporated to serve as pixel-level pretext tasks, accommodating the range of satellite sensors by varying in GSD from 10m to 500m. ESA WorldCover (10 m) \cite{WorldCover2021_v200} and the Sentinel-2 SLC map is predicted from the Sentinel-1 and Sentinel-2 bands, the ESA GlobCover (300 m) \cite{GlobCover2009} is predicted from the Sentinel-3 OLCI bands, and MOD12Q1 map (500 m) \cite{MCD12Q1_v061} is predicted from the Sentinel-3 SLSTR bands.
    \item ERA5-Land (image-level targets): The dataset includes ERA5-Land data based on daily statistics, derived from hourly land variables aggregated daily at 0.1 degrees resolution (approximate 9 km grid spacing). We select a diverse set of 17 variables covering temperature, hydrological cycles, snow cover, and vegetation indices (detailed in Table \ref{tab:variable_description}). This data is used for image-level prediction pretext tasks. 
\end{itemize}

\begin{figure*}
    \centering
    \begin{subfigure}[b]{0.3\linewidth}
        \centering
        \includegraphics[width=\textwidth]{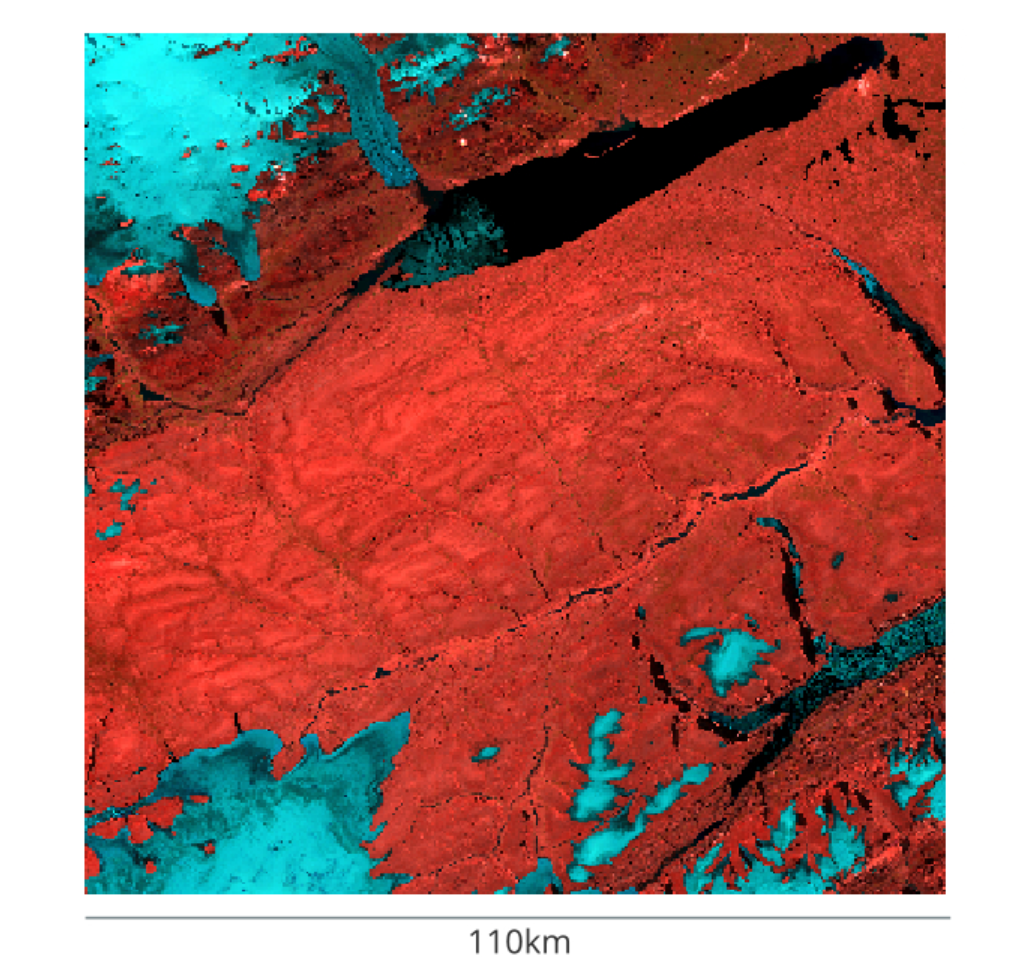}
        \caption{Sentinel-1 MSI 10/20/60 m.}
        \label{fig:sub-a}
    \end{subfigure}
    \hfill 
    \begin{subfigure}[b]{0.3\linewidth}
        \centering
        \includegraphics[width=\textwidth]{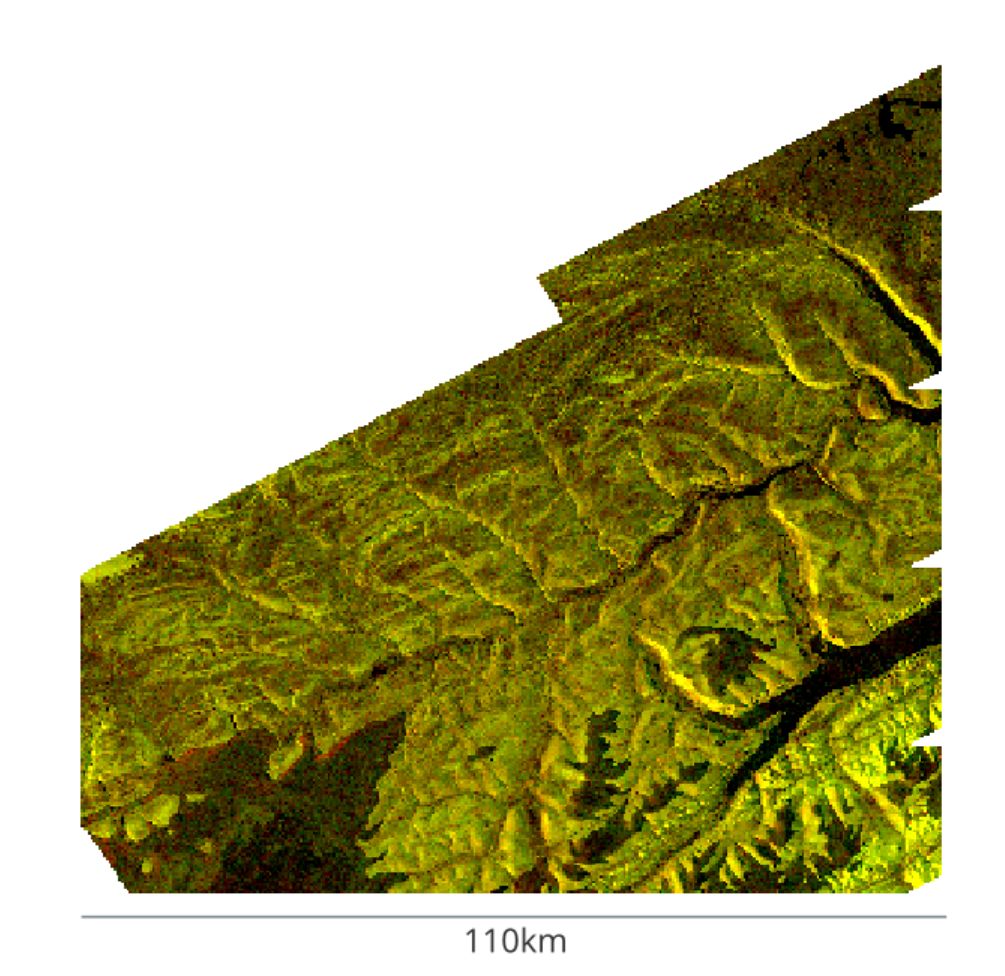}
        \caption{Sentinel-1 SAR 10 m.}
        \label{fig:sub-b}
    \end{subfigure}
    \hfill 
    \begin{subfigure}[b]{0.3\linewidth}
        \centering
        \includegraphics[width=\textwidth]{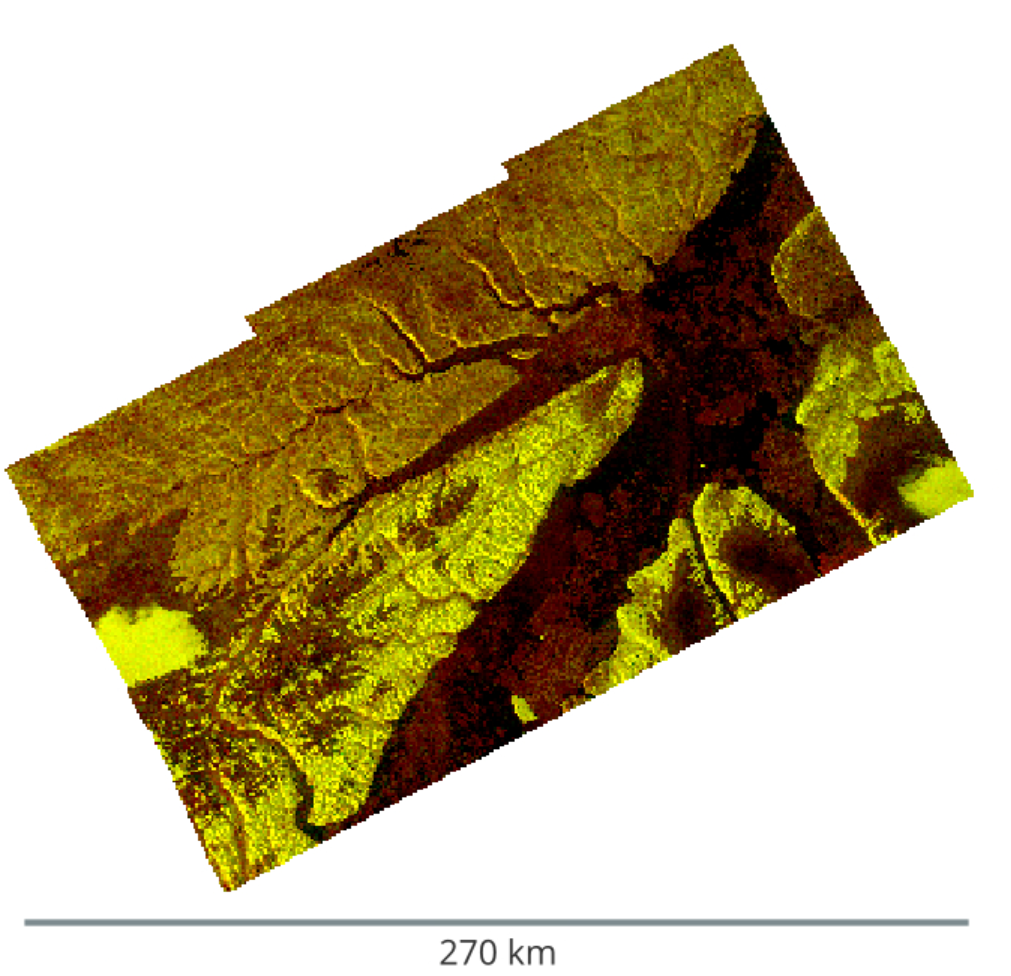}
        \caption{Sentinel-1 SAR 60 m.}
        \label{fig:sub-c}
    \end{subfigure}
    \begin{subfigure}[b]{0.3\linewidth}
        \centering
        \includegraphics[width=\textwidth]{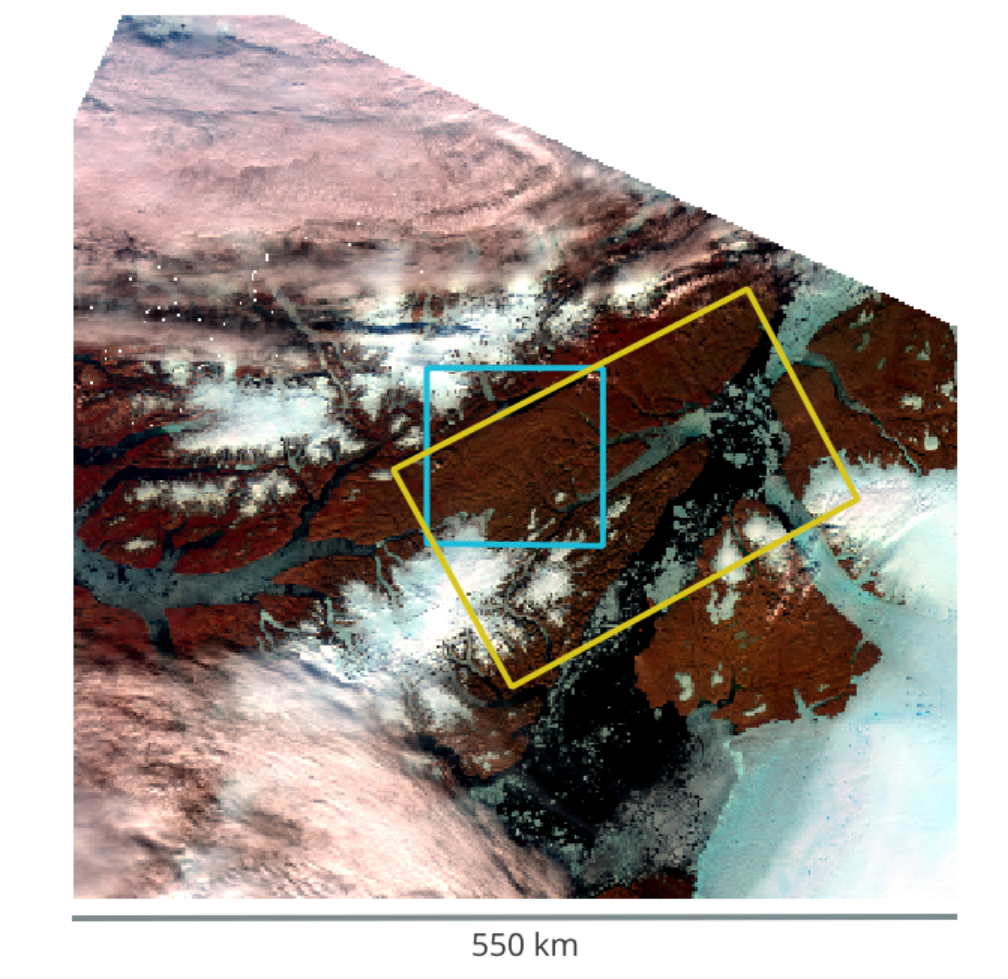}
        \caption{Sentinel-3 OLCI 300 m.}
        \label{fig:sub-d}
    \end{subfigure}
    \hfill 
    \begin{subfigure}[b]{0.3\linewidth}
        \centering
        \includegraphics[width=\textwidth]{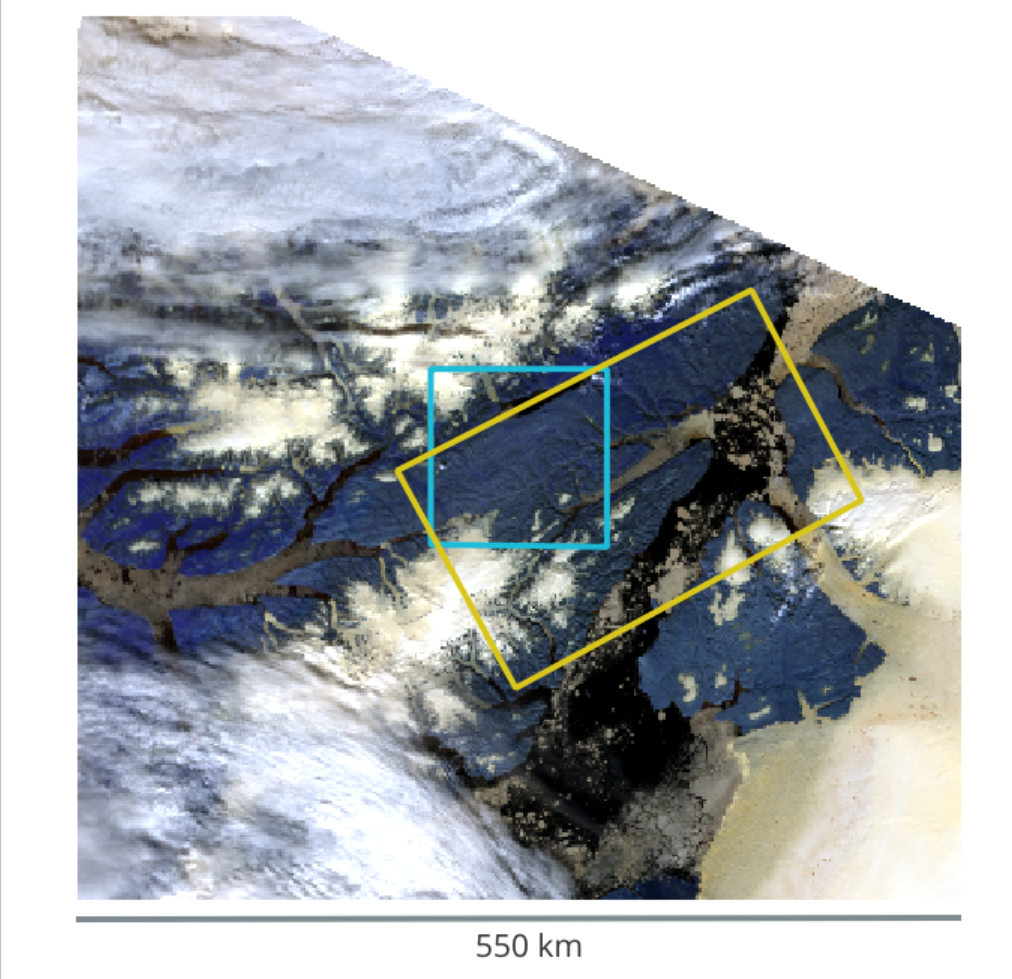}
        \caption{Sentinel-3 SLSTR 500 m. }
        \label{fig:sub-e}
    \end{subfigure}
    \hfill 
    \begin{subfigure}[b]{0.3\linewidth}
        \centering
        \includegraphics[width=\textwidth]{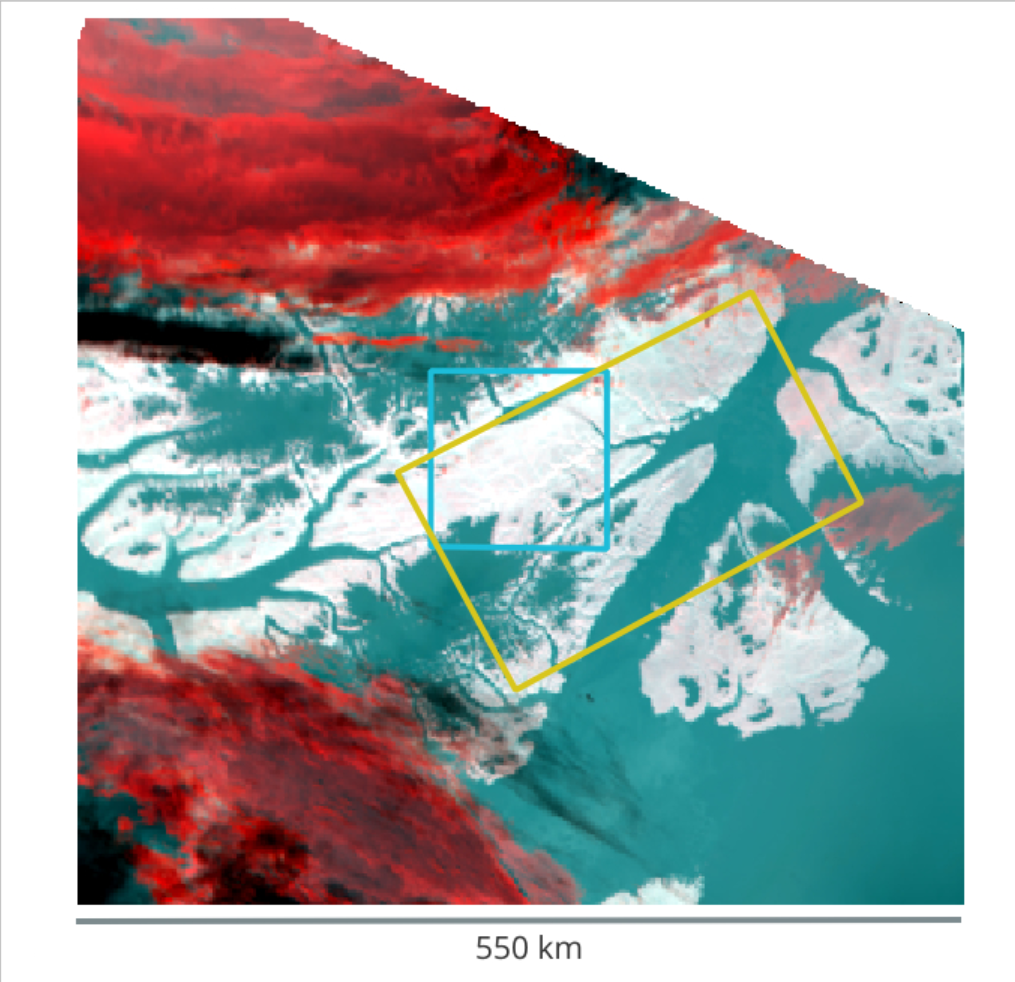}
        \caption{Sentinel-3 SLSTR 1000 m (thermal).}
        \label{fig:sub-f}
    \end{subfigure}
    \hfill 
    \begin{subfigure}[b]{0.3\linewidth}
        \centering
        \includegraphics[width=\textwidth]{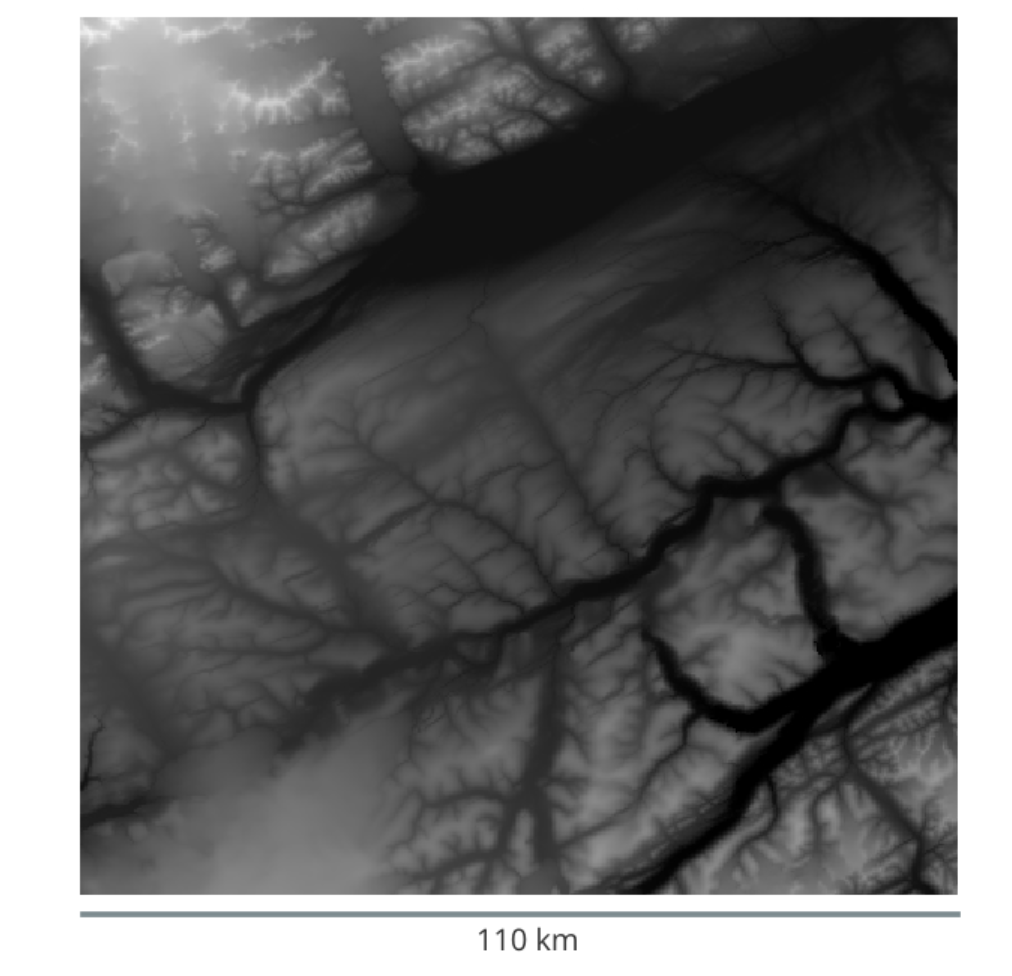}
        \caption{DEM 10 m.}
        \label{fig:sub-g}
    \end{subfigure}
    \hfill 
    \begin{subfigure}[b]{0.3\linewidth}
        \centering
        \includegraphics[width=\textwidth]{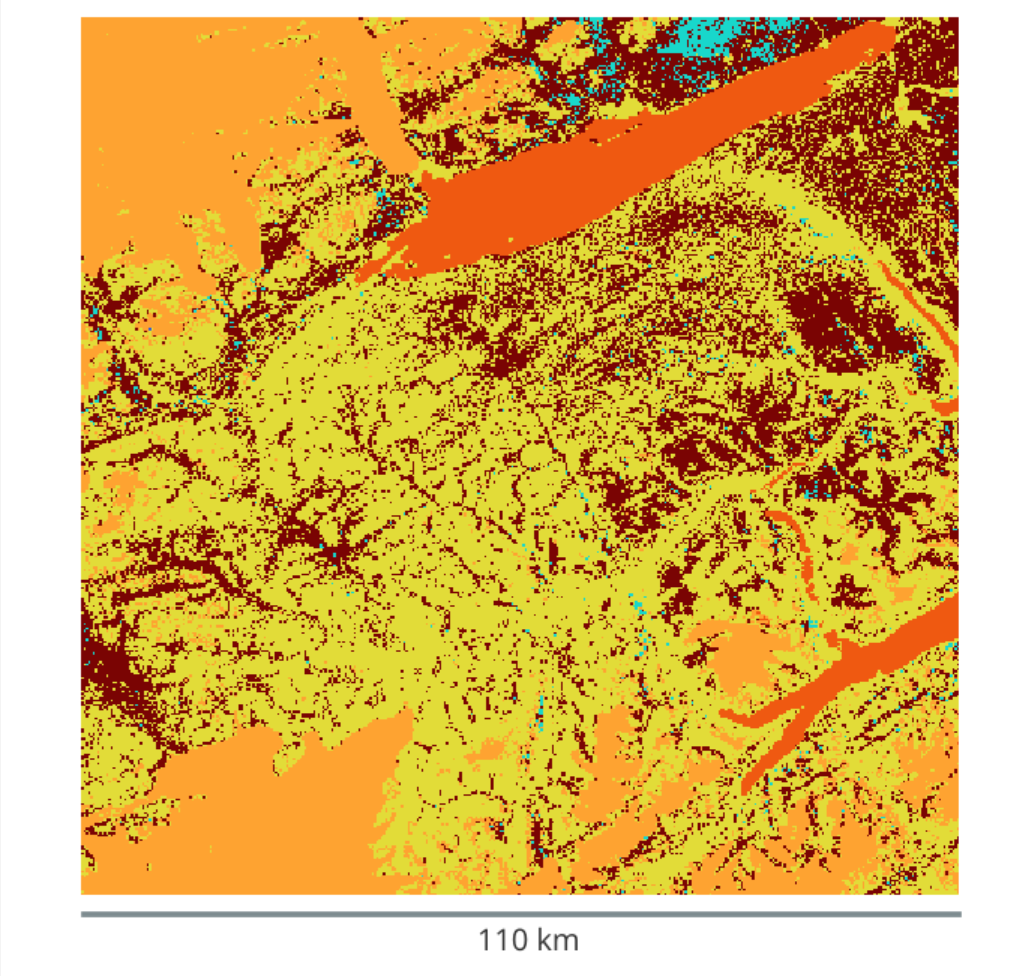}
        \caption{ESA WorldCover map 10 m.}
        \label{fig:sub-h}
    \end{subfigure}
    \hfill 
    \begin{subfigure}[b]{0.3\linewidth}
        \centering
        \includegraphics[width=\textwidth]{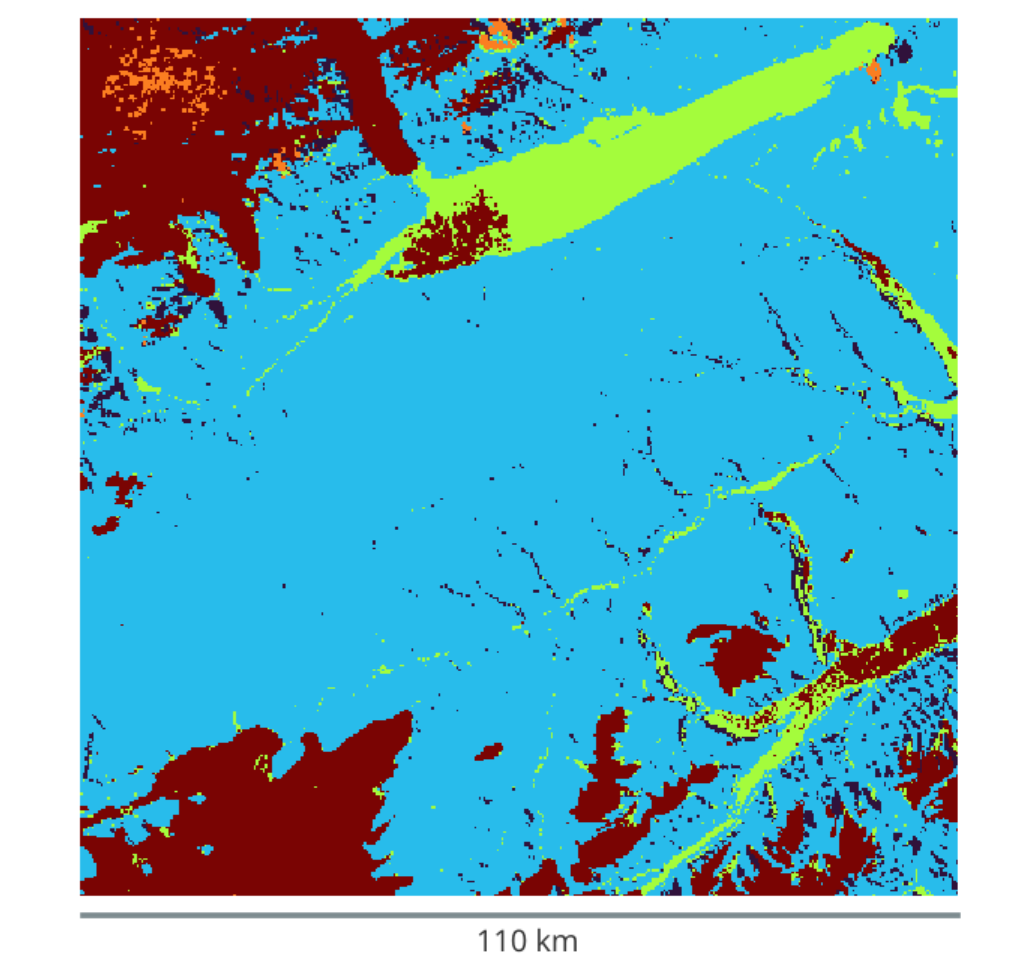}
        \caption{Sentinel-2 SCL map 20 m.}
        \label{fig:sub-i}
    \end{subfigure}
    \hfill 
    \begin{subfigure}[b]{0.3\linewidth}
        \centering
        \includegraphics[width=\textwidth]{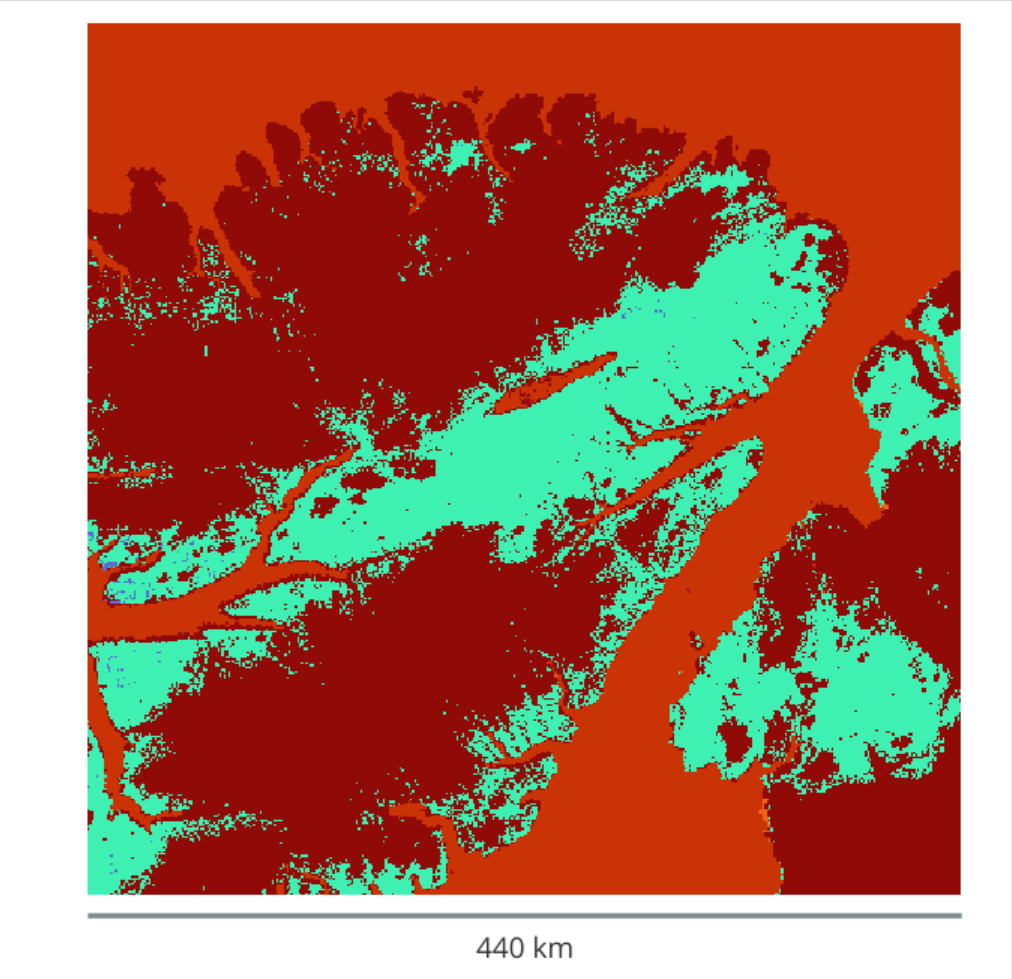}
        \caption{ESA GlobCover map 250 m}
        \label{fig:sub-j}
    \end{subfigure}
    \hfill 
    \begin{subfigure}[b]{0.3\linewidth}
        \centering
        \includegraphics[width=\textwidth]{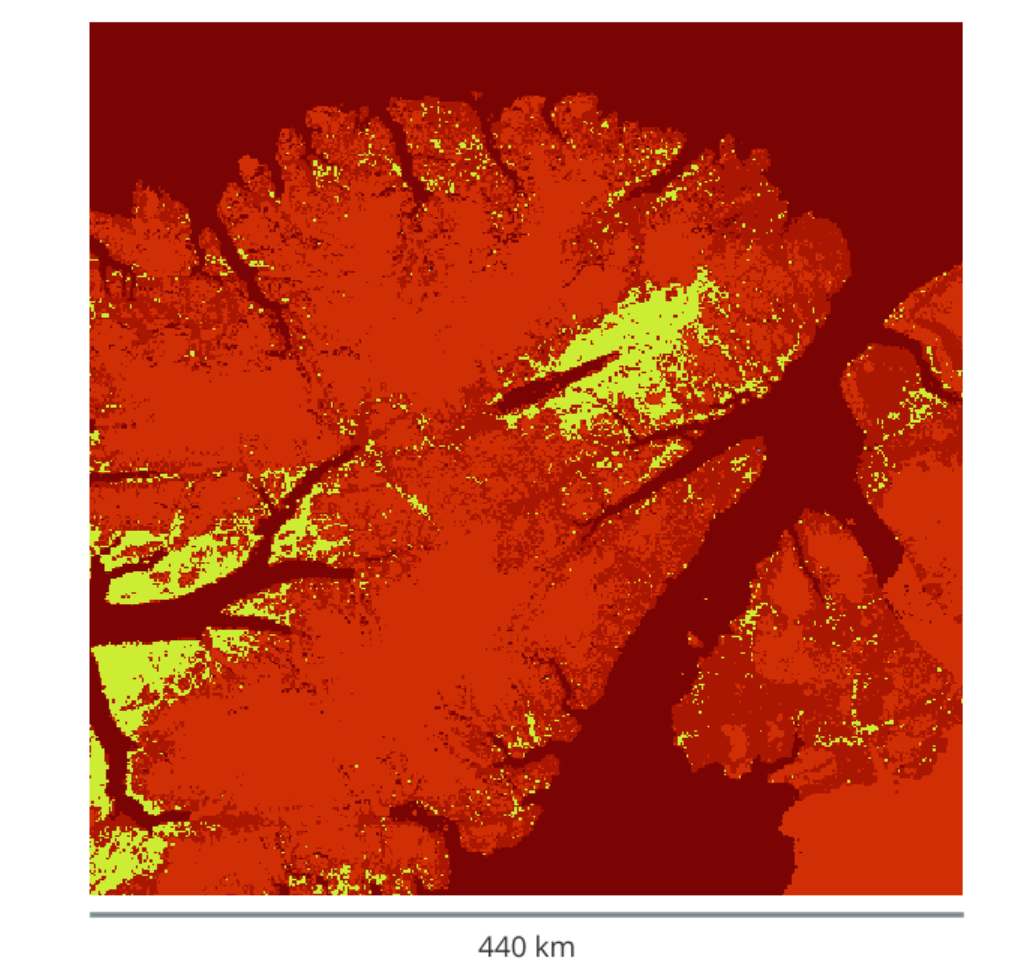}
        \caption{MOD12Q1 map 500 m}
        \label{fig:sub-k}
    \end{subfigure}

    \caption{Example images, from tile T19XDL on 2020-07-17}
    \label{fig:example_tile_data}
\end{figure*}

To qualitatively validate our alignment pipeline, Fig.~\ref{fig:example_tile_data} visualizes a complete sample tuple from the dataset. This visualization highlights the extreme heterogeneity THOR must resolve: the model must reconcile fine-grained textural details from the Sentinel-2 and Sentinel-1 (10 m) inputs with the broad-scale climatic context provided by the Sentinel-3 sensors.

As illustrated by the bounding boxes, the dataset preserves the spatial hierarchy of the sensors. The Sentinel-3 inputs cover a spatial footprint 25 times larger than the Sentinel-2 anchor tile (Figs.~\ref{fig:sub-d} - \ref{fig:sub-f}), ensuring that the model captures large-scale atmospheric and thermal gradients that would be imperceptible in a narrow field-of-view crop. The inclusion of aligned DEM and Land Cover maps (Figs.~\ref{fig:sub-g} - \ref{fig:sub-k}) further confirms that the model receives dense topographic and semantic supervision alongside the raw radiometric data

\begin{table*}[t]
    \centering
    \caption{Description of ERA5-Land variables used to pre-train THOR, and included in THOR Pretrain.}
    \label{tab:variable_description}
    \begin{tabularx}{\textwidth}{l l >{\raggedright\arraybackslash}X}
        \toprule
        \textbf{Variable Name} & \textbf{Unit} & \textbf{Description} \\
        \midrule
        \texttt{volumetric\_soil\_water\_layer\_1} & None & Volumetric soil water fraction for layer 1 (0--7cm). \\
        \texttt{volumetric\_soil\_water\_layer\_4} & None & Volumetric soil water fraction for layer 4 (100--289cm). \\
        \texttt{skin\_temperature} & K & Temperature of the surface of the Earth. \\
        \texttt{dewpoint\_temperature\_2m} & K & Temperature at 2m to which air must be cooled for saturation. \\
        \texttt{temperature\_2m} & K & Air temperature at 2 meters above the surface. \\
        \texttt{soil\_temperature\_level\_1} & K & Soil temperature at layer 1 (0--7cm). \\
        \texttt{soil\_temperature\_level\_4} & K & Soil temperature at layer 4 (100--289cm). \\
        \texttt{snow\_cover} & None & Fraction of grid cell covered by snow. \\
        \texttt{snow\_depth\_water\_equivalent} & m & The depth of water that would result from melting the snow. \\
        \texttt{snowfall\_sum} & m & Accumulated snowfall (water equivalent). \\
        \texttt{snow\_depth} & m & Depth of the snowpack. \\
        \texttt{leaf\_area\_index\_high\_vegetation} & None & Leaf area index fraction for high vegetation (e.g., trees). \\
        \texttt{leaf\_area\_index\_low\_vegetation} & None & Leaf area index fraction for low vegetation (e.g., grass). \\
        \texttt{surface\_pressure} & Pa & Air pressure at the surface. \\
        \texttt{total\_precipitation\_sum} & m & Accumulated total precipitation (rain and snow). \\
        \texttt{surface\_runoff\_sum} & m & Accumulated water flowing over the land surface. \\
        \texttt{total\_evaporation\_sum} & m & Accumulated evaporation from the surface. \\
        \bottomrule
    \end{tabularx}
\end{table*}

\subsection{Stratified sampling strategy}\label{sec:stratified}
The global land cover is not homogeneous, but highly imbalanced. Over 70\% of the globe is covered with oceans, and constructing the dataset using uniformly sampling the Sentinel-2 tiles will result in a large part of ocean tiles. Even if we only sample only tiles covering land, we will get a bias towards forest, desert and shrublands. Since increasing the pretraining data diversity enhances SSL performance \cite{al2024pretraining}, we need to capture the variation of the land cover and sample the Sentinel-2 tiles in a stratified manner. 

\subsubsection{Land cover stratification}
First, we perform a ocean/land split, selecting 80\% of the Sentinel-2 tile from land areas. 

To capture diversity of land areas, the strategy is based on k-means clustering of extracted features \cite{ordonez2024towards, lai2023domain}. We use two data-sources to extract features from : ESA WorldCover maps and ESA Sentinel-2 RGB composite for 2022. Each of them are treated independently.

\begin{itemize}
    \item Feature extraction: For each tile location, we divide the corresponding image data (WorldCover and RGB composite) into $224\times 224$ crops. For ESA WorldCover maps we create a histogram of the 11 classes from each of crop, using bin counts as the feature vector. For the ESA Sentinel-2 2022 RGB composite, we use an ImageNet pre-trained ViT-MAE model to create a 786-dimensional embedding vector for each $224\times 224$ crop.
    \item Clustering and probability: K-means clustering (with 1000 clusters) is applied to group similar crops. The sampling probability for each tile location is determined as the inverse of its cluster size, emphasizing rarity.
    \item Tile selection: Tile sampling probability is the average of all crop probabilities within the tile, resulting in two probabilities: one from WorldCover and one from the Sentinel-2 RGB composite.
\end{itemize}

\subsubsection{Ocean data sampling}
To ensure comprehensive coverage of phenomena in the ocean, sampling probabilities utilize various maps:
\begin{itemize}
    \item World Bank Global Shipping Traffic Density maps are used to calculate the normalized density of ship traffic and oil and gas installations per Sentinel-2 tile.
    \item Areas with a higher probability of containing icebergs and sea ice are defined based on existing maps and observations (e.g., specific longitudes for sea ice, and two large regions in the southern hemisphere for icebergs).
\end{itemize}

\subsection{Final location sampling routine}
The final per-tile sampling probabilities are a weighted combination of the land and ocean diversity scores, with an 80/20 split between land and ocean tiles. The detailed stratification for land and ocean samples is shown in Table~\ref{tab:sampling_probabilities}.

\begin{table*}[h!]
\centering
\caption{Combined PR-tile sampling probabilities}
\begin{tabular}{llc}
\hline
\textbf{Category} & \textbf{Sub-category} & \textbf{Pct. of category} \\ \hline
Land & Uniformly sampled from all land tiles & 15\% \\
(80\%)  & Sampled from ESA WorldCover diversity & 75\% \\
 & Sampled from Sentinel-2 RGB composite & 10\% \\ \hline
Ocean & Uniformly sampled from all ocean tiles & 5\% \\
(20\%) & Uniformly sampled from all coast tiles & 40\% \\
 & Sampled from ship-density probabilities & 10\% \\
 & Sampled from oil \& gas installations-density & 2\% \\
 & Uniformly sampled from sea-ice areas & 30\% \\
 & Sampled from iceberg areas & 13\% \\ \hline
\end{tabular}

\label{tab:sampling_probabilities}
\end{table*}

\subsubsection{Temporal sampling}
\begin{table*}[t!]
    \centering
    \caption{Sampling probabilities versus cloud coverage.}
    \begin{tabular}{l*{10}{c}}
    \hline
    & \multicolumn{10}{c}{\textbf{Categories}} \\ \hline
    Cloud-cover interval [\%] & $>$10 & 10-20 & 20-30 & 30-40 & 40-50 & 50-60 & 60-70 & 70-80 & 80-90 & $<$90 \\
    Sampling probability [\%] & 97.66 & 0.99 & 0.49 & 0.25 & 0.12 & 0.06 & 0.06 & 0.06 & 0.06 & 0.25 \\
    \hline
    \end{tabular}
    \label{tab:sampling_vs_cloud}
\end{table*}

When a Sentinel-2 tile is sampled, the sampling routine selects an image from all available dates. While data constraints necessitate a balance between spatial and temporal coverage, the goal is to obtain an average of two images per tile.  This is implemented by using a Poisson distribution with an expectation of one to determine the number of \textit{additional} images to sample, ensuring at least one image per tile.  

We generally aim to have as low cloud cover as possible, but since the model will encounter clouded images in inference, we want THOR Pretrain to contain clouded images as well. Hence, we assign sampling probabilities for each 10\%-interval of cloud cover, and sample the image within (or as close as possible) to that interval (Tab.~\ref{tab:sampling_vs_cloud}). 

\subsection{Dataset summary}
THOR Pretrain consists of a total of number of tile and date combinations of 18332, with 6273 unique Sentinel-2 tiles and 2926 unique dates, from 2016-01-01 to 2024-05-27. 

\begin{figure}
    \centering
    \includegraphics[width=1.0\linewidth]{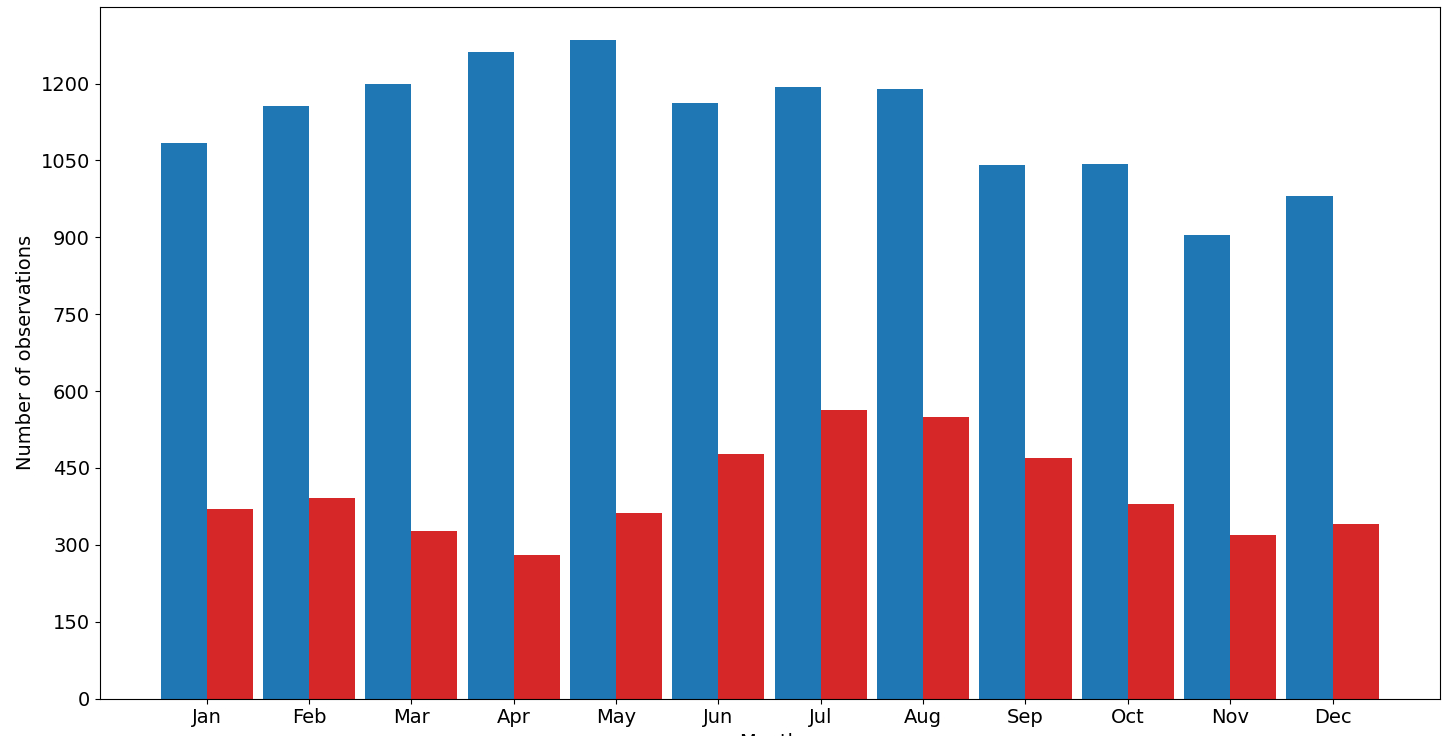}
    \caption{Number of observations per month for northern (blue) and southern (red) hemisphere.}
    \label{fig:montly_histogram}
\end{figure}

Fig.~\ref{fig:montly_histogram} illustrates the monthly distribution of the sampled observations, stratified by hemisphere. The distribution reveals two key characteristics of the dataset that align with the physical realities of optical remote sensing:

The total volume of samples from the Northern Hemisphere Fig.~\ref{fig:montly_histogram} (blue bars) is consistently higher than that of the Southern Hemisphere (red bars). This reflects the Earth’s geographical distribution, where approximately 68\% of the global landmass resides in the Northern Hemisphere. Since our stratified sampling strategy prioritizes land tiles (80\% land / 20\% ocean split), the dataset naturally mirrors this global land distribution.

\begin{table*}[t!]
    \centering
    \caption{Modality co-occurrence matrix (raw counts)}
    \label{tab:cooccurrence_raw}
    \begin{tabular}{lcccccccc}
        \toprule
        \textbf{Modality} & \textbf{S2} & \textbf{S1:10m} & \textbf{S1:60m} & \textbf{S3:OLCI} & \textbf{S3:SLSTR} & \textbf{LC} & \textbf{DEM:10m} & \textbf{DEM:60m} \\
        \midrule
        S2 & 15310 & 3393 & 3528 & 9860 & 10952 & 14776 & 14556 & 14722 \\
        S1 10m & 3393 & 4929 & 4896 & 2554 & 2919 & 4005 & 3966 & 3999 \\
        S1 60m & 3528 & 4896 & 5121 & 2652 & 3032 & 4158 & 4099 & 4150 \\
        S3 OLCI & 9860 & 2554 & 2652 & 11023 & 10574 & 10085 & 9927 & 10046 \\
        S3 SLSTR & 10952 & 2919 & 3032 & 10574 & 12605 & 11366 & 11186 & 11321 \\
        LC & 14776 & 4005 & 4158 & 10085 & 11366 & 16318 & 16095 & 16261 \\
        DEM 10m & 14556 & 3966 & 4099 & 9927 & 11186 & 16095 & 16095 & 16094 \\
        DEM 60m & 14722 & 3999 & 4150 & 10046 & 11321 & 16261 & 16094 & 16261 \\
        \bottomrule
    \end{tabular}
\end{table*}

To validate the multi-modal density of THOR Pretrain, Tab.~\ref{tab:cooccurrence_raw} presents the co-occurrence matrix of all available sensor modalities. This distribution reveals three critical characteristics of the dataset that directly motivated our architectural choices:
\begin{itemize}
    \item \textit{High-volume multi-resolution alignment:} Approximately 10,000 overlapping samples between Sentinel-2 and Sentinel-3 (OLCI/SLSTR) bridge the 10~m -- 1000~m resolution gap. This alignment enables the model to propagate fine-grained optical textures to coarse thermal and atmospheric readings.

    \item \textit{Dense token supervision for radar-optical fusion:} Although 3,400 aligned Sentinel-1/Sentinel-2 pairs appear low in raw count, they represent full $110 \times 110$~km tiles rather than crops, yielding hundreds of millions of pixel-aligned tokens. Combined with stratified sampling for geodiversity, this provides a dense signal for learning radar-optical distributions without the redundancy of uncurated datasets.
    
    \item \textit{Natural sparsity as a regularizer:} Variable sensor availability contrasts with the consistency of static auxiliary variables (land cover, DEM) across  approximately 16,300 locations. This natural sparsity validates our independent per-band projection layers, acting as a regularizer that forces robustness to missing modalities and prevents over-reliance on single sensors.
\end{itemize}

Is is important to note that we sample smaller crops from the full tiles during pre-training, i.e., Fig.~\ref{fig:example_tile_data} is only an illustration of what the modalities available. During pre-training, random image locations in a given tile is sampled, and smaller crops of each available modality corresponding to the the same footprint is extracted. 

\section{THOR foundation model implementation}

\subsection{Band groups}
\begin{table*}[t]
    \centering
    \caption{THOR input band grouping. Input bands are organized into 10 groups based on sensor source and spatial resolution. Note that Sentinel-1 data is split into coarse ($60~\text{m}$) and high-resolution ($10~\text{m}$) streams based on polarization/mode availability in the dataset. The Sentinel-1 IW and EW more are mutually exclusive. $\dagger$ During training the GSD of the SAR is aggregated ("multi-looked") to 10, 20, 30, 60, 120, 180 or 240 m.}
    \label{tab:band_groups}
    \begin{tabularx}{\textwidth}{c l >{\raggedright\arraybackslash}X c}
        \toprule
        \textbf{Group} & \textbf{Sensor} & \textbf{Bands} & \textbf{Default GSD (m)} \\
        \midrule
        1 & Sentinel-2 & Red, Green, Blue, NIR & 10 \\
        2 & Sentinel-2 & RE1, RE2, RE3, RE4, SWIR1, SWIR2 & 20 \\
        3 & Sentinel-2 & CoastAerosol, WaterVapor & 60 \\
        \midrule
        4 & Sentinel-1 & IW-VH, IW-VV, EW-VH, EW-VV  & 10/$60\dagger$ \\
        5 & Sentinel-1 & IW-HV, IW-HH, EW-HV, EW-HH  & 10/$60\dagger$ \\
        \midrule
        6 & Sentinel-3 OLCI & Oa01, Oa02, Oa03, Oa04, Oa05, Oa06, Oa07 & 240 \\
        7 & Sentinel-3 OLCI & Oa08, Oa09, Oa10, Oa11, Oa12, Oa13, Oa14 & 240 \\
        8 & Sentinel-3 OLCI & Oa15, Oa16, Oa17, Oa18, Oa19, Oa20, Oa21 & 240 \\
        \midrule
        9 & Sentinel-3 SLSTR & S1, S2, S3, S4, S5, S6 (reflectance) & 480 \\
        10 & Sentinel-3 SLSTR & S7, S8, S9 (thermal BT) & 960 \\
        \bottomrule
    \end{tabularx}
\end{table*}
To handle the heterogeneous resolutions of the input sensors efficiently, we organize the input bands into 10 distinct groups as detailed in Tab.~\ref{tab:band_groups}. Grouping is primarily determined by the native GSD and the sensor source.

By grouping bands of identical resolution (e.g., Sentinel-2 10~m bands in Group 1, Sentinel-1 10 m bands in Groups 4 \& 5), we allow the encoder to process each group with a patch number proportional to its information density. For instance, the thermal bands from Sentinel-3 (Group 10, 960~m GSD) require significantly fewer tokens than the optical bands from Sentinel-2 (Group 1, 10~m GSD) for the same input image footprint. This grouping strategy is fundamental to our token budget heuristic, ensuring that high-frequency spatial details are preserved where available, while minimizing computational waste on coarser modalities.

\subsection{Multi-looking}
Multi-looking is often applied in SAR applications to reduce speckle noise, a granular distortion inherent to coherent imaging systems like radar \cite{oliver2004understanding}. By averaging independent "looks" (images) of the same scene, the random noise is smoothed out, which improves the image's radiometric quality at the expense of its spatial resolution. While aggregating the, say 10~m GRD pixels to 50~m, achieves a similar result in terms of reducing speckle and lowering resolution, it is technically not referred to as "multi-looking" in strict SAR processing terminology.

THOR has been pretrained using a random "multi-looking" by aggregating pixels to 10~m, 20~m, 30~m, 60~m, 120~m, 180~m or 240~m. 

\subsection{Model configurations}
\begin{table*}[t]
    \centering
    \caption{\textbf{THOR model family configurations.} Hyperparameters for the Tiny, Small, Base, and Large variants. All models support dynamic input resolutions and patch sizes ($4^2$ to $32^2$) during pre-training. The Token budget is an approximate cap enforced during training to manage memory usage across heterogeneous inputs and is set to 1296 for all variants. The learning rate is set as \lstinline{base_lr * (batch_size * num_gpu) / 256}.}
    \label{tab:model_config}
    \resizebox{0.9\textwidth}{!}{%
    \begin{tabular}{lcccccccc}
        \toprule
        \textbf{Model} & \textbf{Layers} & \textbf{Embed dim} & \textbf{Heads} & \textbf{MLP ratio} & \textbf{Params} & \textbf{Base Training LR} & \textbf{Warmup epochs} \\
        \midrule
        THOR-Tiny   & 12 & 192  & 3  & 4 & $\sim$7.6M   & 4e-4 & 10 \\
        THOR-Small  & 12 & 384  & 6  & 4 & $\sim$25.8M  & 4e-4 & 10 \\
        THOR-Base   & 12 & 768  & 12 & 4 & $\sim$94.1M  & 3e-4 & 20 \\
        THOR-Large  & 24 & 1024 & 16 & 4 & $\sim$314.4M & 3e-4 & 40 \\
        \bottomrule
    \end{tabular}
    }
\end{table*}

We train a family of THOR models ranging from Tiny to Large to evaluate scaling laws and deployment versatility. The specific architectural hyperparameters for each variant are provided in Tab.~\ref{tab:model_config}. All models share the same unified encoder-decoder architecture but vary in embedding dimension, number of heads, and depth. Crucially, all variants support the dynamic input resolution ($32^2$ to $1024^2$) and randomized patch sizes ($4^2$ to $32^2$) described in the main text. 

\subsection{Token budget heuristic}
Processing multi-modal data with randomized input sizes and patch sizes can lead to exploding sequence lengths if left unchecked. To address this, we implement a dynamic token budget heuristic, formally described in Algorithm~\ref{alg:token_budget}. The algorithm operates by first sampling a global spatial footprint (C, C) in meters. For each band group $g$, we calculate number of tokens required based on a sampled patch size $P_g$ and the GSD of the band group. If the cumulative number of tokens approaches the pre-defined maximum token budget, the algorithm dynamically adjusts the minimum allowable patch size for subsequent groups or caps the resolution. The ordering of the groups are randomly permuted ensuring no bias in the algorithm. This ensures that every training batch maximizes GPU utilization without causing Out-Of-Memory errors, regardless of the random footprint sampled.

\begin{algorithm*}[t]
\caption{THOR dynamic token budget heuristic (ground-cover based)}
\label{alg:token_budget}
\begin{algorithmic}[1]
\State \textbf{Hyperparameters:}
\State $B_{max} \gets$ Maximum token budget (e.g., 1296)
\State $C_{range} \gets [960, 46080]$ \Comment{Ground-cover range (meters)}
\State $P_{range} \gets [P_{\min}, P_{\max}] = [4, 32]$ \Comment{Patch size range (pixels)}
\State $Groups \gets$ List of sensor groups (e.g., [S1, S2, S3-OLCI, ...])

\Statex
\Function{SamplePatchParameters}{$Groups, C, B_{max}$}
    \State \Comment{$C$ is sampled: $C \sim \mathcal{U}(C_{range})$}
    \State Randomly permute $Groups$ to get $(g_1, \dots, g_G)$
    \State $T_{used} \gets 0$

    \For{each group $g$ in $(g_1, \dots, g_G)$}
        \State $H_g \gets \left\lfloor \dfrac{C}{g.\text{GSD}} \right\rfloor$; \quad $W_g \gets H_g$
        \State $B_{remain} \gets B_{max} - T_{used}$
        \If{$B_{remain} \le 0$}
            \State \textbf{break}
        \EndIf

        \State \Comment{Token range as in implementation (2--32 grid limit)}
        \State $T_{min} \gets \left(\max\!\big(2,\; \lfloor H_g / P_{\max} \rfloor\big)\right)^2$
        \State $T_{max} \gets \left(\min\!\big(32,\; \lceil H_g / P_{\min} \rceil\big)\right)^2$

        \If{$T_{min} > B_{remain}$}
            \State \textbf{continue} \Comment{Not enough budget for this group}
        \EndIf

        \State $T_{target} \gets \min(T_{max}, B_{remain})$
        \State $G_{target} \gets \sqrt{T_{target}}$ \Comment{Target grid size per side}
        \State $P_g \gets \text{clip}\!\left(\left\lfloor \dfrac{H_g}{G_{target}} \right\rfloor,\, P_{\min},\, P_{\max}\right)$

        \State $T_{group} \gets \left\lceil \dfrac{H_g}{P_g} \right\rceil \times \left\lceil \dfrac{W_g}{P_g} \right\rceil$
        \State $T_{used} \gets T_{used} + T_{group}$
    \EndFor

    \State \Return $\{(H_g, W_g, P_g) \mid g \in Groups\ \text{with allocated budget}\}$
\EndFunction
\end{algorithmic}
\end{algorithm*}

\subsection{Loss Details}

\begin{table}[htbp]
\centering
\caption{Loss function weights used in pre-training}
\label{tab:loss_weights}
\begin{tabular}{lcc}
\toprule
\textbf{Loss Component} & \textbf{Lambda} & \textbf{Weight Value} \\
\midrule
Reconstruction & $\lambda_1$ & $1.5$ \\
Contrastive & $\lambda_2$ & $0.1$ \\
ERA5 & $\lambda_4$ & $0.1$ \\
Month & $\lambda_5$ & $0.1$ \\
Coordinates & $\lambda_6$ & $0.1$ \\
Orbit direction & $\lambda_7$ & $0.1$ \\
Incidence angle & $\lambda_8$ & $0.1$ \\
FFT & $\lambda_9$ & $0.01$ \\
\midrule
\multicolumn{3}{l}{\textbf{Map prediction losses:}} \\
\quad SCL & $\lambda_{3,\text{SCL}}$ & $0.05$ \\
\quad World Cover & $\lambda_{3,\text{WC}}$ & $0.1$ \\
\quad Global Canopy & $\lambda_{3,\text{GC}}$ & $0.1$ \\
\quad MCD12Q1 & $\lambda_{3,\text{MCD}}$ & $0.1$ \\
\quad DEM & $\lambda_{3,\text{DEM}}$ & $0.1$ \\
\bottomrule
\end{tabular}
\end{table}

The total loss $\mathcal{L}_{total}$ is a weighted sum of reconstruction, contrastive, and task-specific prediction losses. The specific weights ($\lambda$) assigned to each component are listed in Tab.~\ref{tab:loss_weights}.We prioritize the reconstruction objective ($\lambda_1 = 1.5$) as it is the primary driver of feature learning in the MAE framework. The auxiliary tasks (ERA5, map prediction, orbital regression) are weighted lower (0.05 - 0.1) to act as regularizers and semantic guides without overwhelming the pixel-level reconstruction signal. The FFT loss \cite{kraus_masked_2024} is included with a small weight to stabilize high-frequency feature reconstruction.

\section{Experiments}

\subsection{Extensive Pangaea results}
We provide the complete tabulation of results for the Pangaea benchmark suite \cite{marsocci2024pangaea} across three data availability regimes: 10\% (Tab. \ref{tab:10_perc_training_data_extended}), 50\% (Tab. \ref{tab:50_perc_training_data_extended}), and 100\% (Tab.~\ref{tab:100_perc_training_data_extended}). All THOR family model experiments are with patch size 6, input size of 108 and concatenation of the output features. These experiments validate that THOR provides good performance in low training data regimes. 

In the 10\% regime (Tab. \ref{tab:10_perc_training_data_extended}), THOR-Base performs on par with the current state-of-the-art, TerraMind-Base \cite{jakubik2025terramind}, by on fine-grained segmentation tasks. This confirms that our flexible patching strategy, which allows for dense token representations at inference time, compensates for the lack of training labels by providing a richer signal to the decoder. For the full training dataset, TerraMind show strong  performance (achieving the top rank on average), but THOR-Base remains highly competitive, outperforming the other models on PASTIS and CropMap tasks (Tab.~\ref{tab:100_perc_training_data_extended}).

\begin{table*}[h!]
\caption{Extended Pangaea results with 10\% training data in mIoU. Bold/underline mark best/second-best per column.}
\centering
\resizebox{0.8\textwidth}{!}{%
\begin{tabular}{lccccccccccc}
\toprule
Model & HLS Burns & MADOS & PASTIS & Sen1Floods11 & FBP & DynEarthNet & CropMap & SN7 & AI4Farms & Avg. Rank \\
\midrule
CROMA & 76.44 & 32.44 & 32.80 & 87.22 & 37.39 & 36.08 & 36.77 & 42.15 & 38.48 & 7.44 \\
DOFA & 71.98 & 23.77 & 27.68 & 82.84 & 27.82 & \textbf{39.15} & 29.91 & 46.10 & 27.74 & 11.78 \\
GFM-Swin & 67.23 & 28.19 & 21.47 & 62.57 & 55.58 & 28.16 & 27.21 & 39.48 & 32.88 & 14.11 \\
Prithvi & 77.73 & 21.24 & 33.56 & 86.28 & 29.98 & 32.28 & 27.71 & 36.78 & 35.04 & 11.67 \\
RemoteCLIP & 69.40 & 20.57 & 17.19 & 62.22 & \underline{56.23} & 34.43 & 19.86 & 43.11 & 23.85 & 13.89 \\
SatlasNet & 74.79 & 29.87 & 16.76 & 83.92 & 37.86 & 34.64 & 29.08 & 49.78 & 13.91 & 12.11 \\
Scale-MAE & 75.47 & 21.47 & 22.86 & 64.74 & 48.75 & 35.27 & 13.44 & 49.68 & 26.66 & 12.33 \\
SpectralGPT & \textbf{83.35} & 20.29 & 34.53 & 83.12 & 39.51 & 35.33 & 31.06 & 36.31 & 37.35 & 10.00 \\
S12-MoCo & 73.11 & 19.47 & 32.51 & 79.58 & 35.57 & 32.24 & 36.54 & 49.46 & 37.97 & 12.44 \\
S12-DINO & 75.93 & 23.47 & 36.62 & 84.95 & 34.63 & 32.78 & 38.44 & 41.15 & 37.91 & 9.78 \\
S12-MAE & 76.60 & 18.44 & 31.06 & 84.81 & 35.56 & 30.59 & 35.29 & 40.51 & 23.60 & 13.00 \\
S12-Data2Vec & 74.38 & 17.86 & 33.09 & 81.91 & 37.27 & 33.63 & 34.11 & 40.66 & 22.85 & 13.89 \\
Terramind-B & 77.39 & \textbf{44.06} & \textbf{39.96} & 84.43 & 54.00 & \underline{37.35} & 35.65 & 43.21 & 38.59 & \textbf{4.56} \\
\midrule
UNet Baseline & \underline{79.46} & 24.30 & 29.53 & \textbf{88.55} & 52.58 & 35.59 & 13.88 & 46.08 & 34.84 & 8.11 \\
ViT Baseline & 75.92 & 10.18 & 38.44 & 81.85 & \textbf{56.53} & 35.39 & 27.76 & 36.01 & \textbf{39.20} & 10.11 \\
\midrule
THOR-Tiny & 75.98 & 37.63 & 36.26 & 82.70 & 42.81 & 34.03 & 37.82 & \underline{58.5}2 & 38.56 & 7.11 \\
THOR-Small & 77.29 & \underline{42.64} & 38.48 & 84.21 & 42.81 & 35.31 & \underline{40.39} & \textbf{59.41} & 12.31 & 6.56 \\
THOR-Base & 76.90 & 40.67 & 38.93 & 86.29 & 42.80 & 35.21 & \textbf{42.23} & 55.93 & \underline{38.90} & \underline{4.67} \\
THOR-Large & 75.57 & 36.43 & \underline{39.21} & \underline{87.34} & 43.51 & 36.10 & 36.77 & 55.79 & 18.26 & 6.44 \\
\bottomrule
\end{tabular}
}
\label{tab:10_perc_training_data_extended}
\end{table*}

\begin{table*}[h!]
\caption{Extended Pangaea results with 50\% training data in mIoU. Bold/underline mark best/second-best per column.}
\centering
\resizebox{0.8\textwidth}{!}{%

\begin{tabular}{lcccccccccc}
\toprule
Model & HLS Burns & MADOS & PASTIS & Sen1Floods11 & FBP & DynEarthNet & CropMap & SN7 & AI4Farms & Avg. Rank \\
\midrule
CROMA & \underline{81.52} & 57.68 & 32.33 & \underline{90.57} & 48.01 & \underline{38.30} & 42.20 & 59.31 & 28.19 & \textbf{5.11} \\
DOFA & 78.02 & 55.21 & 28.60 & 88.39 & 36.90 & \textbf{39.20} & 30.93 & 47.06 & 26.69 & 11.78 \\
GFM-Swin & 74.36 & \textbf{63.37} & 20.41 & 71.61 & \underline{63.14} & 31.25 & 31.42 & 59.83 & 28.43 & 10.33 \\
Prithvi & 80.89 & 40.79 & 33.13 & 89.69 & 40.27 & 33.43 & 42.51 & 49.45 & 29.27& 9.33 \\
RemoteCLIP & 74.28 & 53.26 & 17.46 & 71.67 & \textbf{65.92} & 30.91 & 36.3 & 50.83 & 25.11 & 13.11 \\
SatlasNet & 75.97 & 52.24 & 16.78 & 89.45 & 46.04 & 36.34 & 35.29 & \underline{60.74} & 27.08 & 10.00 \\
Scale-MAE & 75.47 & 46.87 & 23.26 & 72.54 & 62.11 & 32.60 & 20.32 & \textbf{61.24} & 26.40 & 12.33 \\
SpectralGPT & 76.40 & \underline{58.00} & 34.61 & 87.52 & 21.71 & 36.52 & 32.09 & 56.28 & 27.46 & 10.56 \\
S12-MoCo & 79.79 & 42.90 & 32.59 & 89.22 & 46.92 & 34.45 & 41.32 & 56.21 & 28.38 & 8.89 \\
S12-DINO & 80.12 & 40.42 & 35.71 & 88.93 & 44.85 & 32.76 & 31.13 & 55.14 & 25.68 & 12.33 \\
S12-MAE & 80.13 & 44.29 & 31.15 & 88.43 & 45.63 & 33.29 & 28.07 & 55.55 & 27.50 & 11.44 \\
S12-Data2Vec & 79.82 & 41.22 & 33.42 & 86.58 & 46.73 & 32.61 & 28.53 & 56.94 & 25.84 & 11.89 \\
\midrule
UNet Baseline & \textbf{82.39} & 43.87 & 30.25 & \textbf{90.91} & 55.42 & 35.14 & 36.30 & 46.82 & \textbf{45.02} & 7.89 \\
ViT Baseline & 78.17 & 28.77 & 38.71 & 86.08 & 57.32 & 37.33 & 39.53 & 49.21 & \underline{38.37} & 9.00 \\
\midrule
THOR-Tiny & 78.22 & 53.87 & 36.40 & 89.29 & 45.20 & 35.00 & 48.58 & 60.03 & 27.67 & 7.44 \\
THOR-Small & 79.14 & 52.14 & 38.20 & 90.42 & 44.99 & 36.64 & 45.41 & 59.46 & 27.95 & 7.11 \\
THOR-Base & 79.15 & 49.43 & \underline{39.50} & 89.05 & 45.39 & 36.66 & \underline{50.81} & 60.24 & 27.76 & 6.67 \\
THOR-Large & 76.69 & 52.30 & \textbf{39.96} & 89.92 & 46.03 & 37.36 & \textbf{54.88} & 59.67 & 27.96 & \underline{5.78} \\
\bottomrule
\end{tabular}
}
\label{tab:50_perc_training_data_extended}
\end{table*}

\begin{table*}[t]
\caption{Extended Pangaea results with 100\% training data in mIoU. Bold/underline mark best/second-best per column.}
\centering
\resizebox{0.8\textwidth}{!}{%
\begin{tabular}{lccccccccccc}
\toprule
Model & HLS Burns & MADOS & PASTIS & Sen1Floods11 & FBP & DynEarthNet & CropMap & SN7 & AI4Farms & Avg. Rank \\
\midrule
CROMA & 82.42 & \underline{67.55} & 32.32 & \underline{90.89} & 51.83 & 38.29 & 49.38 & 59.28 & 25.65 & 7.00 \\
DOFA & 80.63 & 59.58 & 30.02 & 89.37 & 43.18 & \underline{39.29} & 51.33 & 61.84 & 27.07 & 8.33 \\
GFM-Swin & 76.90 & 64.71 & 21.24 & 72.60 & 67.18 & 34.09 & 46.98 & 60.89 & 27.19 & 10.67 \\
Prithvi & \underline{83.62} & 49.98 & 33.93 & 90.37 & 46.81 & 27.86 & 43.07 & 56.54 & 26.86 & 12.11 \\
RemoteCLIP & 76.59 & 60.00 & 18.23 & 74.26 & \textbf{69.19} & 31.78 & 52.05 & 57.76 & 25.12 & 12.56 \\
SatlasNet & 79.96 & 55.86 & 17.51 & 90.30 & 50.97 & 36.31 & 46.97 & 61.88 & 25.13 & 10.67 \\
Scale-MAE & 76.68 & 57.32 & 24.55 & 74.13 & \underline{67.19} & 35.11 & 25.42 & \textbf{62.96} & 21.47 & 12.44 \\
SpectralGPT & 80.47 & 57.99 & 35.44 & 89.07 & 33.42 & 37.85 & 46.95 & 58.86 & 26.75 & 10.67 \\
S12-MoCo & 81.58 & 51.76 & 34.49 & 89.26 & 53.02 & 35.44 & 48.58 & 57.64 & 25.38 & 11.00 \\
S12-DINO & 81.72 & 49.37 & 36.18 & 88.61 & 51.15 & 34.81 & 48.66 & 56.47 & 25.62 & 12.11 \\
S12-MAE & 81.91 & 49.90 & 32.03 & 87.79 & 51.92 & 34.08 & 45.8 & 57.13 & 24.69 & 13.56 \\
S12-Data2Vec & 81.91 & 44.36 & 34.32 & 88.15 & 48.82 & 35.90 & 54.03 & 58.23 & 24.23 & 11.89 \\
TerraMindv1-B & 82.42 & \textbf{69.52} & \underline{40.51} & 90.62 & 59.72 & 37.87 & 55.80 & 60.61 & 28.12 & \textbf{3.56} \\
\midrule
UNet Baseline & \textbf{84.51} & 54.79 & 31.60 & \textbf{91.42} & 60.47 & \textbf{39.46} & 47.57 & \underline{62.09} & \textbf{46.34} & \underline{5.00} \\
ViT Baseline & 81.58 & 48.19 & 38.53 & 87.66 & 59.32 & 36.83 & 44.08 & 52.57 & \underline{38.37} & 11.11 \\
\midrule
THOR-Tiny & 79.34 & 53.82 & 38.02 & 89.35 & 46.41 & 33.59 & 50.39 & 60.61 & 26.36 & 11.11 \\
THOR-Small & 79.26 & 52.66 & 39.54 & 90.14 & 47.25 & 34.84 & \underline{59.49} & 60.01 & 26.91 & 9.11 \\
THOR-Base & 79.65 & 51.48 & \textbf{40.76} & 89.44 & 47.42 & 37.57 & 56.78 & 59.87 & 26.29 & 8.78 \\
THOR-Large & 79.47 & 53.73 & 39.88 & 89.55 & 47.62 & 37.29 & \textbf{60.75} & 59.71 & 26.75 & 8.33 \\
\bottomrule
\end{tabular}
}
\label{tab:100_perc_training_data_extended}
\end{table*}

\paragraph{Feature aggregation strategy}
THOR processes input groups independently, requiring a fusion strategy to combine features before the decoder. We compare mean aggregation (averaging token embeddings across groups) against concatenation (stacking tokens along the channel dimension). As shown in Tab.~\ref{tab:mean_vs_concat_miou_results}, concatenation consistently outperforms mean aggregation, achieving 52.94\% mIoU (vs. 49.90\%) in the 10\% data regime. This may suggest that distinct sensor modalities contain complementary, non-redundant information, and by averaging these features high-frequency modality-specific signals may be "washed out", whereas concatenation preserves the full feature variance often needed for fine-grained segmentation.

\paragraph{Validation of compute-adaptive patching}
A core premise of THOR is that smaller patch sizes yield denser feature maps, improving performance on pixel-level tasks. Tab.~\ref{tab:patch_size_4_6_8_miou_results} validates this hypothesis: reducing the patch size from 8 to 4 results in a significant performance boost, rising from 54.69\% to 58.63\% mIoU on the full dataset.

While smaller patches increase the sequence length (quadratic computational cost), they provide the necessary spatial granularity for segmentation tasks that coarse patches (e.g., $16\times16$) fail to resolve. This confirms that THOR’s randomized patch pre-training successfully enables test-time adaptation to higher resolutions.

\begin{table}[htbp]
\centering
\caption{Mean Pangaea test mIoU by output aggregation method and training data, THOR base model.}
\begin{tabular}{lcc}
\toprule
\textbf{Output Aggregation} & \textbf{10\%} & \textbf{100\%} \\
\midrule
concat & 52.94 & 58.63 \\
mean & 49.90 & 56.02 \\
\bottomrule
\end{tabular}
\label{tab:mean_vs_concat_miou_results}
\end{table}

\begin{table}[htbp]
\centering
\caption{Mean Pangaea test mIoU by patch size and training data, THOR base model.}
\begin{tabular}{lcc}
\toprule
\textbf{Patch Size} & \textbf{10\%} & \textbf{100\%} \\
\midrule
4 & 52.94 & 58.63 \\
6 & 50.82 & 54.36 \\
8 & 52.09 & 54.69 \\
\bottomrule
\end{tabular}
\label{tab:patch_size_4_6_8_miou_results}
\end{table}

\begin{figure}[t]
    \centering
    \includegraphics[width=1.0\columnwidth]{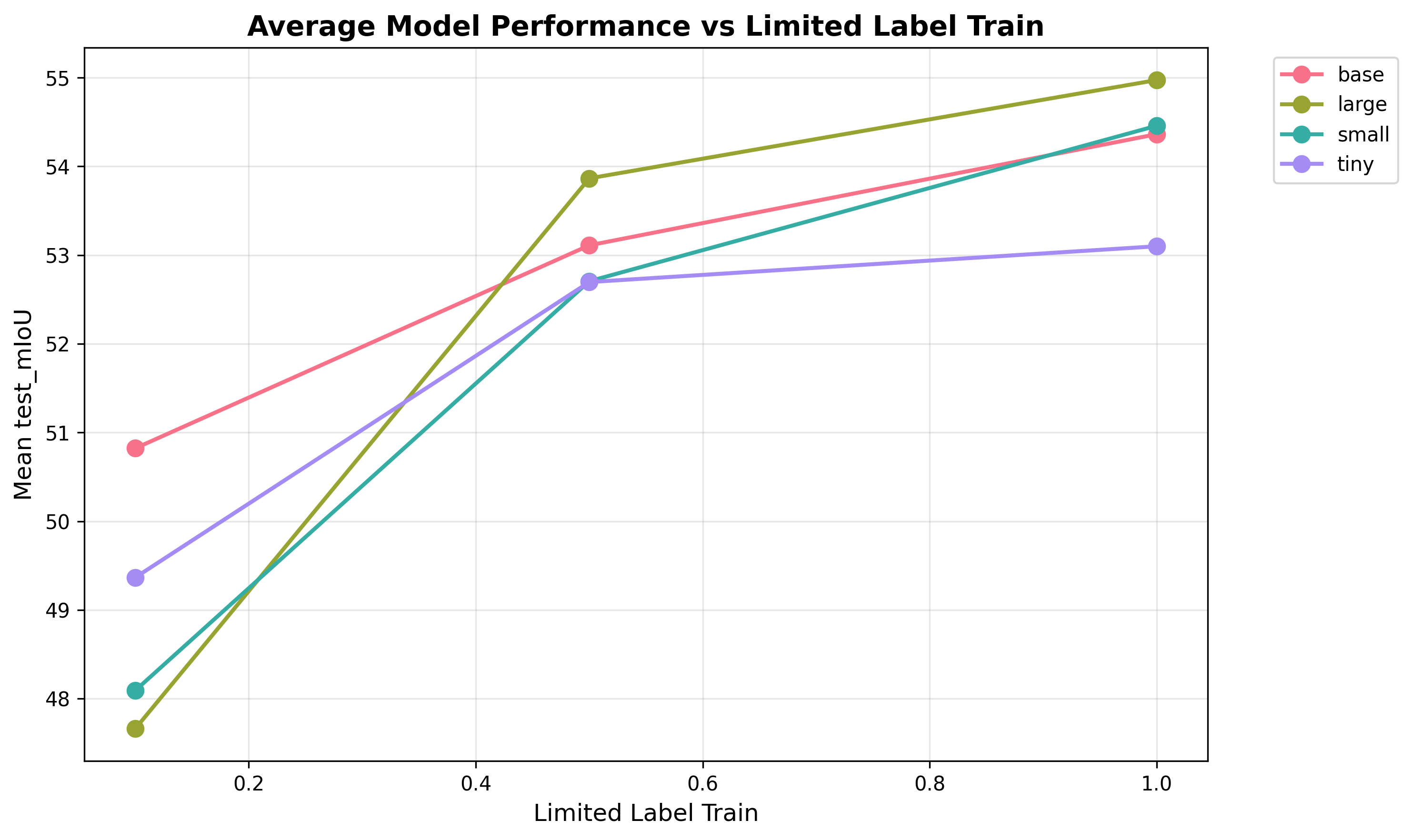}
    \caption{Aggregated mIoU over all Pangaea benchmarks using 10, 50 and 100\% training data for tiny, small, base and large model. Patch size 6, concat feature aggregation and input image size of 108 pixels.}
    \label{fig:pangaea_aggregated_results}
\end{figure}

\begin{figure*}[t]
    \centering
    \includegraphics[width=0.8\textwidth]{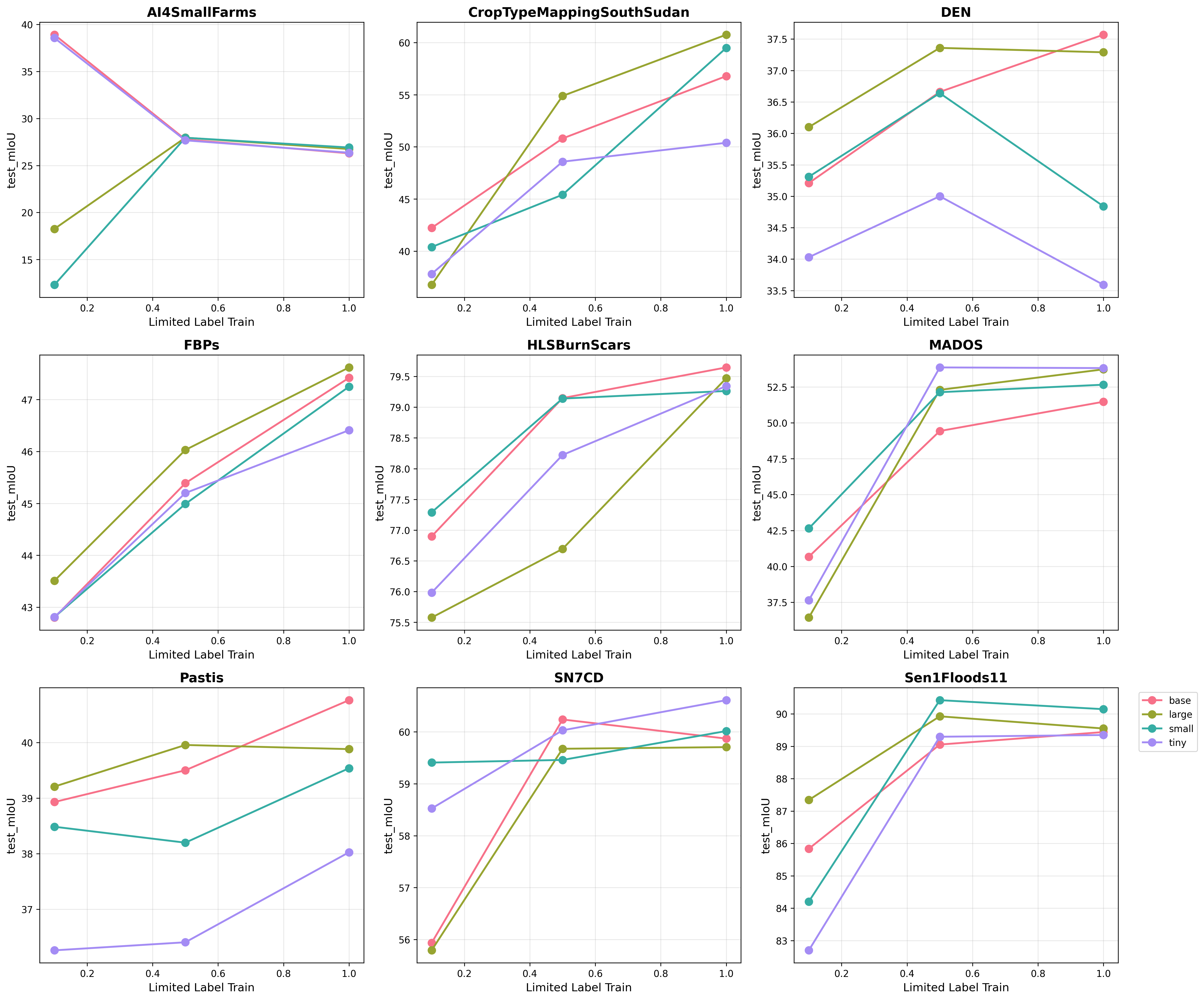}
    \caption{Per dataset mIoU for all Pangaea benchmarks using 10, 50 and 100\% training data for tiny, small, base and large model. Patch size 6, concat feature aggregation and input image size of 108 pixels.}
    \label{fig:pangaea_scaling_per_dataset}
\end{figure*}

\paragraph{Scaling and data efficiency}
Fig.~\ref{fig:pangaea_aggregated_results} Figure S.4 illustrates the scaling behavior of the THOR family (Tiny, Small, Base, Large) across data regimes. We observe a clear "crossover" effect: In data-scarce regimes (10\%), the THOR-Base model is the most robust performer. Notably, THOR-Large underperforms on the 10\% data (lowest starting point in Fig.~\ref{fig:pangaea_aggregated_results}), indicating that massive models may be prone to overfitting when fine-tuning data is insufficient. In data-rich regimes (50-100\%), THOR-Large recovers and surpasses all other variants, validating standard scaling laws where capacity correlates with performance given sufficient supervision. However, the performance depends strongly on the dataset, as observed in Fig.~\ref{fig:pangaea_scaling_per_dataset}, where we show per-dataset performance of each model in the THOR family. 

To further investigate the trade-off between computational cost and downstream performance, we conducted a series of experiments on four single-date Pangaea benchmarks using 10\% of the training data. We compared the performance of a standard UperNet decoder against a lightweight linear probe decoder across varying patch sizes.

\begin{figure*}[htbp]
\centering
\begin{subfigure}[b]{0.48\textwidth}
    \centering
    \includegraphics[width=\textwidth]{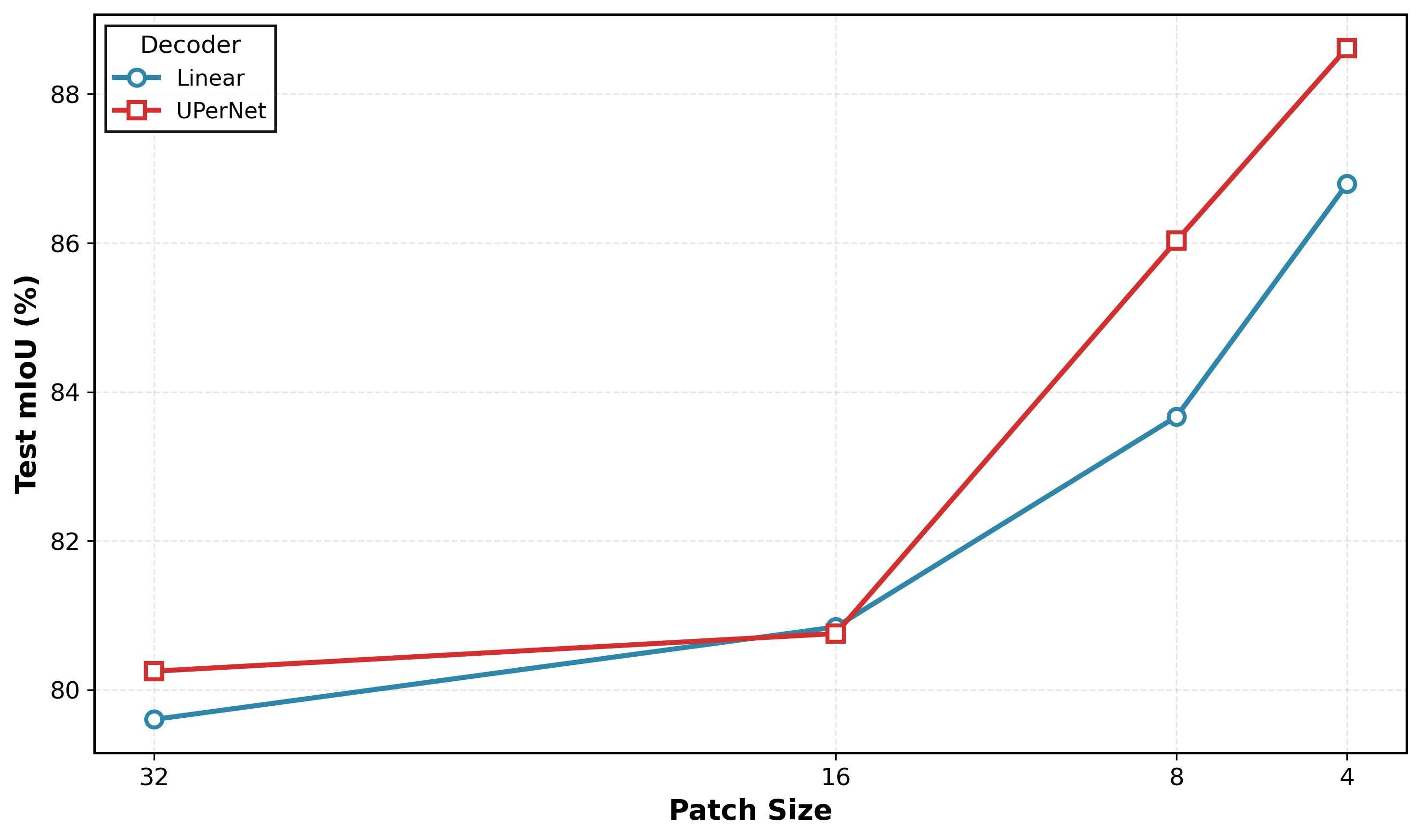}
    \caption{Sen1Floods11 - Patch Size}
    \label{fig:sen1floods11_patch}
\end{subfigure}
\hfill
\begin{subfigure}[b]{0.48\textwidth}
    \centering
    \includegraphics[width=\textwidth]{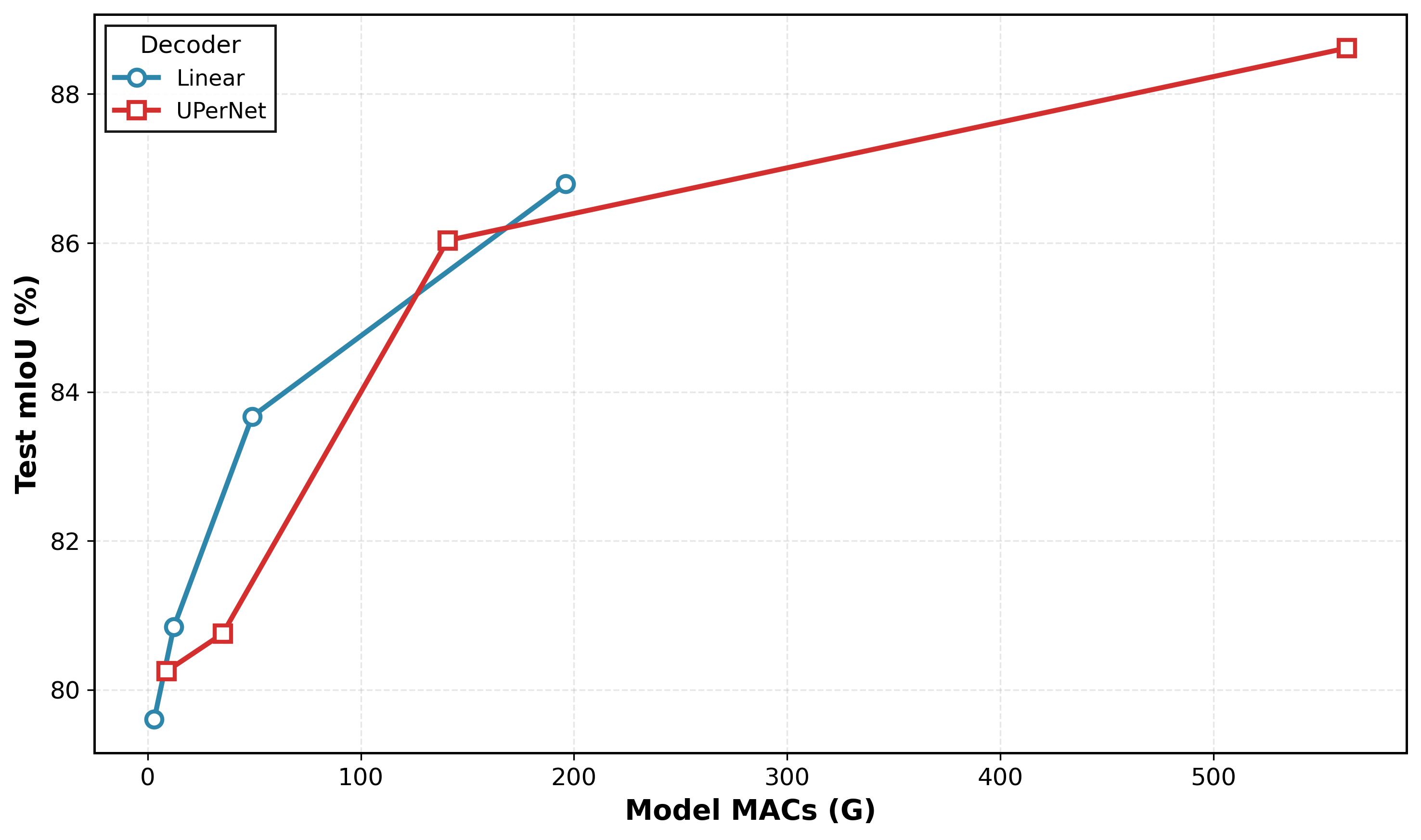}
    \caption{Sen1Floods11 - Model MACs}
    \label{fig:sen1floods11_macs}
\end{subfigure}

\vspace{0.5cm}

\begin{subfigure}[b]{0.48\textwidth}
    \centering
    \includegraphics[width=\textwidth]{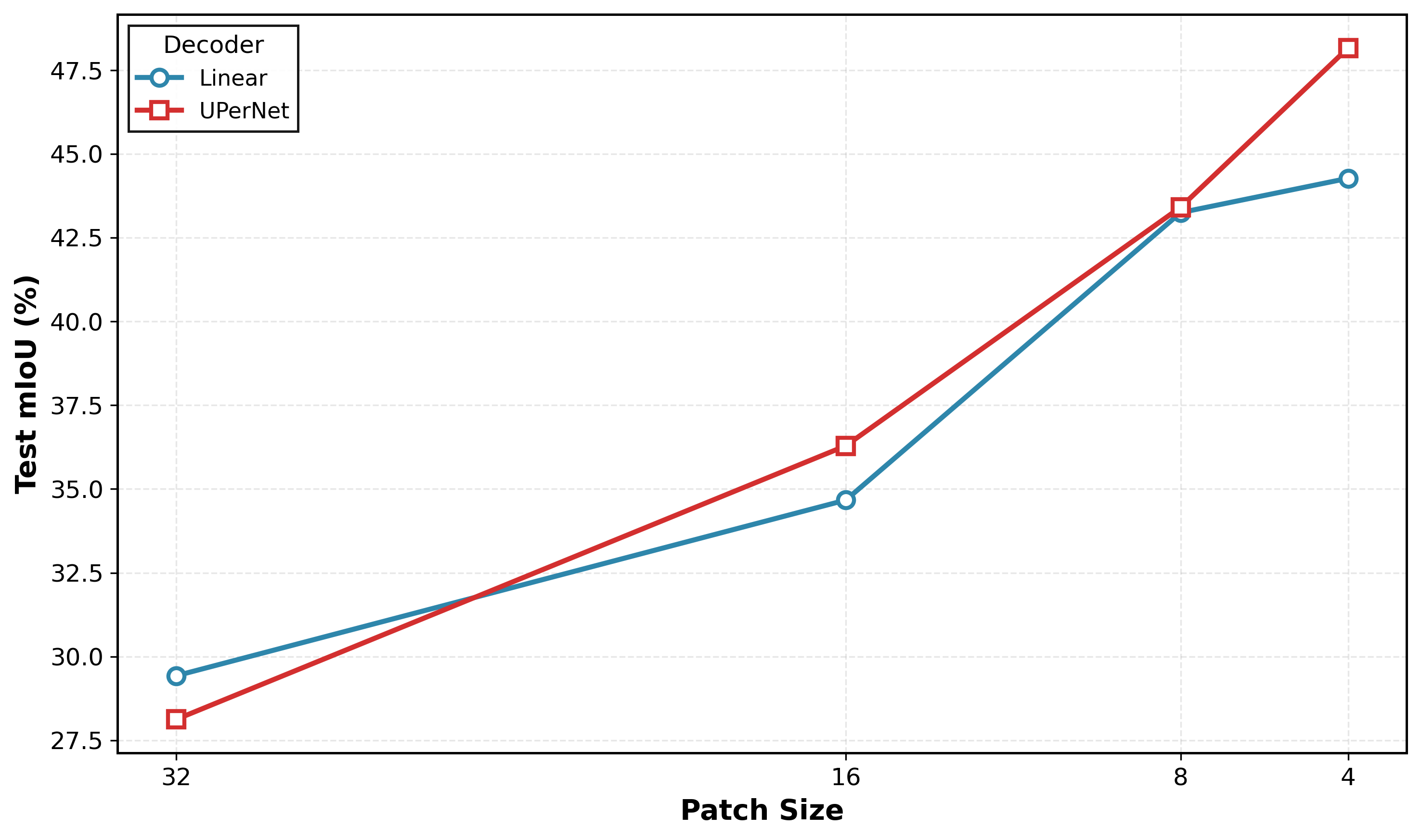}
    \caption{MADOS - Patch Size}
    \label{fig:mados_patch}
\end{subfigure}
\hfill
\begin{subfigure}[b]{0.48\textwidth}
    \centering
    \includegraphics[width=\textwidth]{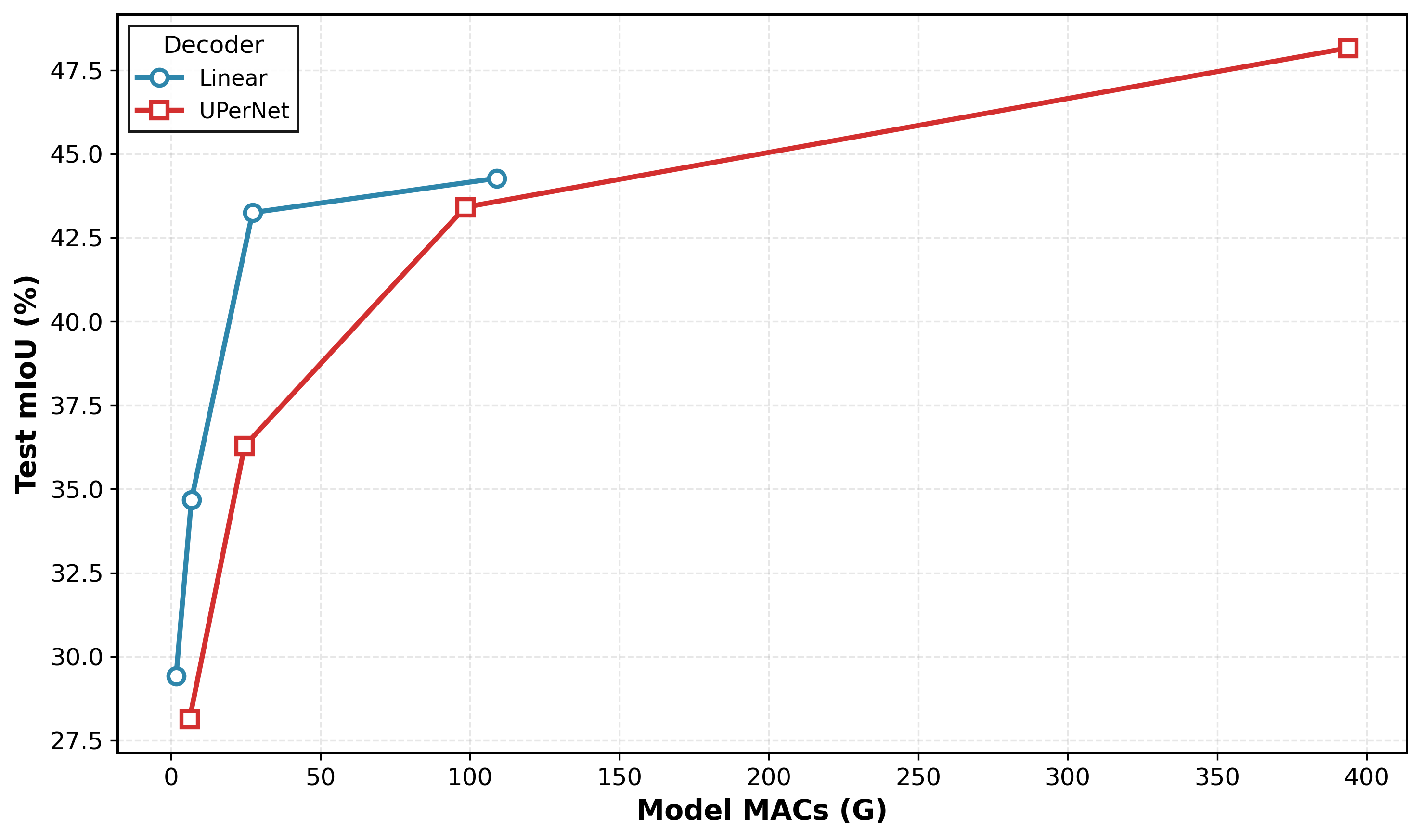}
    \caption{MADOS - Model MACs}
    \label{fig:mados_macs}
\end{subfigure}

\vspace{0.5cm}

\begin{subfigure}[b]{0.48\textwidth}
    \centering
    \includegraphics[width=\textwidth]{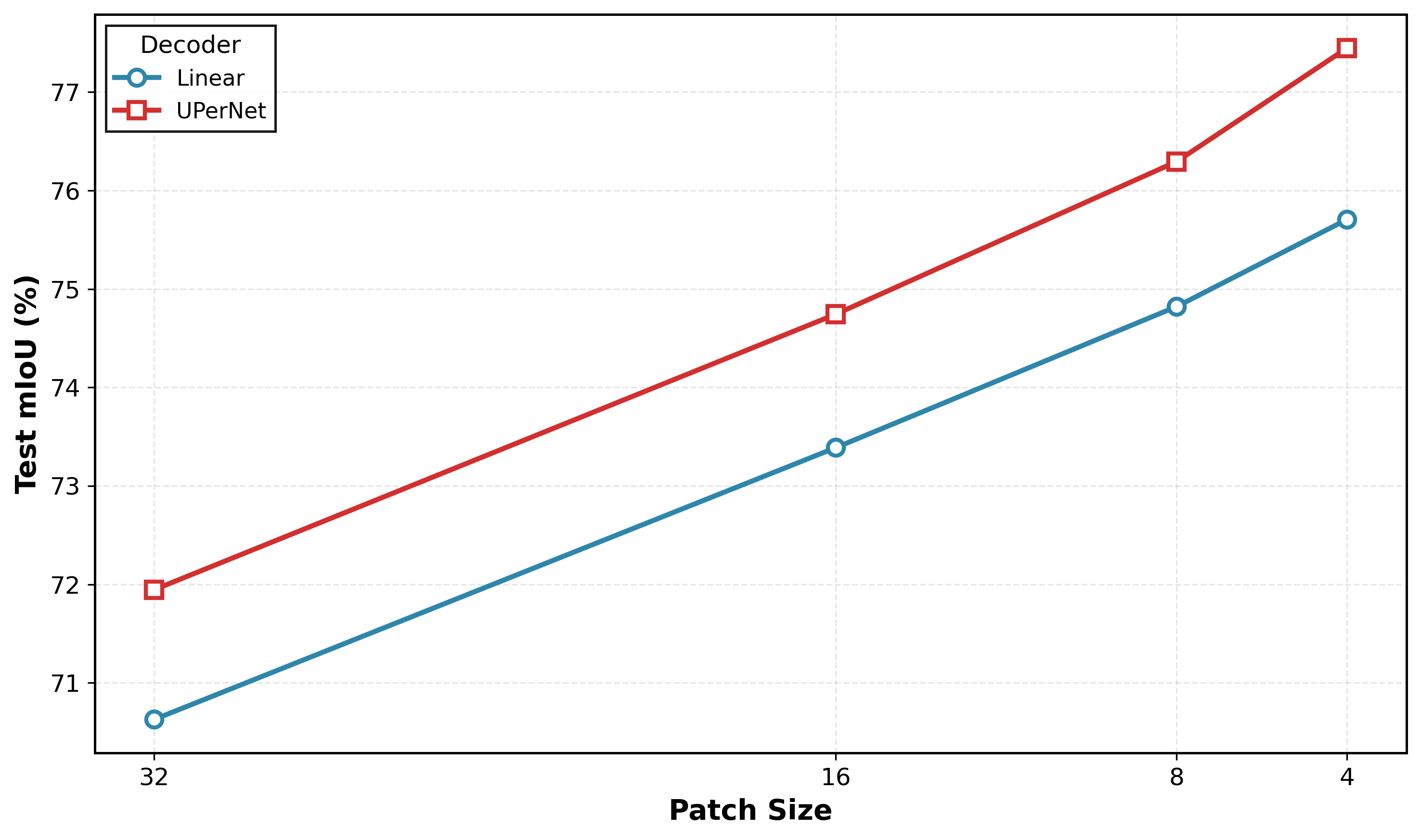}
    \caption{HLS Burns - Patch Size}
    \label{fig:hlsburns_patch}
\end{subfigure}
\hfill
\begin{subfigure}[b]{0.48\textwidth}
    \centering
    \includegraphics[width=\textwidth]{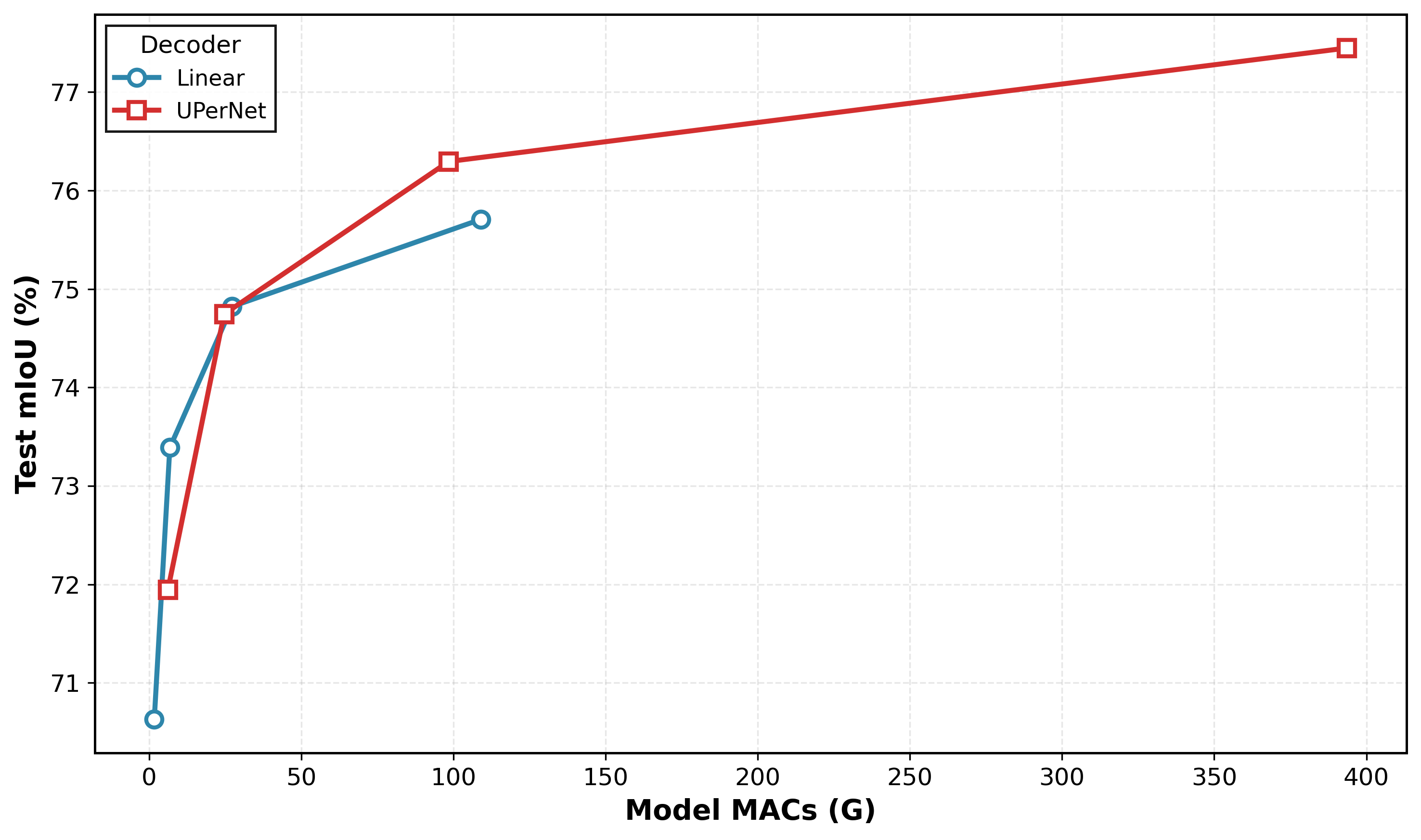}
    \caption{HLS Burns - Model MACs}
    \label{fig:hlsburns_macs}
\end{subfigure}

\vspace{0.5cm}

\begin{subfigure}[b]{0.48\textwidth}
    \centering
    \includegraphics[width=\textwidth]{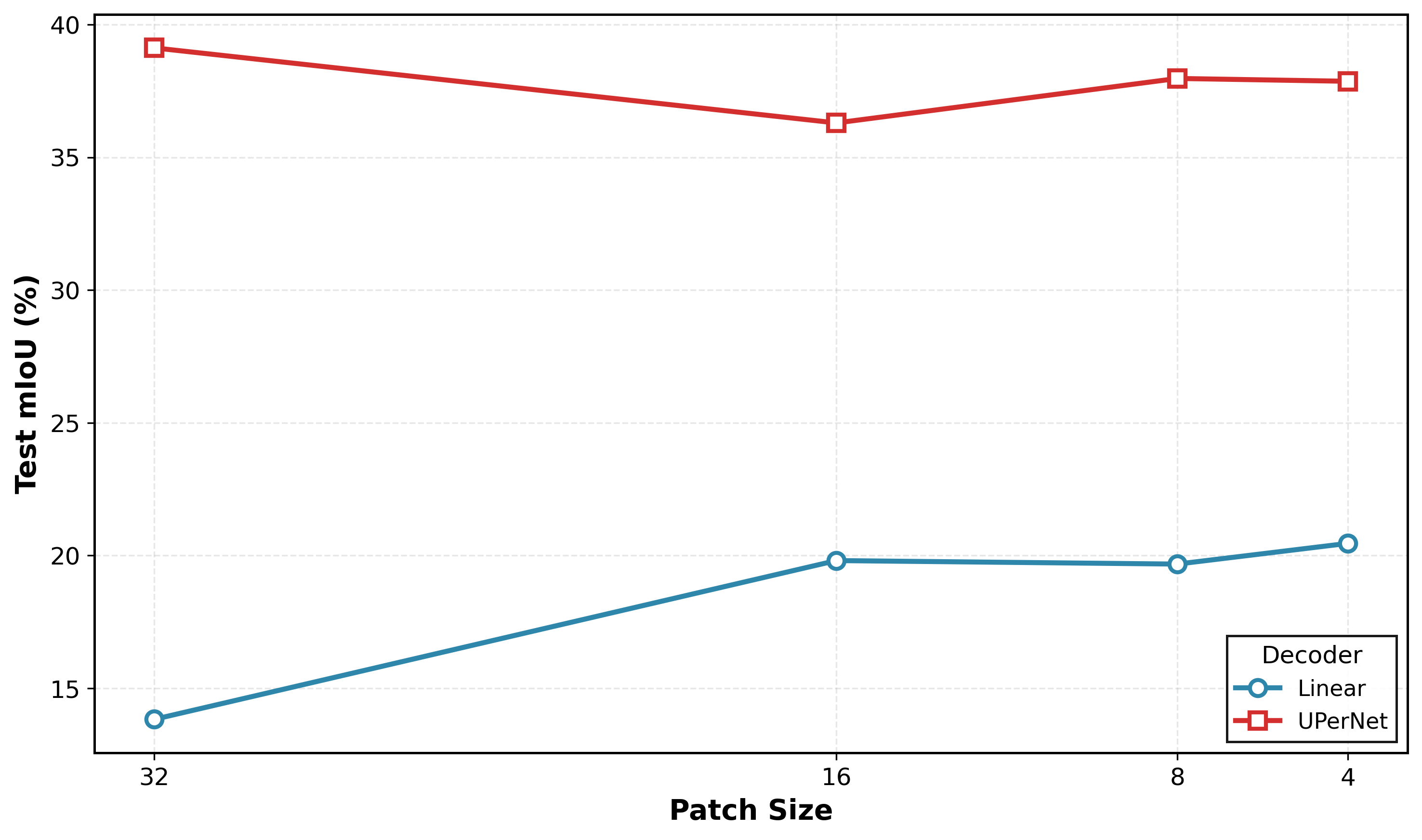}
    \caption{AI4Smallfarms - Patch Size}
    \label{fig:ai4smallfarms_patch}
\end{subfigure}
\hfill
\begin{subfigure}[b]{0.48\textwidth}
    \centering
    \includegraphics[width=\textwidth]{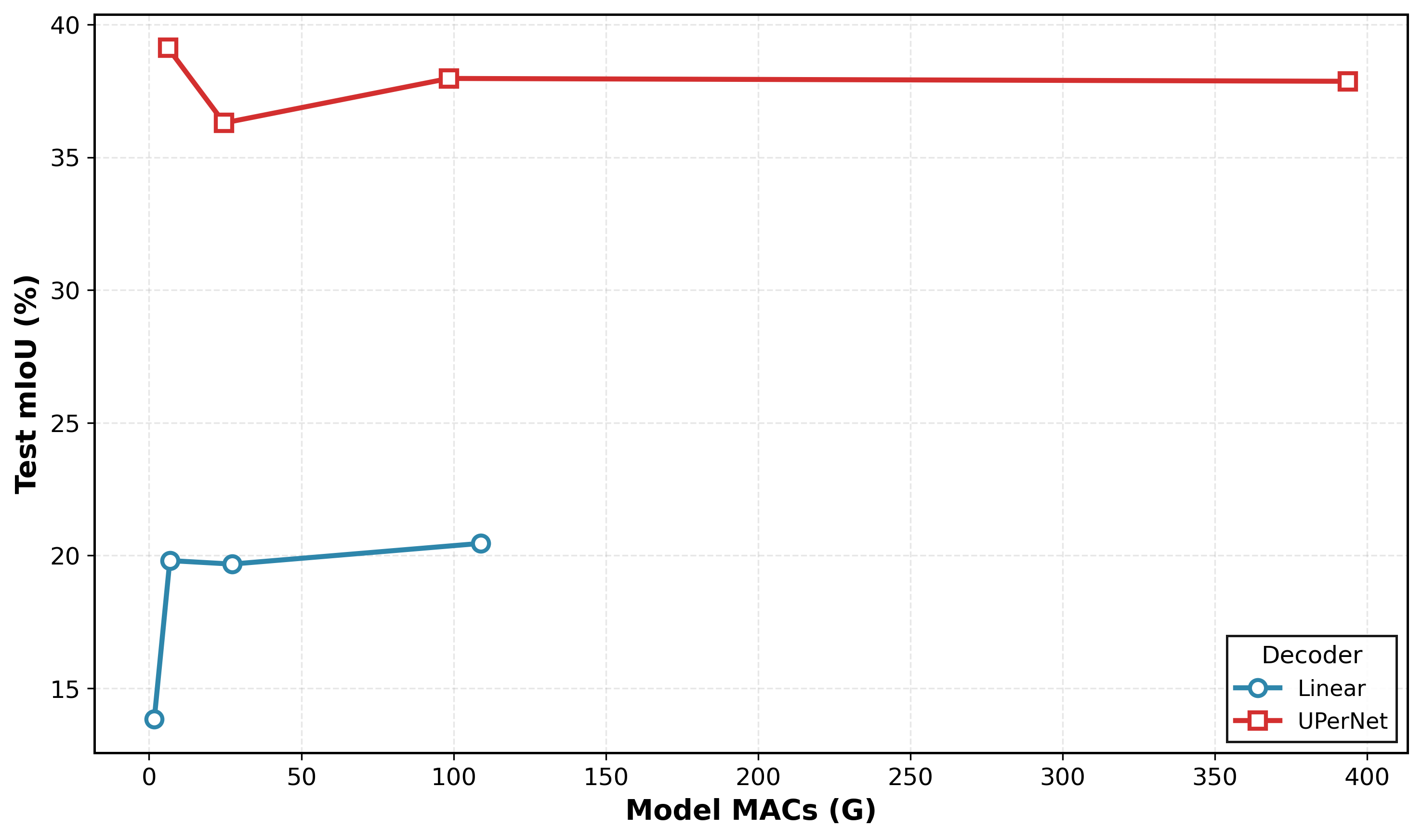}
    \caption{AI4Smallfarms - Model MACs}
    \label{fig:ai4smallfarms_macs}
\end{subfigure}

\caption{Model performance across different datasets. Left column shows test mIoU versus patch size, right column shows the respective test mIoU versus model MACs (G). Using THOR Base frozen encoder and linear decoder $\sim 0.2$ M parameters (blue circles) and UPerNet decoder $\sim 67-109$M parameters (red squares) are compared across four benchmark datasets. Using a fixed input size of 128, and concatenation of feature maps. All experiments run with 10\% training data.}
\label{fig:patch_size_ablation_grid}
\end{figure*}

As illustrated in Figs.~\ref{fig:sen1floods11_patch}, \ref{fig:mados_patch}, \ref{fig:hlsburns_patch}, and \ref{fig:ai4smallfarms_patch}, while the UperNet decoder yields a performance boost in certain configurations, the linear decoder achieves competitive accuracy levels that are frequently on par with the much larger architecture. Critically, when analyzing the computational burden (Figs.~\ref{fig:sen1floods11_macs}, \ref{fig:mados_macs}, and \ref{fig:hlsburns_macs}), the advantage of the linear approach becomes clear. As detailed in Tab.~\ref{tab:extended_snow_results}, the UperNet architecture requires approximately $1000\times$ more parameters than the linear decoder. This drastic reduction in decoder parameter count validates THOR as a true foundation model. The ability of a simple linear probe to match a complex non-linear decoder indicates that the pre-trained encoder produces highly semantic, linearly separable features. It suggests that in data-limited regimes, the heavy UperNet decoder is largely redundant and potentially prone to overfitting, whereas THOR's dense representations may be deployed with minimal adaptation.

\subsection{Snow use-case}
We evaluate the regression capability of THOR on the fractional snow cover task (Tab.~\ref{tab:extended_snow_results}). A linear decoder (using TerraTorch's LinearDecoder) trained on frozen THOR features consistently outperforms the fully supervised UNet baseline (RMSE 12.4). Notably, THOR-Base with a linear decoder achieves the state-of-the-art RMSE of 9.88, marginally surpassing the UPerNet head (RMSE 9.90). Most critically, the linear decoder achieves this performance using only 24.6k parameters, compared to the 22.9M parameters required by the UPerNet head. This ~1000x reduction in decoder complexity demonstrates that THOR's pre-trained representations are linearly separable and semantically rich, requiring minimal adaptation for downstream physical variable mapping.

\begin{table*}{}   
    \centering
    \caption{RMSE snow cover fraction. Image size $128\times 128$ and concatenated the tokens of the 500 m and 1000 m bands.}
    \begin{tabular}{llcccc}
    \toprule
         Decoder & Encoder & Patch size & No. dec. param. &Tot. no. param. & RMSE \\ %
     \midrule
         UNet &  & & & 24.4M & 12.4 \\
         UPerNet & THOR Base & 16x16 & 22.9M & 0.1G & 14.0 \\ 
                 &   & 8x8  &  & & 12.4 \\
                 &   & 4x4 &  & & 9.90 \\
                 & THOR Tiny & 4x4 & 8.6M  & 16.2M &  10.5 \\ 
                 & THOR Small & 4x4 & 12.1M & 37.9M & \textbf{9.69} \\ 
                 & THOR Large & 4x4 & 33.1M & 0.3G & 12.2 \\ 
         Linear decoder & THOR Base & 4x4 & 24.6k & 94.2M & \underline{9.88} \\  
                & THOR Tiny &  4x4 & 6.1k & 7.6M & 11.5 \\ 
                & THOR Small &  4x4 & 12.3k & 25.8M & 10.9 \\ 
                & THOR Large &  4x4 & 32.8k & 0.3G & 10.3 \\ 
     \bottomrule
    \end{tabular}
    \label{tab:extended_snow_results}
\end{table*}

\subsection{ERA5 Land analysis}
\begin{figure*}[t]
    \centering
    \includegraphics[width=1.0\textwidth]{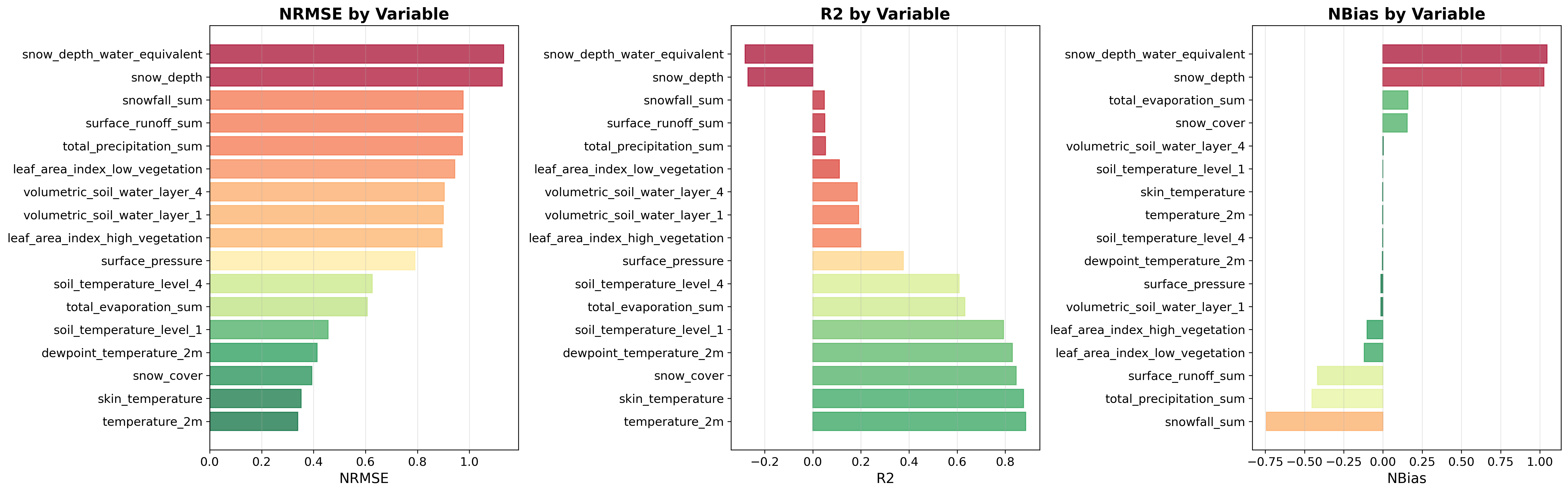}
    \caption{Comparison of NRMSE, R², and normalized bias for 17 ERA5 variables predicted from satellite embeddings.  Color scale ranges from green (good performance) to red (poor performance), with bias colored to highlight deviations from zero.}
    \label{fig:era5_metrics}
\end{figure*}

\begin{figure*}[t]
    \centering
    \includegraphics[width=0.95\textwidth]{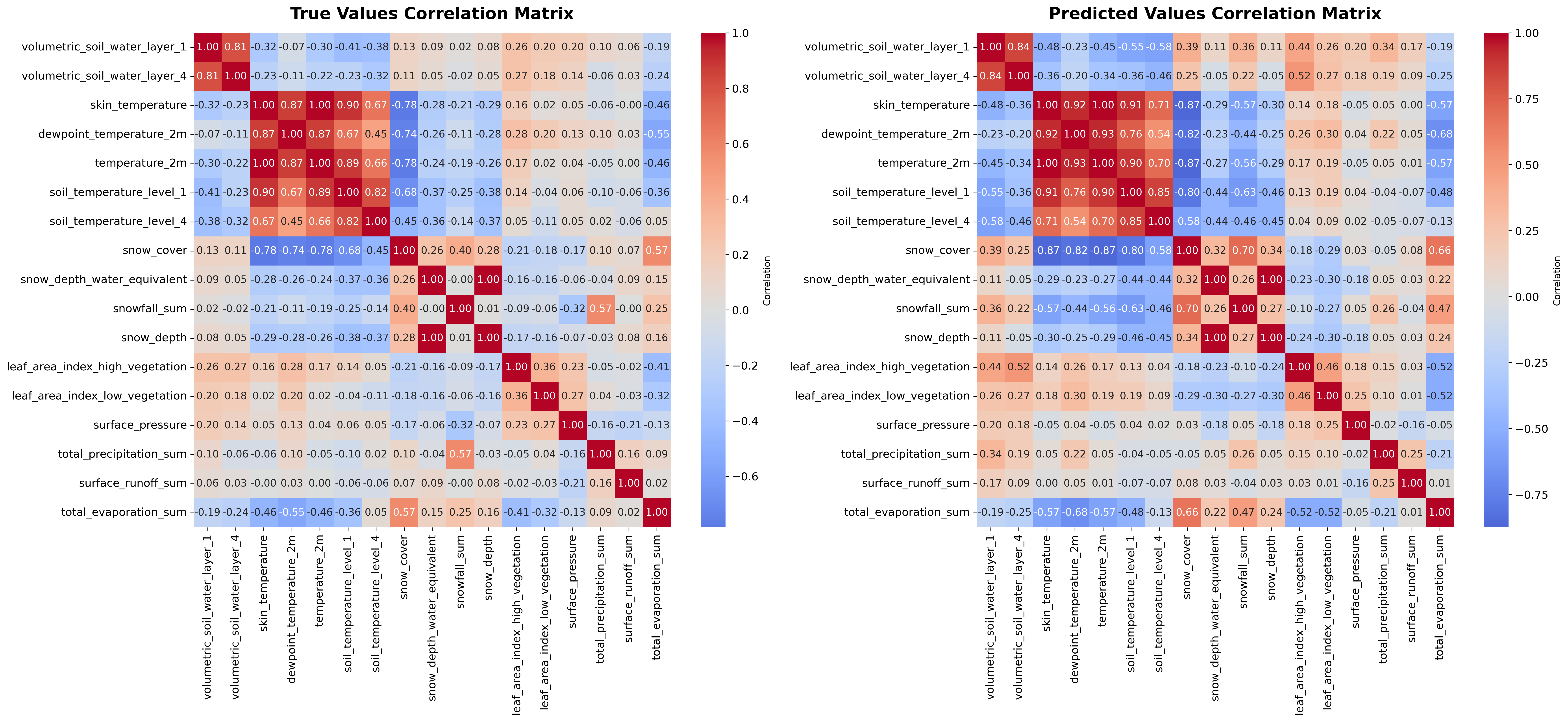}
    \caption{Comparison of inter-variable correlations for ground truth (left) and model-predicted (right) ERA5 variables.}
    \label{fig:era5_correlation_matrix}
\end{figure*}

\begin{figure*}[t]
    \centering
    \includegraphics[width=1.0\textwidth]{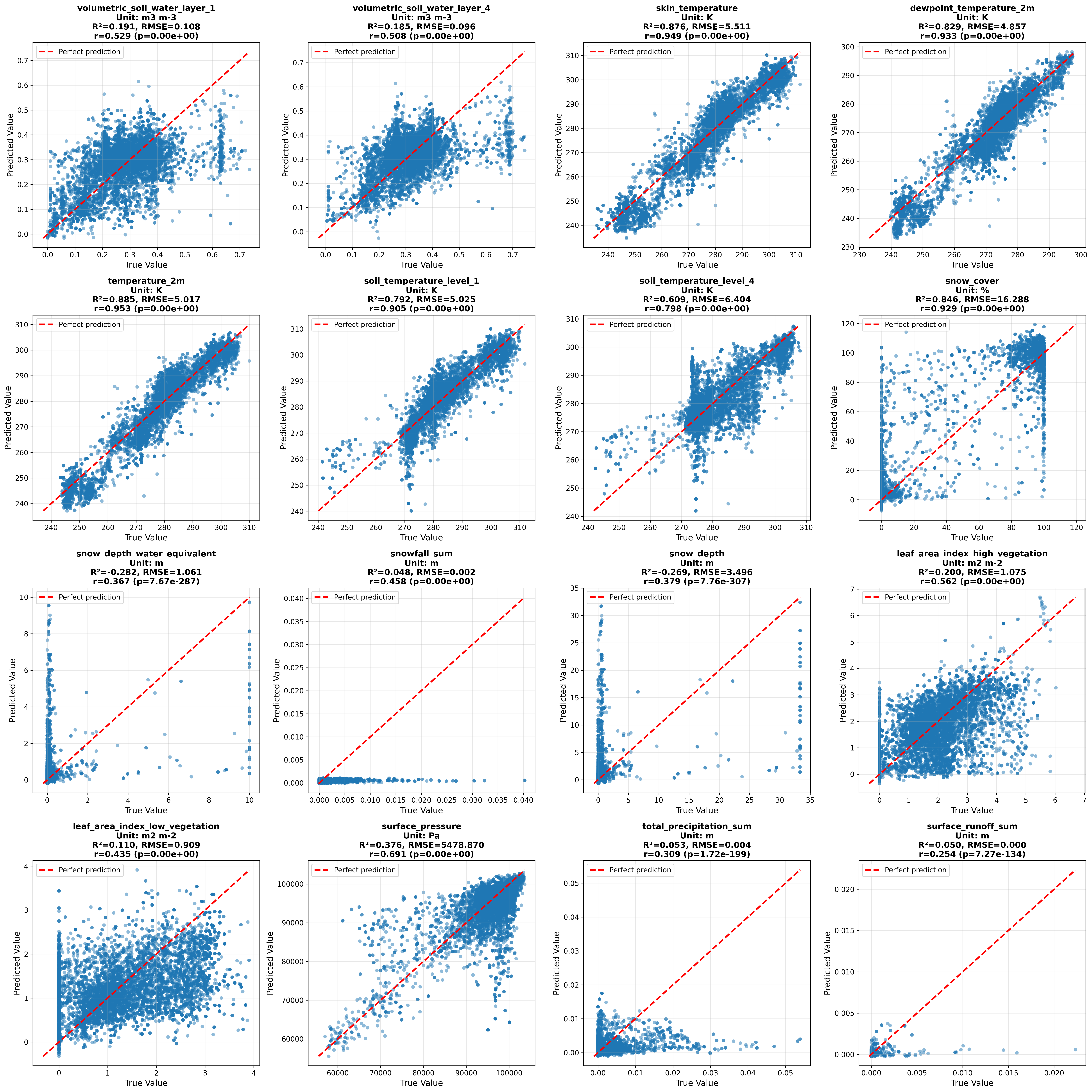}
    \caption{Model predictions plotted against ground truth observations for 17 ERA5 variables. Red dashed lines indicate perfect predictions.}
    \label{fig:era5_scatter}
\end{figure*}

To validate the climate-awareness of the frozen encoder, we analyze the performance of the linear probe on the holdout set against the ground truth ERA5-Land daily statistics variables by sampling random crops with a ground cover of 11520~m and extracting Sentinel-3 OLCI and SLSTR.. Fig.~\ref{fig:era5_scatter} presents scatter plots for these targets, revealing a clear distinction in performance: thermodynamic state variables (e.g., \verb+temperature_2m+, \verb+surface_pressure+) exhibit strong linearity and tight clustering ($R^2 > 0.8$), whereas stochastic, accumulated phenomena (e.g., \verb+snow_depth+, \verb+total_precipitation+) remain challenging to regress from instantaneous optical/SAR snapshots. This trend is quantified in Fig. \ref{fig:era5_metrics}, which shows low NRMSE and high $R^2$ for thermal and vegetation indices, contrasting with higher error rates for hydrological variables. However, the structural fidelity of the learned representation is confirmed in the correlation matrix of the predicted ERA5-Land values closely mirrors that of the ground truth , demonstrating that THOR successfully captures the physical inter-dependencies between these climatic variables (such as the coupling between soil moisture and temperature) (Fig.~\ref{fig:era5_correlation_matrix}). This suggests that the encoder moves beyond visual texture matching to embed the broad climatological context required for downstream climate applications.

\renewcommand{\refname}{Supplementary references} 
\putbib

\end{bibunit}


\begin{thebibliography}{32}
\providecommand{\natexlab}[1]{#1}
\providecommand{\url}[1]{\texttt{#1}}
\expandafter\ifx\csname urlstyle\endcsname\relax
  \providecommand{\doi}[1]{doi: #1}\else
  \providecommand{\doi}{doi: \begingroup \urlstyle{rm}\Url}\fi

\bibitem[{Arino} et~al.(2012){Arino}, {Ramos Perez}, {Kalogirou}, {Bontemps}, {Defourny}, and {Van Bogaert}]{GlobCover2009}
Olivier {Arino}, Jose~Julio {Ramos Perez}, Vasileios {Kalogirou}, Sophie {Bontemps}, Pierre {Defourny}, and Eric {Van Bogaert}.
\newblock {Global Land Cover Map for 2009 (GlobCover 2009)}, 2012.

\bibitem[Assran et~al.(2023)Assran, Duval, Misra, Bojanowski, Vincent, Rabbat, LeCun, and Ballas]{assran2023jepa}
Mahmoud Assran, Quentin Duval, Ishan Misra, Piotr Bojanowski, Pascal Vincent, Michael Rabbat, Yann LeCun, and Nicolas Ballas.
\newblock Self-supervised learning from images with a joint-embedding predictive architecture.
\newblock In \emph{Proceedings of the IEEE/CVF Conference on Computer Vision and Pattern Recognition}, pages 15619--15629, 2023.

\bibitem[Astruc et~al.(2025)Astruc, Gonthier, Mallet, and Landrieu]{astruc2025anysat}
Guillaume Astruc, Nicolas Gonthier, Clement Mallet, and Loic Landrieu.
\newblock {AnySat}: One earth observation model for many resolutions, scales, and modalities.
\newblock In \emph{Proceedings of the Computer Vision and Pattern Recognition Conference}, pages 19530--19540, 2025.

\bibitem[Beyer et~al.(2023)Beyer, Izmailov, Kolesnikov, Caron, Kornblith, Zhai, Minderer, Tschannen, Alabdulmohsin, and Pavetic]{beyer2023flexivit}
Lucas Beyer, Pavel Izmailov, Alexander Kolesnikov, Mathilde Caron, Simon Kornblith, Xiaohua Zhai, Matthias Minderer, Michael Tschannen, Ibrahim Alabdulmohsin, and Filip Pavetic.
\newblock {FlexiViT}: {One} model for all patch sizes.
\newblock In \emph{Proceedings of the IEEE/CVF Conference on Computer Vision and Pattern Recognition}, pages 14496--14506, 2023.

\bibitem[Cong et~al.(2022)Cong, Khanna, Meng, Liu, Rozi, He, Burke, Lobell, and Ermon]{cong2022satmae}
Yezhen Cong, Samar Khanna, Chenlin Meng, Patrick Liu, Erik Rozi, Yutong He, Marshall Burke, David Lobell, and Stefano Ermon.
\newblock Satmae: Pre-training transformers for temporal and multi-spectral satellite imagery.
\newblock \emph{Advances in Neural Information Processing Systems}, 35:\penalty0 197--211, 2022.

\bibitem[Dosovitskiy et~al.(2020)Dosovitskiy, Beyer, Kolesnikov, Weissenborn, Zhai, Unterthiner, Dehghani, Minderer, Heigold, Gelly, et~al.]{dosovitskiy2020image}
Alexey Dosovitskiy, Lucas Beyer, Alexander Kolesnikov, Dirk Weissenborn, Xiaohua Zhai, Thomas Unterthiner, Mostafa Dehghani, Matthias Minderer, Georg Heigold, Sylvain Gelly, et~al.
\newblock An image is worth 16x16 words: Transformers for image recognition at scale.
\newblock \emph{arXiv preprint arXiv:2010.11929}, 2020.

\bibitem[Friedl and Sulla{-}Menashe(2022)]{MCD12Q1_v061}
Mark~A. Friedl and Damien Sulla{-}Menashe.
\newblock {MODIS/Terra{+}Aqua Land Cover Type Yearly L3 Global 500 m SIN Grid V061 (MCD12Q1)}.
\newblock NASA LP DAAC, 2022.

\bibitem[Fuller et~al.(2023)Fuller, Millard, and Green]{fuller2023croma}
Anthony Fuller, Koreen Millard, and James Green.
\newblock {CROMA}: {Remote} sensing representations with contrastive radar-optical masked autoencoders.
\newblock \emph{Advances in Neural Information Processing Systems}, 36:\penalty0 5506--5538, 2023.

\bibitem[Hansen et~al.(2013)Hansen, Potapov, Moore, Hancher, Turubanova, Tyukavina, Thau, Stehman, Goetz, Loveland, et~al.]{hansen2013forest}
Matthew~C Hansen, Peter~V Potapov, Rebecca Moore, Matt Hancher, Svetlana~A Turubanova, Alexandra Tyukavina, David Thau, Stephen~V Stehman, Scott~J Goetz, Thomas~R Loveland, et~al.
\newblock High-resolution global maps of 21st-century forest cover change.
\newblock \emph{science}, 342\penalty0 (6160):\penalty0 850--853, 2013.

\bibitem[He et~al.(2022)He, Chen, Xie, Li, Doll{\'a}r, and Girshick]{he2022masked}
Kaiming He, Xinlei Chen, Saining Xie, Yanghao Li, Piotr Doll{\'a}r, and Ross Girshick.
\newblock Masked autoencoders are scalable vision learners.
\newblock In \emph{Proceedings of the IEEE/CVF conference on computer vision and pattern recognition}, pages 16000--16009, 2022.

\bibitem[Hollmann et~al.(2013)Hollmann, Merchant, Saunders, Downy, Buchwitz, Cazenave, Chuvieco, Defourny, de~Leeuw, Forsberg, et~al.]{hollmann2013esa}
Rainer Hollmann, Chris~J Merchant, Roger Saunders, Catherine Downy, Michael Buchwitz, Anny Cazenave, Emilio Chuvieco, Pierre Defourny, Gerrit de Leeuw, Ren{\'e} Forsberg, et~al.
\newblock The {ESA} climate change initiative: Satellite data records for essential climate variables.
\newblock \emph{Bulletin of the American Meteorological Society}, 94\penalty0 (10):\penalty0 1541--1552, 2013.

\bibitem[Irvin et~al.(2023)Irvin, Tao, Zhou, Ma, Nashold, Liu, and Ng]{irvin2023usat}
Jeremy Irvin, Lucas Tao, Joanne Zhou, Yuntao Ma, Langston Nashold, Benjamin Liu, and Andrew~Y Ng.
\newblock {USat}: {A} unified self-supervised encoder for multi-sensor satellite imagery.
\newblock \emph{arXiv preprint arXiv:2312.02199}, 2023.

\bibitem[Jakubik et~al.(2025)Jakubik, Yang, Blumenstiel, Scheurer, Sedona, Maurogiovanni, Bosmans, Dionelis, Marsocci, Kopp, et~al.]{jakubik2025terramind}
Johannes Jakubik, Felix Yang, Benedikt Blumenstiel, Erik Scheurer, Rocco Sedona, Stefano Maurogiovanni, Jente Bosmans, Nikolaos Dionelis, Valerio Marsocci, Niklas Kopp, et~al.
\newblock Terramind: Large-scale generative multimodality for earth observation.
\newblock \emph{arXiv preprint arXiv:2504.11171}, 2025.

\bibitem[Kraus et~al.(2024)Kraus, Kenyon-Dean, Saberian, Fallah, McLean, Leung, Sharma, Khan, Balakrishnan, Celik, Beaini, Sypetkowski, Cheng, Morse, Makes, Mabey, and Earnshaw]{kraus_masked_2024}
Oren Kraus, Kian Kenyon-Dean, Saber Saberian, Maryam Fallah, Peter McLean, Jess Leung, Vasudev Sharma, Ayla Khan, Jia Balakrishnan, Safiye Celik, Dominique Beaini, Maciej Sypetkowski, Chi~Vicky Cheng, Kristen Morse, Maureen Makes, Ben Mabey, and Berton Earnshaw.
\newblock Masked {Autoencoders} for {Microscopy} are {Scalable} {Learners} of {Cellular} {Biology}, 2024.
\newblock arXiv:2404.10242 [cs].

\bibitem[Li et~al.(2025)Li, Li, Ghamisi, and Hong]{li2025fleximo}
Xuyang Li, Chenyu Li, Pedram Ghamisi, and Danfeng Hong.
\newblock {FlexiMo}: A flexible remote sensing foundation model.
\newblock \emph{arXiv preprint arXiv:2503.23844}, 2025.

\bibitem[Longepe et~al.(2025)Longepe, Alemohammad, Anghelea, Brunschwiler, Camps-Valls, Cavallaro, Chanussot, Delgado, Demir, Dionelis, Fraccaro, Jungbluth, Kennedy, Marsocci, Ramasubramanian, Ramos-Pollan, Roy, Sümbül, Tuia, Zhu, and Ramachandran]{Longepe_2025}
Nicolas Longepe, Hamed Alemohammad, Anca Anghelea, Thomas Brunschwiler, Gustau Camps-Valls, Gabriele Cavallaro, Jocelyn Chanussot, Jose~Manuel Delgado, Begüm Demir, Nikolaos Dionelis, Paolo Fraccaro, Anna Jungbluth, Robert~E. Kennedy, Valerio Marsocci, Muthukumaran Ramasubramanian, Raul Ramos-Pollan, Sujit Roy, Gencer Sümbül, Devis Tuia, Xiao~Xiang Zhu, and Rahul Ramachandran.
\newblock Earth action in transition: Highlights from the 2025 esa-nasa international workshop on ai foundation models for eo.
\newblock 2025.

\bibitem[Marsocci et~al.(2024)Marsocci, Jia, Bellier, Kerekes, Zeng, Hafner, Gerard, Brune, Yadav, Shibli, et~al.]{marsocci2024pangaea}
Valerio Marsocci, Yuru Jia, Georges~Le Bellier, David Kerekes, Liang Zeng, Sebastian Hafner, Sebastian Gerard, Eric Brune, Ritu Yadav, Ali Shibli, et~al.
\newblock Pangaea: A global and inclusive benchmark for geospatial foundation models.
\newblock \emph{arXiv preprint arXiv:2412.04204}, 2024.

\bibitem[McCabe et~al.(2017)McCabe, Rodell, Alsdorf, Miralles, Uijlenhoet, Wagner, Lucieer, Houborg, Verhoest, Franz, et~al.]{mccabe2017hydrology}
Matthew~F McCabe, Matthew Rodell, Douglas~E Alsdorf, Diego~G Miralles, Remko Uijlenhoet, Wolfgang Wagner, Arko Lucieer, Rasmus Houborg, Niko~EC Verhoest, Trenton~E Franz, et~al.
\newblock The future of earth observation in hydrology.
\newblock \emph{Hydrology and earth system sciences}, 21\penalty0 (7):\penalty0 3879--3914, 2017.

\bibitem[Mu\~noz Sabater et~al.(2021)Mu\~noz Sabater, Dutra, Agust\'{\i}-Panareda, Albergel, Arduini, Balsamo, Boussetta, Choulga, Harrigan, Hersbach, Martens, Miralles, Piles, Rodr\'{\i}guez-Fern\'andez, Zsoter, Buontempo, and Th\'epaut]{MunozSabater2021_ERA5Land_ESSD}
J. Mu\~noz Sabater, E. Dutra, A. Agust\'{\i}-Panareda, C. Albergel, G. Arduini, G. Balsamo, S. Boussetta, M. Choulga, S. Harrigan, H. Hersbach, B. Martens, D.~G. Miralles, M. Piles, N.~J. Rodr\'{\i}guez-Fern\'andez, E. Zsoter, C. Buontempo, and J.-N. Th\'epaut.
\newblock Era5-land: a state-of-the-art global reanalysis dataset for land applications.
\newblock \emph{Earth System Science Data}, 13\penalty0 (9):\penalty0 4349--4383, 2021.

\bibitem[Mu{\~n}oz{-}Sabater(2019)]{ERA5Land_dataset}
Joaqu{\'\i}n Mu{\~n}oz{-}Sabater.
\newblock {ERA5{-}Land hourly data from 1950 to present}.
\newblock Copernicus Climate Change Service (C3S) Climate Data Store, 2019.

\bibitem[Nedungadi et~al.(2024)Nedungadi, Kariryaa, Oehmcke, Belongie, Igel, and Lang]{nedungadi2024mmearth}
Vishal Nedungadi, Ankit Kariryaa, Stefan Oehmcke, Serge Belongie, Christian Igel, and Nico Lang.
\newblock {MMEarth}: {Exploring} multi-modal pretext tasks for geospatial representation learning.
\newblock In \emph{European Conference on Computer Vision}, pages 164--182. Springer, 2024.

\bibitem[Reed et~al.(2023)Reed, Gupta, Li, Brockman, Funk, Clipp, Keutzer, Candido, Uyttendaele, and Darrell]{reed2023scale}
Colorado~J Reed, Ritwik Gupta, Shufan Li, Sarah Brockman, Christopher Funk, Brian Clipp, Kurt Keutzer, Salvatore Candido, Matt Uyttendaele, and Trevor Darrell.
\newblock Scale-mae: A scale-aware masked autoencoder for multiscale geospatial representation learning.
\newblock In \emph{Proceedings of the IEEE/CVF International Conference on Computer Vision}, pages 4088--4099, 2023.

\bibitem[Reichstein et~al.(2019)Reichstein, Camps-Valls, Stevens, Jung, Denzler, Carvalhais, and Prabhat]{reichstein2019deep}
Markus Reichstein, Gustau Camps-Valls, Bjorn Stevens, Martin Jung, Joachim Denzler, Nuno Carvalhais, and F Prabhat.
\newblock Deep learning and process understanding for data-driven earth system science.
\newblock \emph{Nature}, 566\penalty0 (7743):\penalty0 195--204, 2019.

\bibitem[Rolf et~al.(2024)Rolf, Klemmer, Robinson, and Kerner]{rolf2024position}
Esther Rolf, Konstantin Klemmer, Caleb Robinson, and Hannah Kerner.
\newblock Position: Mission critical--satellite data is a distinct modality in machine learning.
\newblock In \emph{Forty-first International Conference on Machine Learning}, 2024.

\bibitem[Ruiping et~al.(2024)Ruiping, Kun, Shaohua, Jian, and Zhen]{ruiping2024vitupernet}
Yang Ruiping, Liu Kun, Xu Shaohua, Yin Jian, and Zhang Zhen.
\newblock Vit-upernet: a hybrid vision transformer with unified-perceptual-parsing network for medical image segmentation.
\newblock \emph{Complex \& Intelligent Systems}, 10\penalty0 (3):\penalty0 3819--3831, 2024.

\bibitem[Szwarcman et~al.(2024)Szwarcman, Roy, Fraccaro, G{\'\i}slason, Blumenstiel, Ghosal, de~Oliveira, Almeida, Sedona, Kang, et~al.]{szwarcman2024prithvi}
Daniela Szwarcman, Sujit Roy, Paolo Fraccaro, {\TH}orsteinn~El{\'\i} G{\'\i}slason, Benedikt Blumenstiel, Rinki Ghosal, Pedro~Henrique de Oliveira, Joao Lucas de~Sousa Almeida, Rocco Sedona, Yanghui Kang, et~al.
\newblock Prithvi-eo-2.0: A versatile multi-temporal foundation model for earth observation applications.
\newblock \emph{arXiv preprint arXiv:2412.02732}, 2024.

\bibitem[Tseng et~al.(2025)Tseng, Fuller, Reil, Herzog, Beukema, Bastani, Green, Shelhamer, Kerner, and Rolnick]{tseng2025galileo}
Gabriel Tseng, Anthony Fuller, Marlena Reil, Henry Herzog, Patrick Beukema, Favyen Bastani, James~R Green, Evan Shelhamer, Hannah Kerner, and David Rolnick.
\newblock Galileo: {Learning} global and local features in pretrained remote sensing models.
\newblock \emph{arXiv preprint arXiv:2502.09356}, 2025.

\bibitem[Wang et~al.(2022)Wang, Albrecht, Braham, Mou, and Zhu]{wang2022self}
Yi Wang, Conrad~M Albrecht, Nassim Ait~Ali Braham, Lichao Mou, and Xiao~Xiang Zhu.
\newblock Self-supervised learning in remote sensing: A review.
\newblock \emph{IEEE Geosci. Remote Sensing Mag.}, 10\penalty0 (4):\penalty0 213--247, 2022.

\bibitem[Wang et~al.(2024)Wang, Albrecht, and Zhu]{wang_multi-label_2024}
Yi Wang, Conrad~M. Albrecht, and Xiao~Xiang Zhu.
\newblock Multi-{Label} {Guided} {Soft} {Contrastive} {Learning} for {Efficient} {Earth} {Observation} {Pretraining}, 2024.
\newblock arXiv:2405.20462 [cs] version: 1.

\bibitem[Wang et~al.(2025)Wang, Xiong, Liu, Stewart, Dujardin, Bountos, Zavras, Gerken, Papoutsis, Leal-Taixé, and Zhu]{wang_copernicus_fm_2025}
Yi Wang, Zhitong Xiong, Chenying Liu, Adam~J. Stewart, Thomas Dujardin, Nikolaos~Ioannis Bountos, Angelos Zavras, Franziska Gerken, Ioannis Papoutsis, Laura Leal-Taixé, and Xiao~Xiang Zhu.
\newblock Towards a {Unified} {Copernicus} {Foundation} {Model} for {Earth} {Vision}, 2025.
\newblock arXiv:2503.11849 [cs].

\bibitem[Xiong et~al.(2024)Xiong, Wang, Zhang, Stewart, Hanna, Borth, Papoutsis, Saux, Camps-Valls, and Zhu]{xiong_dofa_neural_2024}
Zhitong Xiong, Yi Wang, Fahong Zhang, Adam~J. Stewart, Joëlle Hanna, Damian Borth, Ioannis Papoutsis, Bertrand~Le Saux, Gustau Camps-Valls, and Xiao~Xiang Zhu.
\newblock Neural {Plasticity}-{Inspired} {Foundation} {Model} for {Observing} the {Earth} {Crossing} {Modalities}, 2024.
\newblock arXiv:2403.15356 [cs].

\bibitem[Zanaga et~al.(2021)Zanaga, Van De~Kerchove, De~Keersmaecker, Souverijns, Brockmann, Quast, Wevers, Grosu, Paccini, Vergnaud, Cartus, Santoro, Fritz, Georgieva, Lesiv, Carter, Herold, Li, Tsendbazar, Ramoino, and Arino]{WorldCover2020_v100}
Daniele Zanaga, Ruben Van De~Kerchove, Wanda De~Keersmaecker, Niels Souverijns, Carsten Brockmann, Ralf Quast, Jan Wevers, Alex Grosu, Audrey Paccini, Sylvain Vergnaud, Oliver Cartus, Maurizio Santoro, Steffen Fritz, Ivelina Georgieva, Myroslava Lesiv, Sarah Carter, Martin Herold, Linlin Li, Nandin-Erdene Tsendbazar, Fabrizio Ramoino, and Olivier Arino.
\newblock Esa worldcover 10 m 2020 v100, 2021.

\end{thebibliography}


\begin{thebibliography}{11}
\providecommand{\natexlab}[1]{#1}
\providecommand{\url}[1]{\texttt{#1}}
\expandafter\ifx\csname urlstyle\endcsname\relax
  \providecommand{\doi}[1]{doi: #1}\else
  \providecommand{\doi}{doi: \begingroup \urlstyle{rm}\Url}\fi

\bibitem[{Al Kader Hammoud} et~al.(2024){Al Kader Hammoud}, Das, Pizzati, Torr, Bibi, and Ghanem]{al2024pretraining}
Hasan~Abed {Al Kader Hammoud}, Tuhin Das, Fabio Pizzati, Philip~HS Torr, Adel Bibi, and Bernard Ghanem.
\newblock On pretraining data diversity for self-supervised learning.
\newblock In \emph{European Conference on Computer Vision}, pages 54--71. Springer, 2024.

\bibitem[{Arino} et~al.(2012){Arino}, {Ramos Perez}, {Kalogirou}, {Bontemps}, {Defourny}, and {Van Bogaert}]{GlobCover2009}
Olivier {Arino}, Jose~Julio {Ramos Perez}, Vasileios {Kalogirou}, Sophie {Bontemps}, Pierre {Defourny}, and Eric {Van Bogaert}.
\newblock {Global Land Cover Map for 2009 (GlobCover 2009)}, 2012.

\bibitem[Friedl and Sulla{-}Menashe(2022)]{MCD12Q1_v061}
Mark~A. Friedl and Damien Sulla{-}Menashe.
\newblock {MODIS/Terra{+}Aqua Land Cover Type Yearly L3 Global 500 m SIN Grid V061 (MCD12Q1)}.
\newblock NASA LP DAAC, 2022.

\bibitem[Jakubik et~al.(2025)Jakubik, Yang, Blumenstiel, Scheurer, Sedona, Maurogiovanni, Bosmans, Dionelis, Marsocci, Kopp, et~al.]{jakubik2025terramind}
Johannes Jakubik, Felix Yang, Benedikt Blumenstiel, Erik Scheurer, Rocco Sedona, Stefano Maurogiovanni, Jente Bosmans, Nikolaos Dionelis, Valerio Marsocci, Niklas Kopp, et~al.
\newblock Terramind: Large-scale generative multimodality for earth observation.
\newblock \emph{arXiv preprint arXiv:2504.11171}, 2025.

\bibitem[Kraus et~al.(2024)Kraus, Kenyon-Dean, Saberian, Fallah, McLean, Leung, Sharma, Khan, Balakrishnan, Celik, Beaini, Sypetkowski, Cheng, Morse, Makes, Mabey, and Earnshaw]{kraus_masked_2024}
Oren Kraus, Kian Kenyon-Dean, Saber Saberian, Maryam Fallah, Peter McLean, Jess Leung, Vasudev Sharma, Ayla Khan, Jia Balakrishnan, Safiye Celik, Dominique Beaini, Maciej Sypetkowski, Chi~Vicky Cheng, Kristen Morse, Maureen Makes, Ben Mabey, and Berton Earnshaw.
\newblock Masked {Autoencoders} for {Microscopy} are {Scalable} {Learners} of {Cellular} {Biology}, 2024.
\newblock arXiv:2404.10242 [cs].

\bibitem[Lai et~al.(2023)Lai, Ahmed, Vijay, Jaroensri, Loo, Vyawahare, Agarwal, Jamil, Matias, Corrado, et~al.]{lai2023domain}
Jeremy Lai, Faruk Ahmed, Supriya Vijay, Tiam Jaroensri, Jessica Loo, Saurabh Vyawahare, Saloni Agarwal, Fayaz Jamil, Yossi Matias, Greg~S Corrado, et~al.
\newblock Domain-specific optimization and diverse evaluation of self-supervised models for histopathology.
\newblock \emph{arXiv preprint arXiv:2310.13259}, 2023.

\bibitem[Marsocci et~al.(2024)Marsocci, Jia, Bellier, Kerekes, Zeng, Hafner, Gerard, Brune, Yadav, Shibli, et~al.]{marsocci2024pangaea}
Valerio Marsocci, Yuru Jia, Georges~Le Bellier, David Kerekes, Liang Zeng, Sebastian Hafner, Sebastian Gerard, Eric Brune, Ritu Yadav, Ali Shibli, et~al.
\newblock Pangaea: A global and inclusive benchmark for geospatial foundation models.
\newblock \emph{arXiv preprint arXiv:2412.04204}, 2024.

\bibitem[Mets{\"a}m{\"a}ki et~al.(2015)Mets{\"a}m{\"a}ki, Pulliainen, Salminen, Luojus, Wiesmann, Solberg, B{\"o}ttcher, Hiltunen, and Ripper]{metsamaki2015globsnow}
Sari Mets{\"a}m{\"a}ki, Jouni Pulliainen, Miia Salminen, Kari Luojus, Andreas Wiesmann, Rune Solberg, Kristin B{\"o}ttcher, Mwaba Hiltunen, and Elisabeth Ripper.
\newblock {Introduction to GlobSnow Snow Extent} products with considerations for accuracy assessment.
\newblock \emph{Remote Sensing of Environment}, 156:\penalty0 96--108, 2015.

\bibitem[Oliver and Quegan(2004)]{oliver2004understanding}
Chris Oliver and Shaun Quegan.
\newblock \emph{Understanding synthetic aperture radar images}.
\newblock SciTech Publishing, 2004.

\bibitem[Ordonez et~al.(2024)Ordonez, Wade, Ravaut, and Waldeland]{ordonez2024towards}
A Ordonez, D Wade, C Ravaut, and AU Waldeland.
\newblock Towards a foundation model for seismic interpretation.
\newblock In \emph{85th EAGE Annual Conference \& Exhibition (including the Workshop Programme)}, pages 1--5. European Association of Geoscientists \& Engineers, 2024.

\bibitem[Zanaga et~al.(2022)Zanaga, Van De~Kerchove, Daems, De~Keersmaecker, Brockmann, Kirches, Wevers, Cartus, Santoro, Fritz, Lesiv, Herold, Tsendbazar, Xu, Ramoino, and Arino]{WorldCover2021_v200}
Daniele Zanaga, Ruben Van De~Kerchove, Dirk Daems, Wanda De~Keersmaecker, Carsten Brockmann, Grit Kirches, Jan Wevers, Oliver Cartus, Maurizio Santoro, Steffen Fritz, Myroslava Lesiv, Martin Herold, Nandin-Erdene Tsendbazar, Panpan Xu, Fabrizio Ramoino, and Olivier Arino.
\newblock Esa worldcover 10 m 2021 v200, 2022.

\end{thebibliography}
\end{document}